\documentclass[reprint,amsmath,amssymb,aps,prb,longbibliography]{revtex4-2}
\usepackage{graphicx} 
\usepackage{subcaption} 
\usepackage[labelsep=period,font=small,labelsep=period]{caption} 
\usepackage{ragged2e} 
\usepackage{float} 

\usepackage{latexsym, amsmath,amssymb,bm,euscript} 
\usepackage{xcolor} 
\usepackage[colorlinks=true,linkcolor=blue,citecolor=blue,urlcolor=blue]{hyperref}
\usepackage{ulem}

\usepackage[T1]{fontenc}

\begin{document}

\title{Fermi-liquid-like phase driven by next-nearest-neighbor couplings in a lightly doped kagome-lattice $t$-$J$ model}

\author{Xu-Yan Jia$^1$}
\author{Fan Yang$^2$}
\email{yangfan\_blg@bit.edu.cn}
\author{D. N. Sheng$^3$}
\email{donna.sheng1@csun.edu}
\author{Shou-Shu Gong$^{4,5}$}
\email{shoushu.gong@gbu.edu.cn}

\affiliation{$^1$School of Physics, Beihang University, Beijing 100191, China}
\affiliation{$^2$School of Physics, Beijing Institute of Technology, Beijing 100081, China}
\affiliation{$^3$Department of Physics and Astronomy, California State University Northridge, Northridge, California 91330, USA}
\affiliation{$^4$School of Physical Sciences, Great Bay University, Dongguan 523000, China}
\affiliation{$^5$Great Bay Institute for Advanced Study, Dongguan 523000, China}

\newcommand{\donna}[1]{\textcolor{blue}{#1}}

\date{\today}

\begin{abstract}
Due to the interplay between charge fluctuation and geometry frustration, the doped kagome-lattice Mott insulator is a fascinating platform to realize exotic quantum states.
Through the state-of-the-art density matrix renormalization group calculation, we explore the quantum phases of the lightly doped kagome-lattice $t$-$J$ model in the presence of the next-nearest-neighbor electron hopping $t_2$ and spin interaction $J_2$.
On the $L_y = 3$ cylinder ($L_y$ is the number of unit cells along the circumference direction), we establish a quantum phase diagram with tuning $t_2 > 0$ and $J_2 > 0$, showing an emergent Fermi-liquid-like phase driven by increased $t_2$ and $J_2$, at the neighbor of the previously identified charge density wave (CDW) phase.
Compared with the CDW phase, the charge order is significantly suppressed in the Fermi-liquid-like phase, and most correlation functions are greatly enhanced with power-law decay.
In particular, we find the absence of hole pairing and a strong three-sublattice magnetic correlation.
On the wider $L_y = 4$ cylinder, this Fermi-liquid-like phase persists at low doping levels, strongly suggesting that this state might be stable in the two-dimensional kagome system. 
\end{abstract}

\maketitle

\section{Introduction}

The kagome-lattice systems realize
an ideal platform to study exotic quantum states, including quantum spin liquid (QSL), unconventional superconductivity, and density wave orders~\cite{kagome-RV-nature_reviews_physics-2023,kagome-RV-nature-2022,kagome-AV3Sb5-National_Science_Review-2022,kagome-AV3Sb5-nature_Review_material-2024,kagome-Co3Sn2S2-Reviews_in_physics-2022,QSL-RV-Leon-2010}. 
In addition to lattice frustration, the electronic band structure of the kagome lattice features a topological band-touching point, van Hove singularity, and a flat band. 
The van Hove singularity and the flat band possess a high density of states, which may give novel quantum states even in the presence of a weak interaction~\cite{Flatband-HTC-1,Flatband-FQHE,Flatband-Wignercrystal,vHS-kagome-1,vHS-kagome-2,vHS-kagome-3,vHS-kagome-4}. 
Due to the interplay between band topology, geometry frustration, and electronic correlation, kagome-lattice materials have been extensively explored to investigate novel quantum phenomena and intertwined orders~\cite{kagome-Co3Sn2S2-sci-2019-2,kagome-Co3Sn2S2-sci-2019,kagome-CsV3Sb5-nature-2021,kagome-CsV3Sb5-nature-2021-2,kagome-Mn3Sn-nature-2015,kagome-TbMn6Sn6-nature-2020}.

QSLs have been one of the central subjects in the study of kagome-lattice systems. Various spin-$1/2$ kagome spin models and candidate materials exhibiting spin-liquid-like behaviors have been extensively studied (see the review articles~\cite{kagome-Herbertsmithite-RMP-2016,savary_2017,zhou_2017,broholm_2020}).
For the kagome model with only the nearest-neighbor (NN) Heisenberg interaction, various numerical results have consistently found a QSL ground state, but the nature of this QSL remains an outstanding issue~\cite{kagome-J1-VMC-2007,kagome-J1-VMC-2011,kagome-J1-VMC-2013,kagome-J1-VMC-2014,kagome-J1-DMRG-Sci-2011,jiang_2008,kagome-J1-DMRG-PRL-2012,liao_2017,kagome-J1-iDMRG-2017,lauchli_2019,kagome-J1J2J3-mean_field-2012,zhu_2018,sun_2024,pinaki_2025,kagome_spin_model_zw}. 
Interestingly, some additional perturbative interactions can easily stabilize a time-reversal-symmetry-breaking chiral spin liquid, which is an analog of the $\nu = 1/2$ Laughlin fractional quantum Hall state in spin systems~\cite{kalmeyer_1987}.
This chiral spin liquid can be obtained by introducing additional second- and third-neighbor exchange couplings~\cite{kagome-J1J2szJ3sz-heyinchen-2014,kagome-J1J2J3-gss-2014,kagome-J1J2J3-ssg-2015}, or a uniform scalar chiral interaction for the three spins of each triangle~\cite{kagome-J1Jchihz-Ncom-2014,kagome-J1J2J3/J1Jchi-VMC-2015}.

With doping, the interplay between charge and spin degrees of freedom may give rise to other novel quantum states.
Anderson's famous proposal to dope a QSL provides a promising framework for understanding unconventional superconductivity (SC) in cuprate superconductors~\cite{anderson_1987}. 
Furthermore, doping a CSL may even lead to a $d+id$-wave topological SC through the condensation of paired fractional quasiparticles~\cite{CSL-Laughlin-1988,CSL-xiaogang-1989,CSL-Fisher-1989}.
In recent years, the $d$-wave and $d+id$-wave SC have been identified by means of the unbiased density matrix renormalization group (DMRG) calculations in doped square-~\cite{Square-tt'JJ'-JiangHC-2021,Square-tt'JJ'-gss-2021,Square-tt'J-JiangShengtao-2021,Square-tt'JJ'-gss-2024,chen_2025,jiang_2023,jiang_2024} and triangular-lattice Mott insulators~\cite{jiang_2020,huang_2022,triangular-tt'JJ'-HuangYX-2023}, respectively.
These fascinating discoveries naturally stimulate broad interests in doped kagome systems, which may provide new insights into the formation of unconventional superconductivity.

Theoretical studies of doped kagome systems started from the doped Heisenberg model, i.e., the kagome-lattice $t$-$J$ model. In variational Monte Carlo (VMC) studies, while early works found a valence bond crystal at the doping range $\delta \lesssim 0.18$~\cite{kagome-tJ-VMC-2011,kagome-tJ-VMC-2013}, considering the SU(2)-gauge rotation in the mean-field Hamiltonian discovered a chiral noncentrosymmetric nematic superconducting state with lower energy at small doping ratio $\delta \lesssim 0.02$~\cite{kagome-tJ-VMC-2021}.
However, DMRG calculations on the cylinder system obtained an insulating charge density wave (CDW) state at $\delta \lesssim 0.11$~\cite{kagome-tJ-Jiang-2017,kagome-tJlike-PCheng-2021}. 
A recent projected entangled simplex state simulation also reported a CDW phase in the lightly doped regime~\cite{kagome-tJ-GuZC-2024}.
These results suggest the absence of hole pairing in the lightly doped kagome $t$-$J$ model, raising a big challenge to the emergence of SC.
On the other hand, inspired by the well established chiral spin liquid in kagome spin systems, the lightly doped $t$-$J$-$J_{\chi}$ model has also been studied using DMRG simulation~\cite{kagome-tJlike-PCheng-2021}, yet the CDW phase is still robust.

To suppress CDW order, recent DMRG results of square and triangular $t$-$J$ models have revealed that the next-nearest-neighbor (NNN) hopping $t_2$ and spin exchange $J_2$ can weaken charge order and enhance SC~\cite{Square-tt'JJ'-JiangHC-2021,Square-tt'JJ'-gss-2021,Square-tt'J-JiangShengtao-2021,Square-tt'JJ'-gss-2024,chen_2025,jiang_2023,jiang_2020,huang_2022,triangular-tt'JJ'-HuangYX-2023,triangular-tt'JJ'-JiangHC-2021}. 
Interestingly, a previous DMRG study on the kagome $t$-$J$ model also found enhanced pairing correlations by introducing a positive NNN hopping $t_2$~\cite{kagome-tJlike-PCheng-2021}, suggesting a potential SC phase.
However, the nature of this phase remains unclear, and it is uncertain whether the enhanced pairing correlations could develop a strong quasi-long-range order.

In this work, we employ the DMRG method to investigate the extended $t$-$J$ model that incorporates both the NN ($t_1$ and $J_1$) and NNN ($t_2$ and $J_2$) couplings on kagome cylinders. 
For a doping level $\delta = 1/18$ on the YC6 cylinder ($L_y = 3$), we present a phase diagram with $t_2/t_1 > 0$ and $J_2/J_1 > 0$ as tuning parameters.
While no superconducting state is observed, we find that introducing $t_2$ and $J_2$ can suppress the existing CDW order. 
Notably, all measured correlation functions exhibit significant enhancement, including single-particle, SC pairing, spin-spin, and density-density correlation functions. 
Across a broad range of $t_2$ and $J_2$, we identify a Fermi-liquid-like phase characterized by the power-law decay of single-particle correlation and the absence of hole pairing. 
Systematic studies of doping dependence on the YC6 cylinder reveal that this Fermi-liquid-like phase remains stable within the doping range of $1/36$ to $1/18$, while further increasing the doping to $1/9$ drives the system to another CDW state. 
To examine finite-size effects, we also study the wider YC8 cylinder ($L_y = 4$), where this Fermi-liquid-like state persists in the lightly doped regime, confirming the robustness of our findings.
This kagome system may realize a transition from an insulating CDW to a Fermi liquid in slightly doped two-dimensional Mott insulator.

The paper is organized as follows. In Sec. II, we introduce the model Hamiltonian and the details of DMRG calculations. In Sec. III, we present the DMRG results on the YC6 cylinder. In Sec. IV, we study the doping dependence of the Fermi-liquid-like phase. In Sec. V, we study the Fermi-liquid-like phase on the YC8 cylinder. Sec. VI is devoted to the summary and discussion.

\section{MODEL AND METHOD}
\begin{figure}[t]
	\centering
	\begin{subfigure}[b]{0.48\textwidth}
		\raggedleft
		\includegraphics[width=0.85\textwidth]{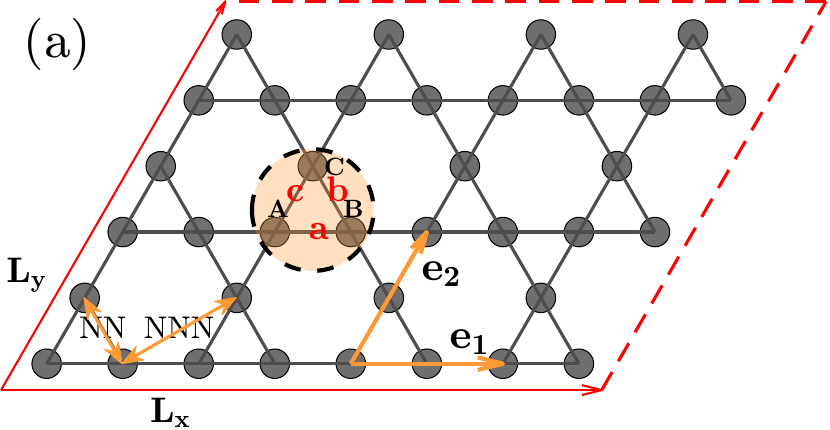} 
	\end{subfigure}
    \begin{subfigure}[b]{0.48\textwidth}
	   \includegraphics[width=\textwidth]{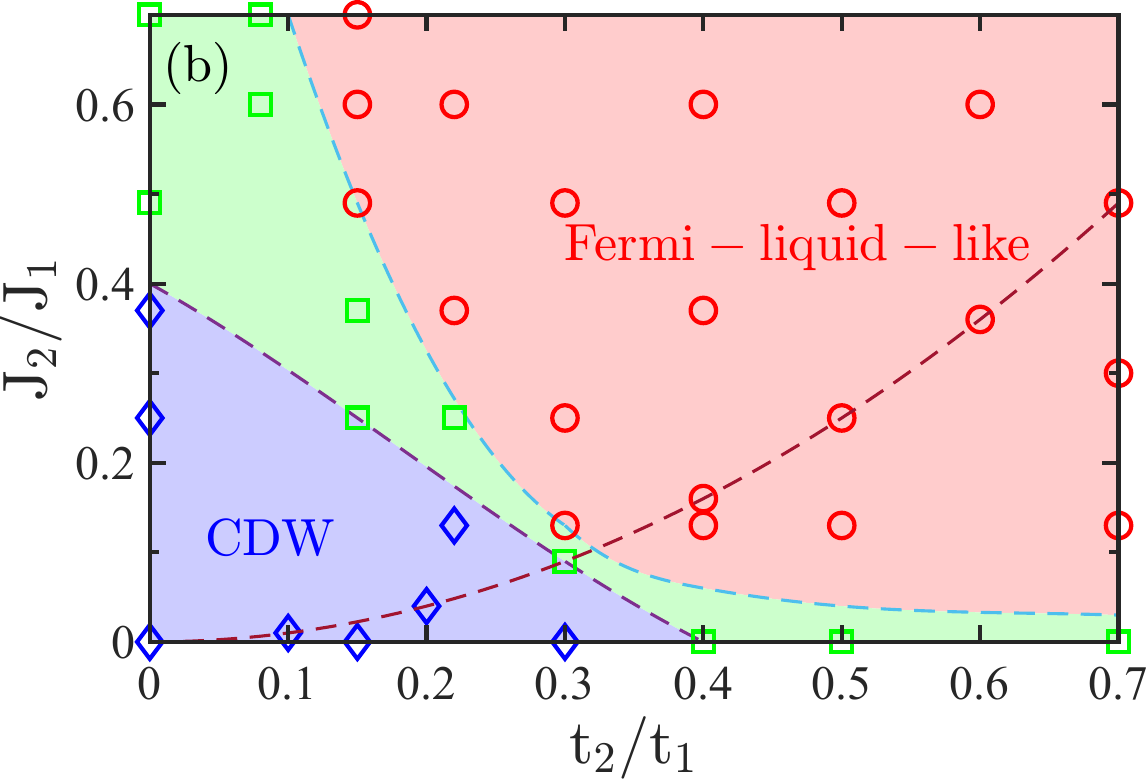}
    \end{subfigure}
	\caption{\justifying Schematic figure of kagome-lattice $t$-$J$ model and quantum phase diagram of the YC6 system by tuning $t_2$ and $J_2$. (a) The studied $t$-$J$ model on the YC6 kagome cylinder, where the electrons and doped holes live at the vertices (solid circles). The model has both the nearest-neighbor and next-nearest-neighbor hoppings ($t_1$ and $t_2$), as well as corresponding spin exchange interactions ($J_1$ and $J_2$). The periodic boundary conditions and open boundary conditions are imposed, respectively, along the directions specified by the lattice vectors, ${\bf e}_2$ and ${\bf e}_1$. Each unit cell (denoted by the small triangle in the shaded region) has three sites ($A$, $B$, and $C$) and three bonds ($a$, $b$, and $c$). $L_x$ and $L_y$ denote the numbers of unit cells along the ${\bf e}_1$ and ${\bf e}_2$ directions, respectively. (b) Quantum phase diagram of the kagome-lattice $t$-$J$ model obtained in the parameter region $0 \leq t_2/t_1 \leq 0.7$ and $0 \leq J_2 /J_1 \leq 0.7$, at the given doping ratio $\delta = 1/18$ and $t_1/J_1 = 3$. Besides the charge density wave (CDW) phase identified previously~\cite{kagome-tJ-Jiang-2017,kagome-tJlike-PCheng-2021}, we find a Fermi-liquid-like phase. The phase boundaries are determined by examining charge density profile and correlation functions. The green region indicates an intermediate region, in which most of physical quantities are similar to those in the Fermi-liquid-like phase but some quantities change slowly with tuning couplings.} 
    \label{lattice_diagram}
\end{figure}

The Hamiltonian of the extended $t$-$J$ model is defined as
\begin{align*}
 H  =-\!\sum\limits_{\{ ij\},\sigma}t_{ij}(\hat{c}_{i,\sigma}^\dagger \hat{c}_{j,\sigma}\!+\!\mathrm{H.c.})+\sum\limits_{\{ij\}}J_{ij}(\mathbf{\hat{S}}_i\cdot \mathbf{\hat{S}}_j\!-\!\frac{1}{4}\hat{n}_{i} \hat{n}_{j}),
\end{align*}
where $\hat{c}^\dagger_{i,\sigma}$ and $\hat{c}_{i,\sigma}$ are the creation and annihilation operators for the electron with spin $\sigma$ ($\sigma = \pm 1/2$) at the site $i$, $\hat{\mathbf{S}}_i$ is the spin-$1/2$ operator, and $\hat{n}_i \equiv \sum_\sigma \hat{c}^\dagger_{i,\sigma} \hat{c}_{i,\sigma}$ is the electron number operator. 
For each site, the $t$-$J$ model requires the no-double-occupancy constraint.
We consider the NN and NNN electron hoppings ($t_1$ and $t_2$) and spin-exchange interactions ($J_1$ and $J_2$). 
We choose $t_1/J_1 = 3$ to mimic a large Hubbard $U$, and tune $t_2$ and $J_2$ separately in the parameter region $0 \leq t_2/t_1 \leq 0.7$ and $0 \leq J_2 /J_1 \leq 0.7$.
In this work, we focus on the lightly doped case with the doping level $\delta = 1/36 - 1/18$, where $\delta = N_h / N$ ($N_h$ is the number of doped holes and $N$ is the total number of sites).
For some typical coupling parameters, we also extend the doping ratio to $\delta = 1/9$.

We solve the ground state of the model using DMRG calculations~\cite{DMRG-White-1992} on a finite-width cylinder.
The cylindric geometry is shown in Fig.~\ref{lattice_diagram}(a), where $\mathbf{e}_1$ and $\mathbf{e}_2$ denote the two unit vectors. 
We consider the kagome Y-cylinder (YC) that has periodic (open) boundary conditions along the $\mathbf{e}_2$ ($\mathbf{e}_1$) direction.
We denote the number of unit cells along the $\mathbf{e}_2$ and $\mathbf{e}_1$ directions as $L_y$ and $L_x$, respectively, and we refer to the cylinder as YC$2L_y$, which has the total number of sites $N = 3 L_y \times L_x $. 
In this work, we focus mainly on the YC6 cylinders with $L_y = 3$ and $L_x = 32-40$.
For some typical parameter points, we also examine YC8.
By combining the charge $U(1)$ and spin $SU(2)$ symmetries, we keep the bond dimensions up to $D=20000$ $SU(2)$ multiplets, which are equivalent to about $60000$ $U(1)$ states. 
In this work, the DMRG truncation error is controlled below $4 \times 10^{-6}$, giving accurate results.

\section{DMRG results of the YC6 system}

\subsection{Quantum phase diagram}

We first summarize our DMRG results for the YC6 cylinder at $\delta = 1/18$, as demonstrated by the phase diagram Fig.~\ref{lattice_diagram}(b).
With growing $t_2/t_1$ and $J_2/J_1$, the system features a CDW phase (purple) that has been reported~\cite{kagome-tJ-Jiang-2017,kagome-tJlike-PCheng-2021} and an emergent Fermi-liquid-like phase (red).
These two phases can be distinguished by the different charge density distributions and decay behaviors of correlation functions. 
The CDW phase is characterized by a pronounced CDW pattern with wavelength $\lambda = 1/(9\delta)$, as well as the exponential decay of various correlation functions with very short correlation lengths, including single-particle, SC pairing, and spin correlations.  
In contrast, in the Fermi-liquid-like phase, the charge density modulation is either significantly suppressed or nearly absent in the bulk of the lattice. 
Meanwhile, the correlation functions are greatly enhanced. 
The spin correlations show a three-sublattice structure, which should originate from the ${\bf k} = (0,0)$ magnetic order of the $J_1$-$J_2$ kagome Heisenberg model~\cite{kagome-J1J2-Schollwock-2015,kagome-J1J2J3-ssg-2015}.
In particular, both the single-particle and pairing correlations exhibit a power-law decay, and the pairing correlations are very close to the square of single-particle correlations, characterizing the absence of hole pairing.

The key finding of this study is the identification of the Fermi-liquid-like phase, which spans a wide range of $t_2$ and $J_2$ values.
In our studied doping range and system size, this phase remains stable within $\delta = 1/36 - 1/18$. 
Crucially, the emergence of the Fermi-liquid-like phase relies on the combined effects of $t_2$ and $J_2$.
Between these two phases, if $t_2$ and $J_2$ are mismatched with $J_2/J_1$ much different from $(t_2/t_1)^2$, the changes of some physical quantities are not synchronous, resulting in an intermediate region (the green region in Fig.~\ref{lattice_diagram}).

\subsection{Charge density profile}

\begin{figure}[h] 
	\centering
	\begin{subfigure}[b]{0.48\textwidth}
		\includegraphics[width=\textwidth]{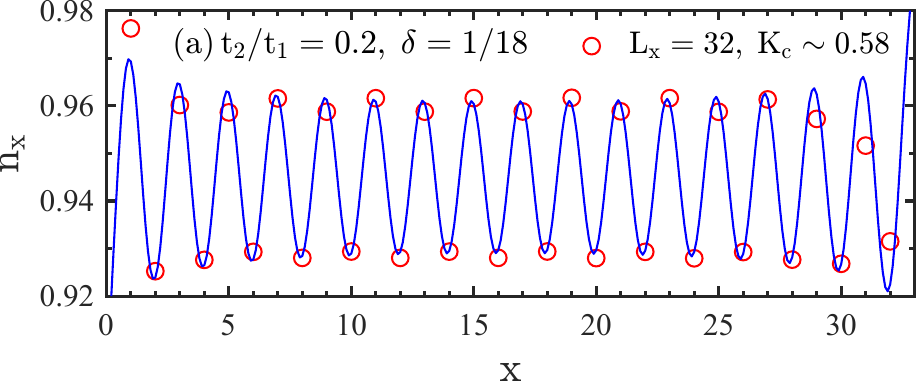} 
	\end{subfigure}
	\begin{subfigure}[b]{0.48\textwidth}
		\includegraphics[width=\textwidth]{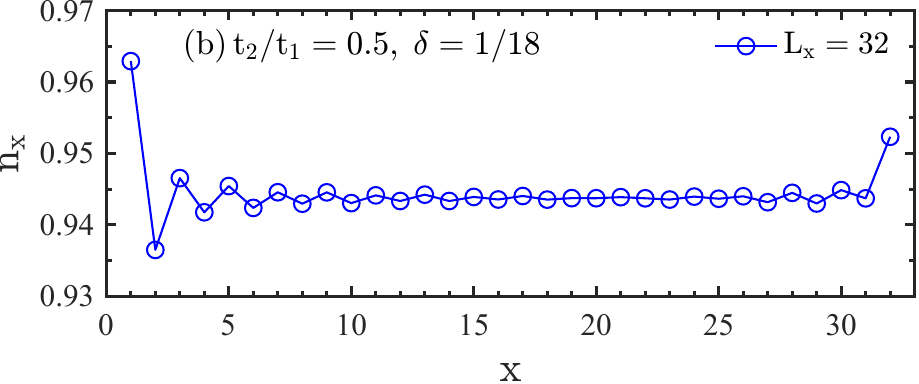} 
    \end{subfigure}
	\begin{subfigure}[b]{0.48\textwidth}
		\includegraphics[width=\textwidth]{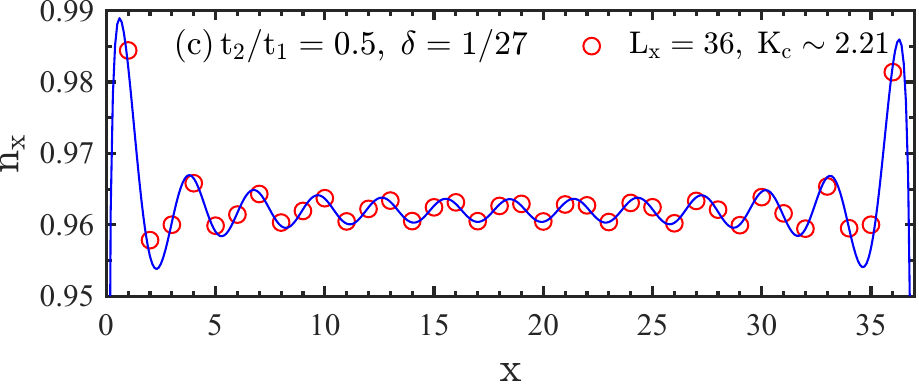} 
    \end{subfigure}
	\begin{subfigure}[b]{0.48\textwidth}
		\includegraphics[width=\textwidth]{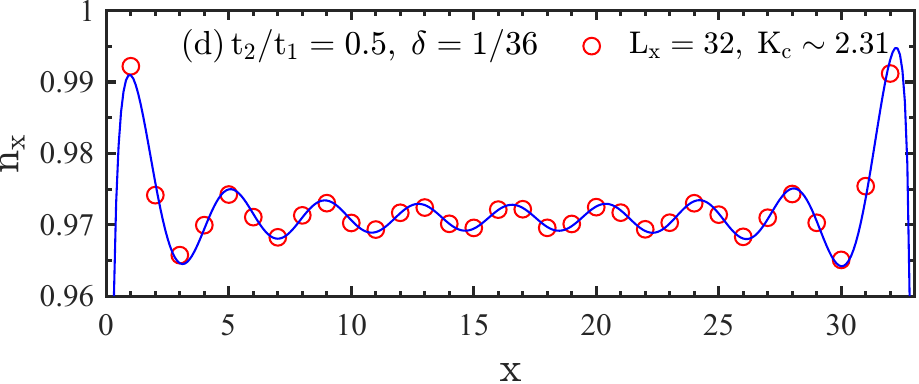} 
    \end{subfigure}
	\caption{\justifying Charge density profile on the YC6 cylinder. The averaged charge density of the unit cell in each column $n_x$ is defined as $n_x = \frac{1}{3L_y} \sum_{y=1}^{L_y} \sum_{i=1}^{3} \langle \hat{n}_{x,y,i} \rangle $. (a) $L_x = 32$ cylinder with $t_2/t_1 = 0.2$, $J_2/J_1=0.04$, and $\delta = 1/18$ in the CDW phase. (b)-(d) show the results in the Fermi-liquid-like phase with $t_2/t_1 = 0.5$, $J_2/J_1=0.25$, and $\delta = 1/18$, $1/27$, and $1/36$ respectively. The blue lines in (a), (c), (d) are the fitting curves to the function $n_x = n_0 + A_\text{CDW} \cos(Qx + \phi)$, where $A_\text{CDW} = A_0 [x^{-K_c/2} + (L_x + 1 - x)^{-K_c/2}]$ and $Q$ are the CDW amplitude and wave vector, respectively. $\phi$ is a phase shift.}
    \label{charge}
\end{figure}

We first show the distribution of charge density in Fig.~\ref{charge}. 
Since the charge density obtained in our calculation is uniform along the $\mathbf{e}_2$ direction, we define the average charge density of the unit cell in each column as $n_x=\frac{1}{3L_y}\sum_{y=1}^{L_y}\sum_{i=1}^{3}\langle \hat{n}_{x,y,i}\rangle$, where $x$ ($y$) is the column (row) number and $i$ denotes the three sites in each unit cell. 
On the YC6 cylinder, we observe a CDW order with wavelength $\lambda = 1/(9\delta)$ in the CDW phase, as shown in Fig.~\ref{charge}(a) at $t_2 / t_1 = 0.2$, $J_2 / J_1 = 0.04$ and $\delta = 1/18$.
This charge density profile has a period of two unit cells in the ${\bf e}_1$ direction, giving one hole per CDW period on average, which agrees with the result at $t_2 = J_2 = 0$ and is consistent with no pairing~\cite{kagome-tJ-Jiang-2017}.

In the Fermi-liquid-like phase with growing $t_2$ and $J_2$, the charge density modulation is significantly suppressed, as demonstrated in Figs.~\ref{charge}(b)-\ref{charge}(d) in the doping range $\delta = 1/36 - 1/18$.
The suppression of CDW driven by the growing NNN couplings has also been observed in the square~\cite{Square-tt'JJ'-JiangHC-2021,Square-tt'JJ'-gss-2021,Square-tt'J-JiangShengtao-2021,Square-tt'JJ'-gss-2024,chen_2025,jiang_2023} and triangular~\cite{jiang_2020,huang_2022,triangular-tt'JJ'-HuangYX-2023} $t$-$J$ models, where a quasi-long-range SC order emerges simultaneously.

\subsection{Correlation functions}

\begin{figure}[h] 
	\centering
	\begin{subfigure}[b]{0.48\textwidth}
		\includegraphics[width=0.494\textwidth]{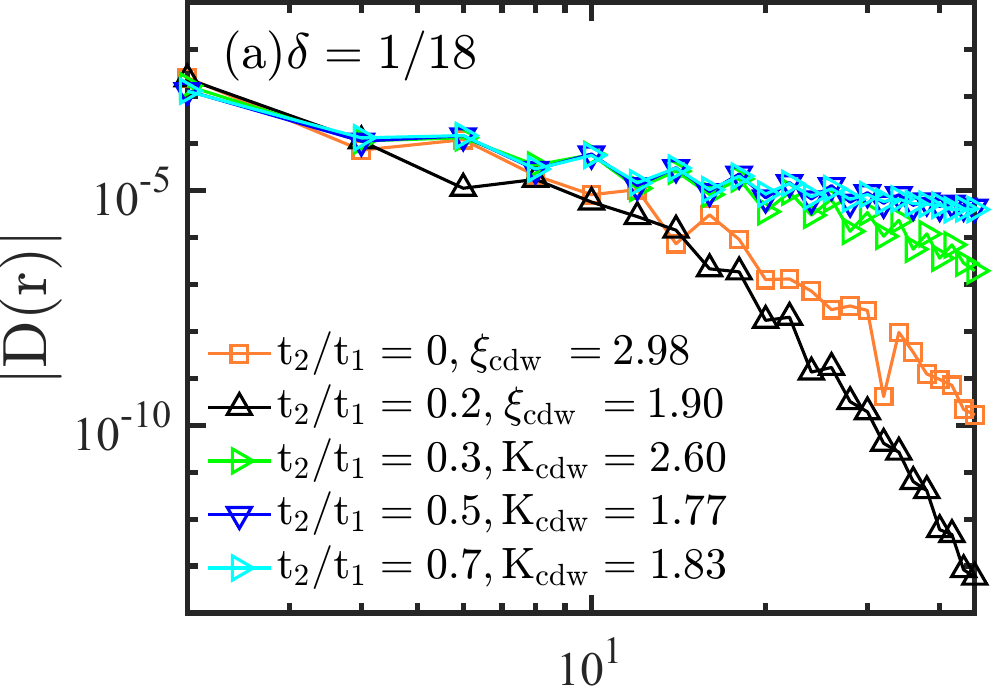} 
		\includegraphics[width=0.494\textwidth]{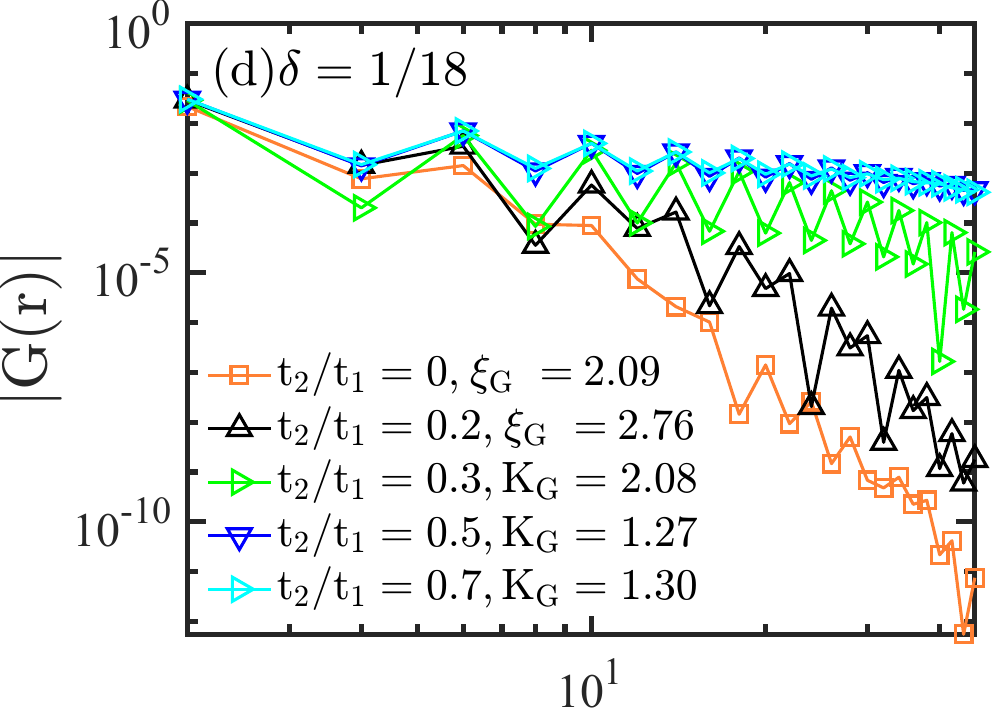} 
	\end{subfigure}
	\begin{subfigure}[b]{0.48\textwidth}
		\includegraphics[width=0.494\textwidth]{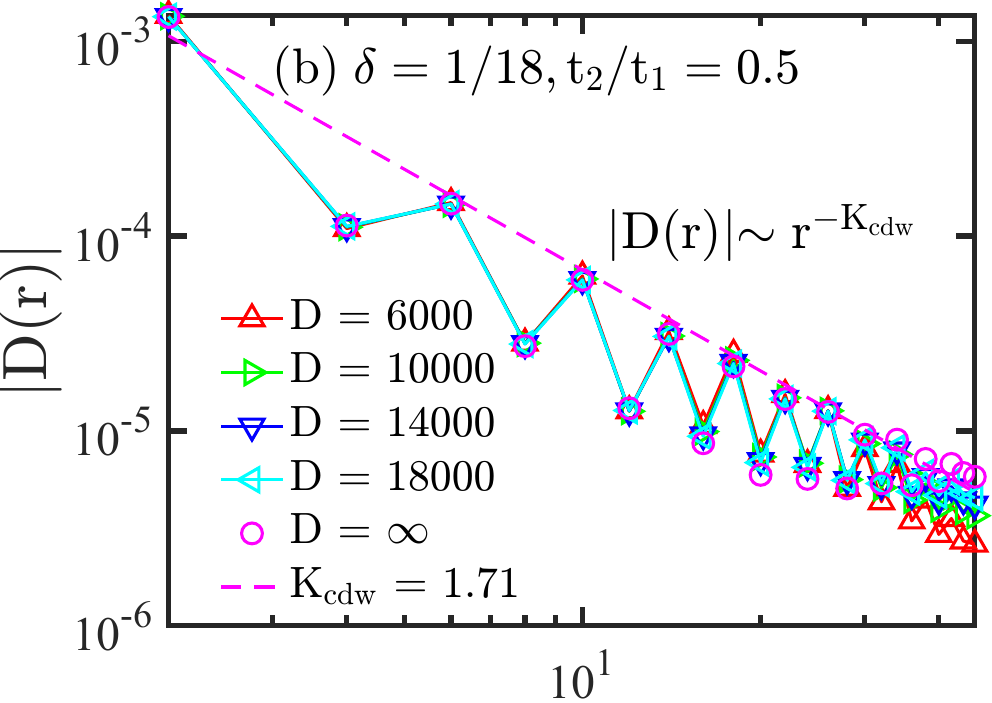} 
		\includegraphics[width=0.494\textwidth]{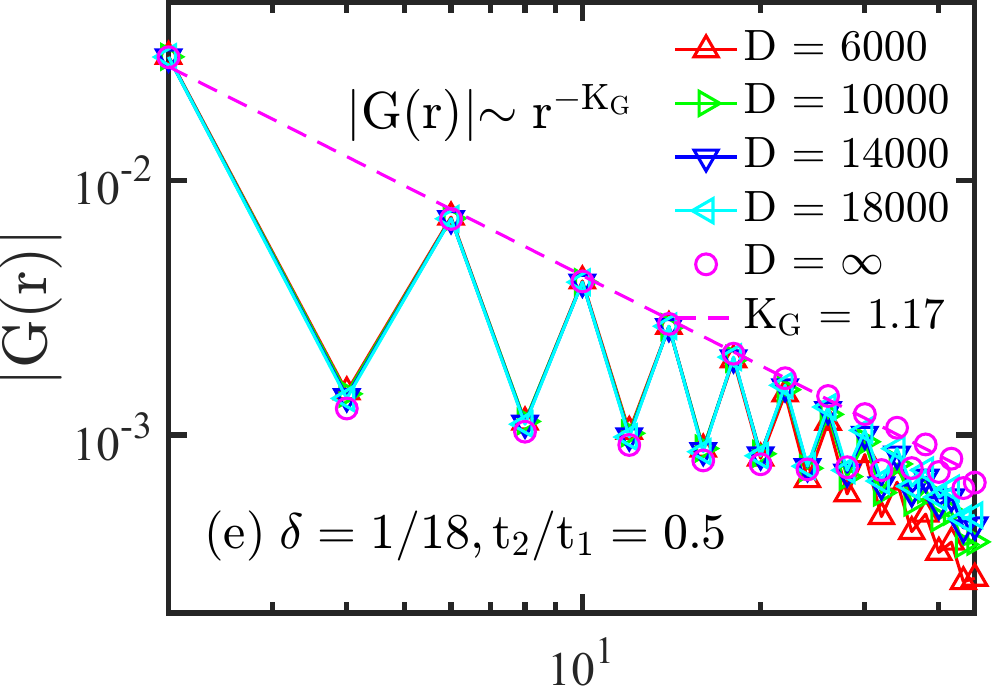} 
	\end{subfigure}
	\begin{subfigure}[b]{0.48\textwidth}
		\includegraphics[width=0.494\textwidth]{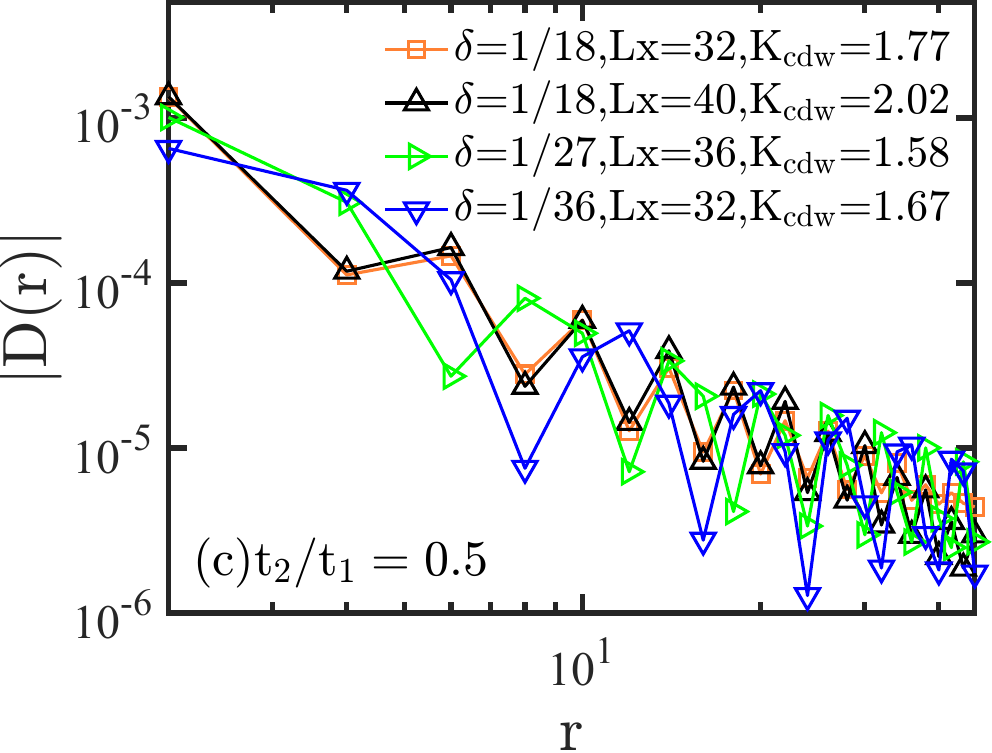} 
		\includegraphics[width=0.494\textwidth]{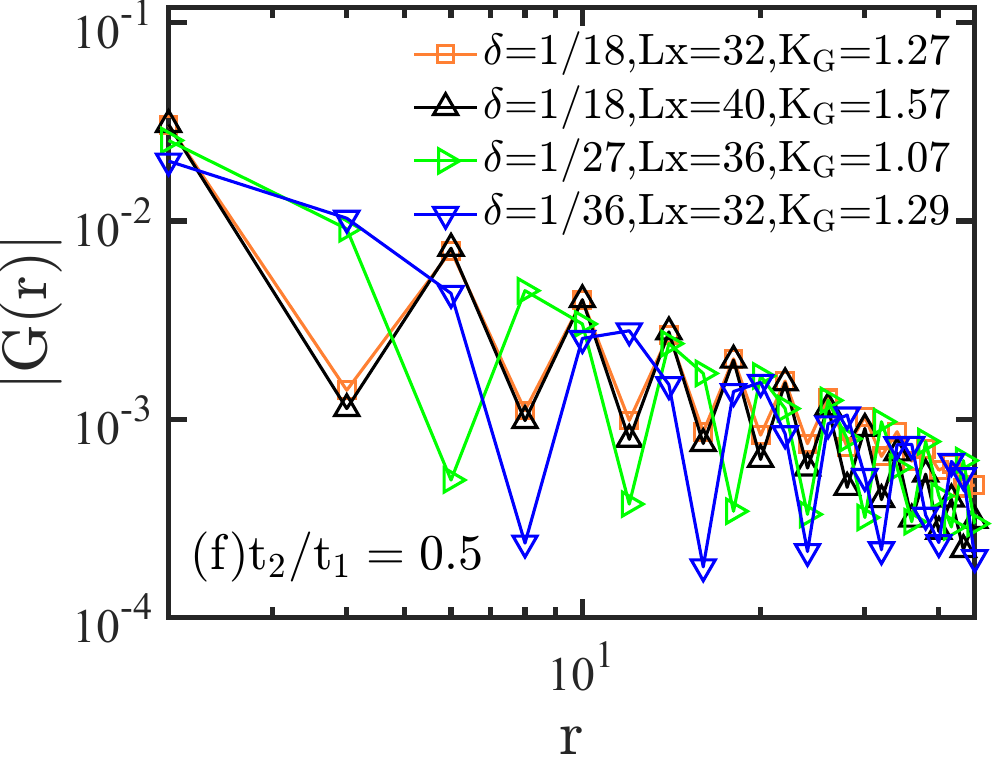} 
	\end{subfigure}
	\caption{\justifying Density correlation function $D(r)$ and single-particle Green’s function $G(r)$ on the YC6 cylinder. (a) $D(r)$ for different $t_2/t_1$ values along the line $(t_2/t_1)^2 = J_2/J_1$ with $\delta = 1/18$. $\xi_{cdw}$ and $K_{cdw}$ are the fitting exponents in exponential decay function and power-law decay function, respectively. (b) $D(r)$ at $t_2/t_1=0.5$ and $J_2/J_1=0.25$, for different bond dimensions in the range of $D = 6000-18000$. The dashed line denotes the power-law fitting of the extrapolated $D \rightarrow \infty $ results. (c) $D(r)$ at $t_2/t_1 = 0.5$ and $J_2/J_1=0.25$ for different doping levels and system sizes. (d)-(f) Similar plots for the single-particle Green’s function $G(r)$.}
    \label{D_and_G}
\end{figure}

\begin{figure}[h] 
	\centering
	\begin{subfigure}[b]{0.48\textwidth}
		\includegraphics[width=0.494\textwidth]{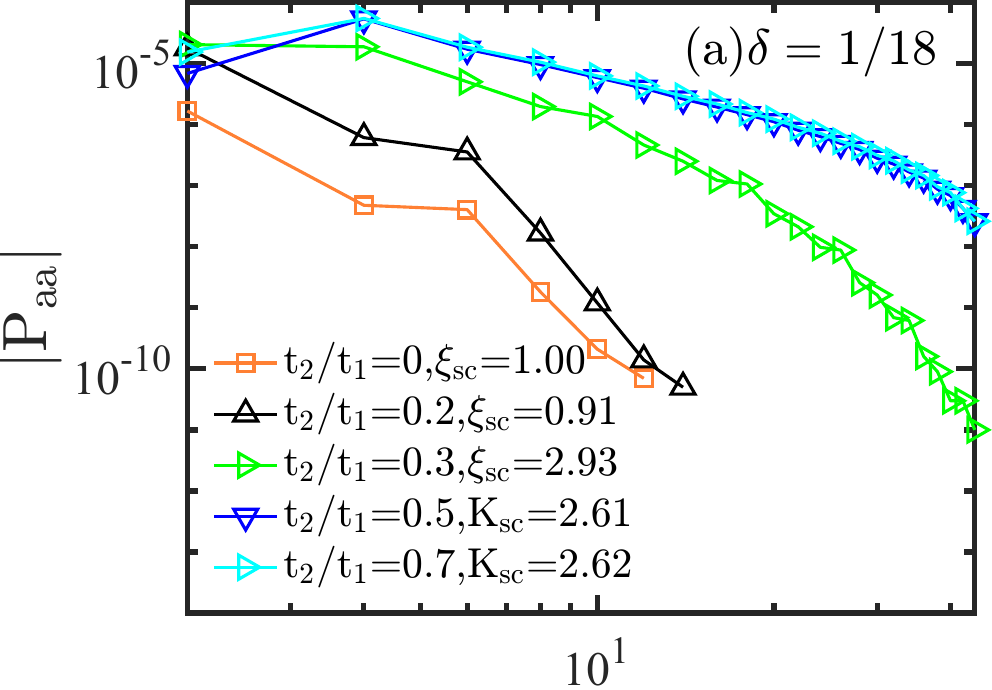} 
		\includegraphics[width=0.494\textwidth]{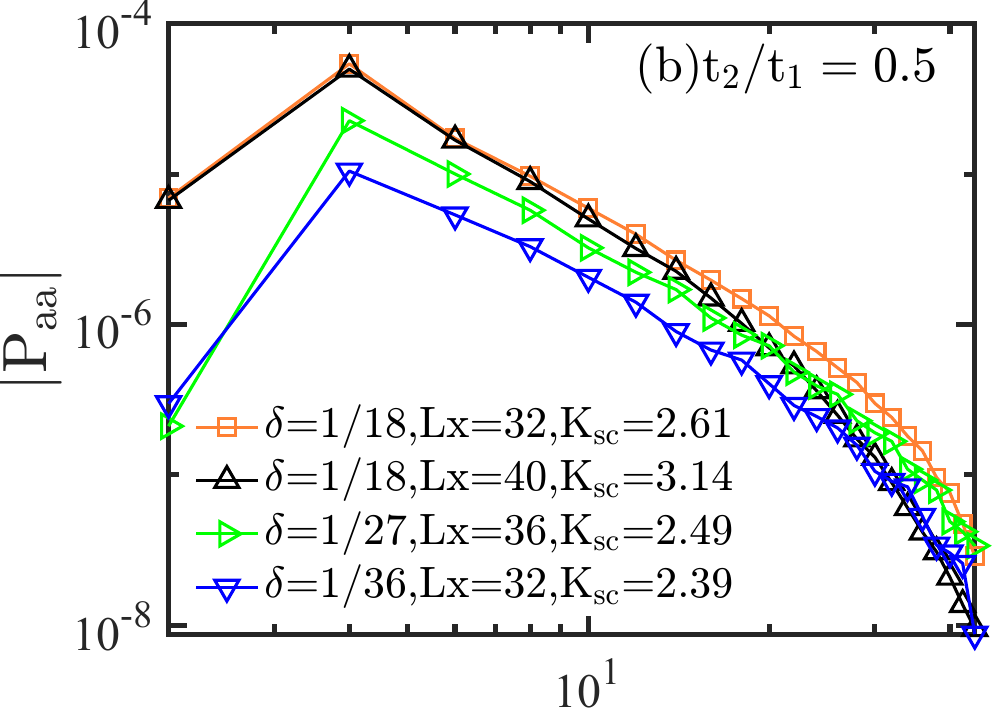} 
	\end{subfigure}
	\begin{subfigure}[b]{0.48\textwidth}
		\includegraphics[width=0.494\textwidth]{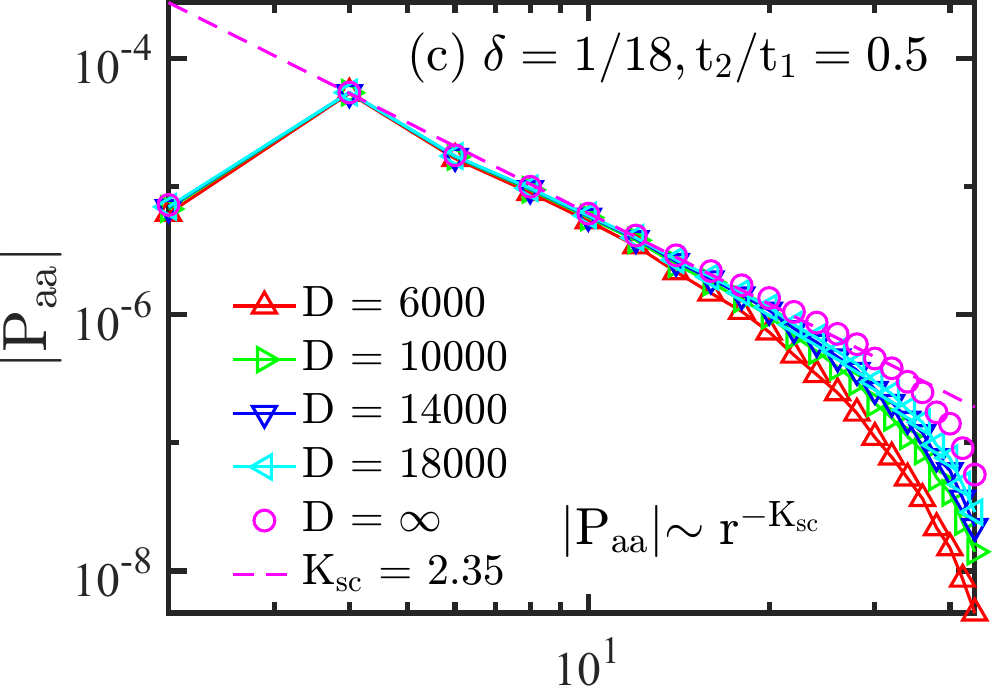} 
		\includegraphics[width=0.494\textwidth]{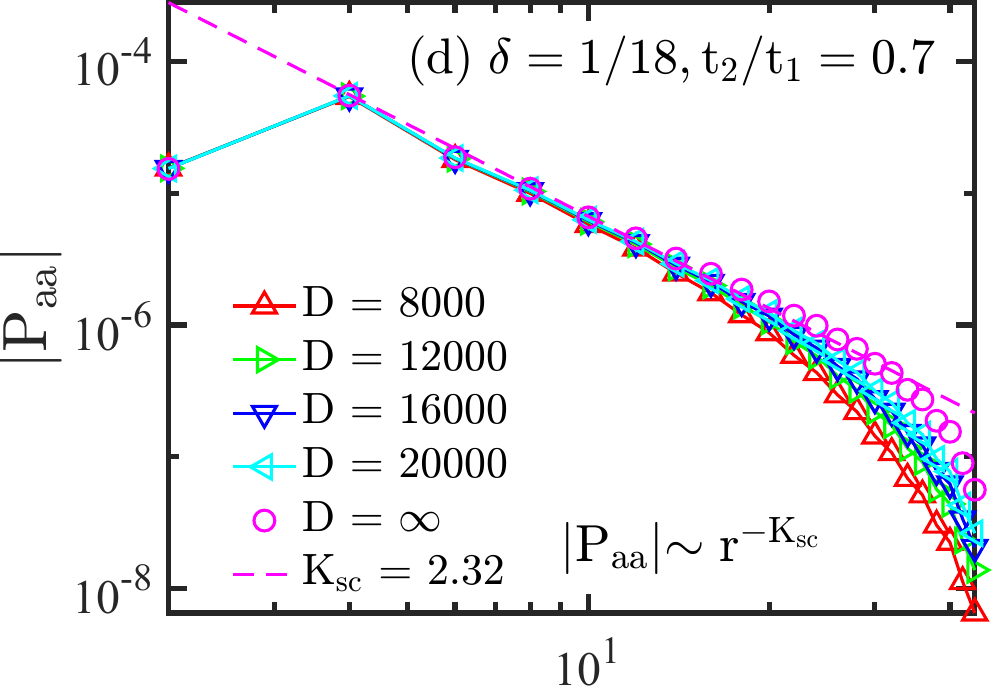} 
	\end{subfigure}
	\begin{subfigure}[b]{0.48\textwidth}
		\includegraphics[width=0.494\textwidth]{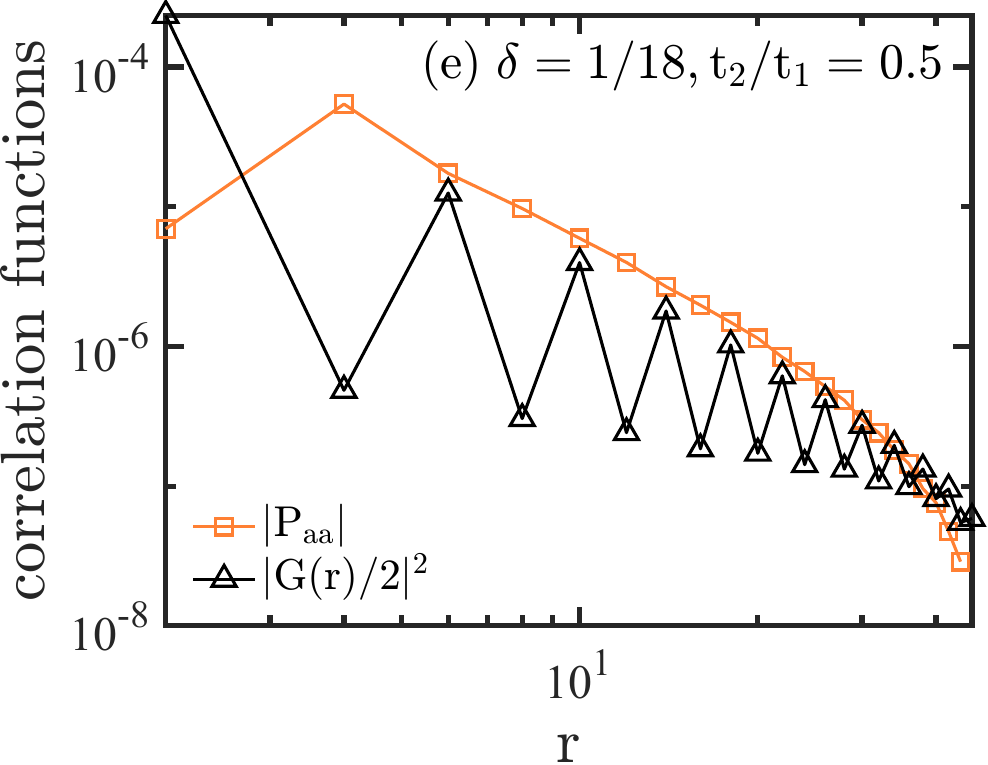} 
		\includegraphics[width=0.494\textwidth]{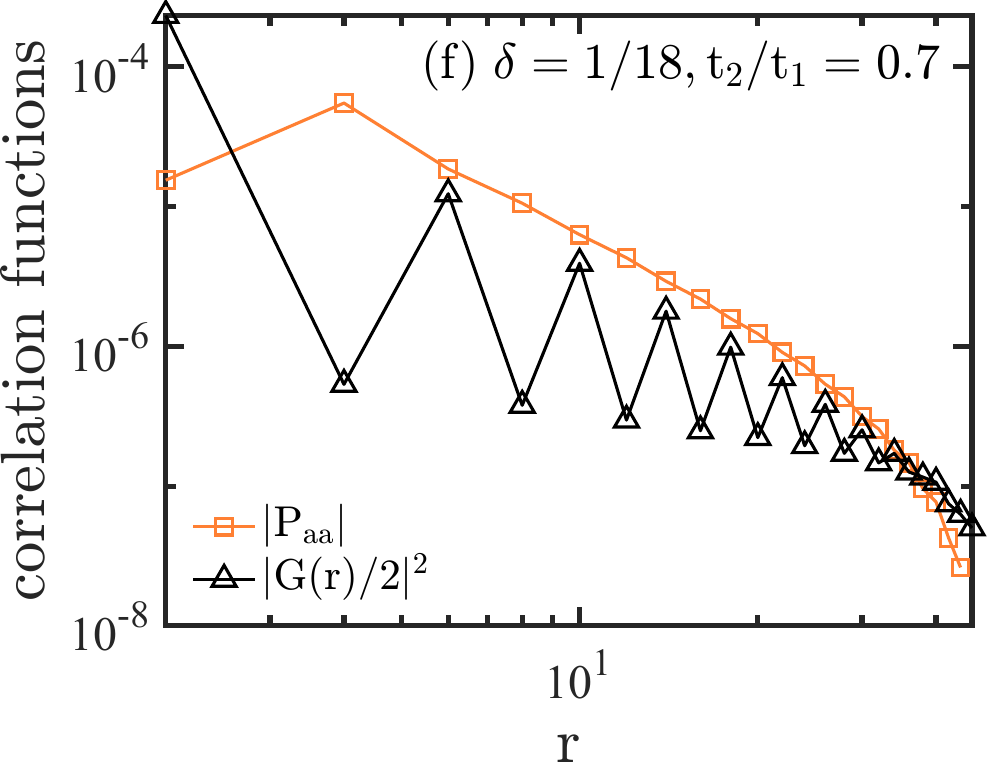} 
	\end{subfigure}
	\caption{\justifying Double-logarithmic plot of SC pairing correlation $P_{aa}$ on the YC6 cylinder. (a) $P_{aa}$ for different $t_2/t_1$ values along the line $(t_2/t_1)^2 = J_2/J_1$ with $\delta = 1/18$. (b) $P_{aa}$ for different doping levels and system sizes at $t_2/t_1 = 0.5$, $J_2/J_1=0.25$ and $\delta = 1/18$. (c) and (d) show the results for different bond dimensions at $\delta = 1/18$ for $t_2/t_1 = 0.5$, $J_2/J_1=0.25$, and $t_2/t_1 = 0.7$, $J_2/J_1=0.49$, respectively. (e) and (f) compare pairing correlation $P_{aa}$ with the the square of single-particle Green’s function $(G(r)/2)^2$ at $\delta = 1/18$ for $t_2/t_1 = 0.5$, $J_2/J_1=0.25$, and $t_2/t_1 = 0.7$, $J_2/J_1=0.49$, respectively.}
    \label{SC}
\end{figure}

In this subsection, we present the DMRG results of the correlation functions with growing $t_2 / t_1$ and $J_2 / J_1 = (t_2 / t_1)^2$, as shown by the dashed line in Fig.~\ref{lattice_diagram}(b). 
In the CDW phase, our data agree with the results reported in previous studies~\cite{kagome-tJ-Jiang-2017,kagome-tJlike-PCheng-2021}.

We first discuss the density-density correlation function $D(r)=\langle \hat{n}_{i_0} \hat{n}_{i_0+r} \rangle - \langle \hat{n}_{i_0} \rangle \langle \hat{n}_{i_0+r} \rangle$ and single-particle Green's function $G(r) = \sum_\sigma\langle \hat{c}_{i_0,\sigma}^\dagger \hat{c}_{i_0+r,\sigma}\rangle$, where $i_0$ denotes the reference site at the $1/4$ length of the cylinder and $r$ is the distance from $i_0$ along the $\mathbf{e}_1$ direction. 
In the CDW phase, since the charge density profile $n_x$ shows a static charge order [Fig.~\ref{charge}(a)], our defined $D(r)$ describes the fluctuations and decays exponentially, as shown for $t_2/t_1 = 0$ and $0.2$ [Fig.~\ref{D_and_G}(a)].
The single-particle Green's function $G(r)$ also decays exponentially [Fig.~\ref{D_and_G}(d)].

In the Fermi-liquid-like phase, $n_x$ is close to uniform in the bulk and $D(r)$ characterizes the charge correlation.
It is quite clear that $D(r)$ decays algebraically in the Fermi-liquid-like phase, with the power exponent $K_{\rm cdw} \sim 2$ characterizing a very weak quasi-long-range charge order [Fig.~\ref{D_and_G}(a)].
Meanwhile, $G(r)$ also becomes power-law decay with the exponent $K_{\rm G} \sim 1$, indicating the gapless single-particle excitations [Fig.~\ref{D_and_G}(d)]. 
For these two quantities, the DMRG results converge quite well as evidenced by the consistent results at the bond dimensions from $6000$ to $18000$ [Figs.~\ref{D_and_G}(b) and \ref{D_and_G}(e)].
The $D = \infty$ results are obtained by extrapolating the finite-$D$ data for each distance $r$ (see the details in Appendix~\ref{app-1}).
We have also checked the systems at different $L_x$ and doping levels, showing that the decay behaviors of the two quantities are quite stable, as shown in Figs.~\ref{D_and_G}(c) and \ref{D_and_G}(f).

\begin{figure}[h] 
	\centering
	\begin{subfigure}[b]{0.48\textwidth}
		\includegraphics[width=0.494\textwidth]{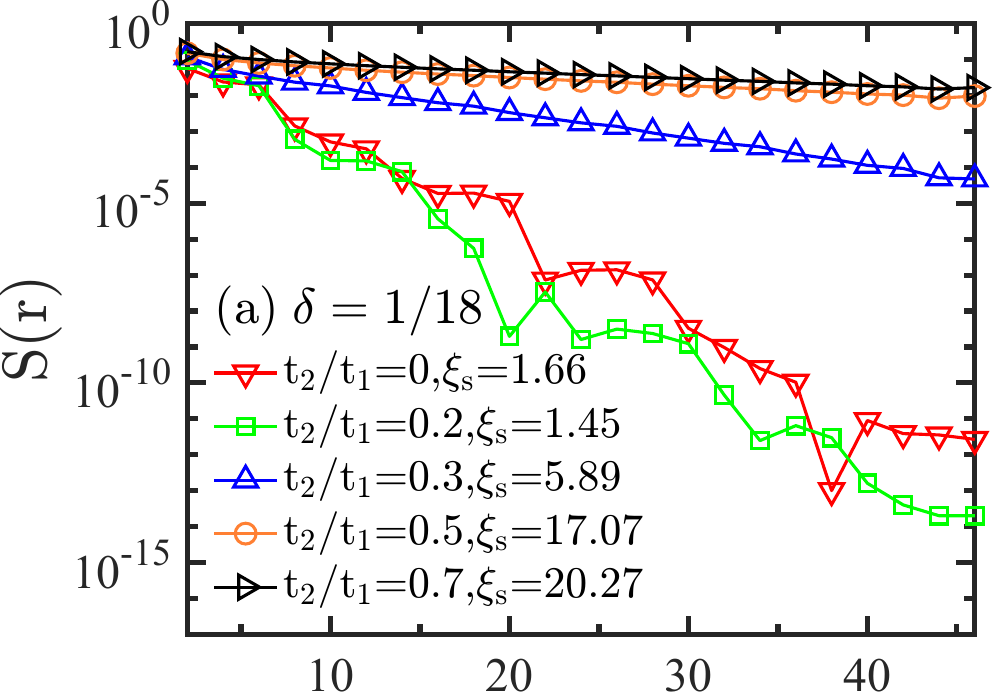} 
		\includegraphics[width=0.494\textwidth]{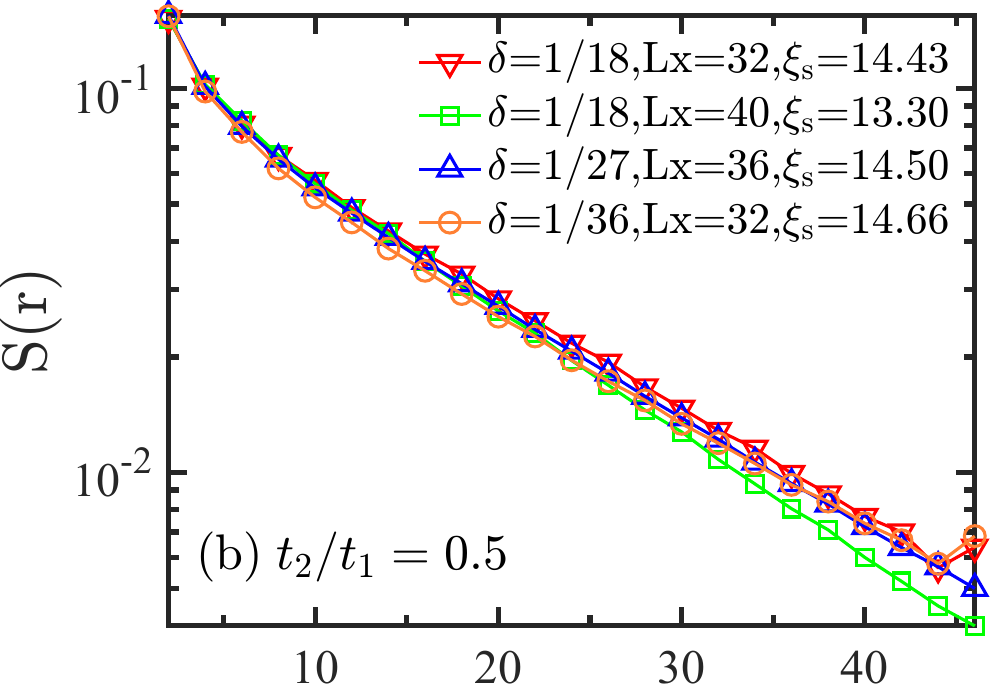} 
	\end{subfigure}
	\begin{subfigure}[b]{0.48\textwidth}
		\includegraphics[width=0.494\textwidth]{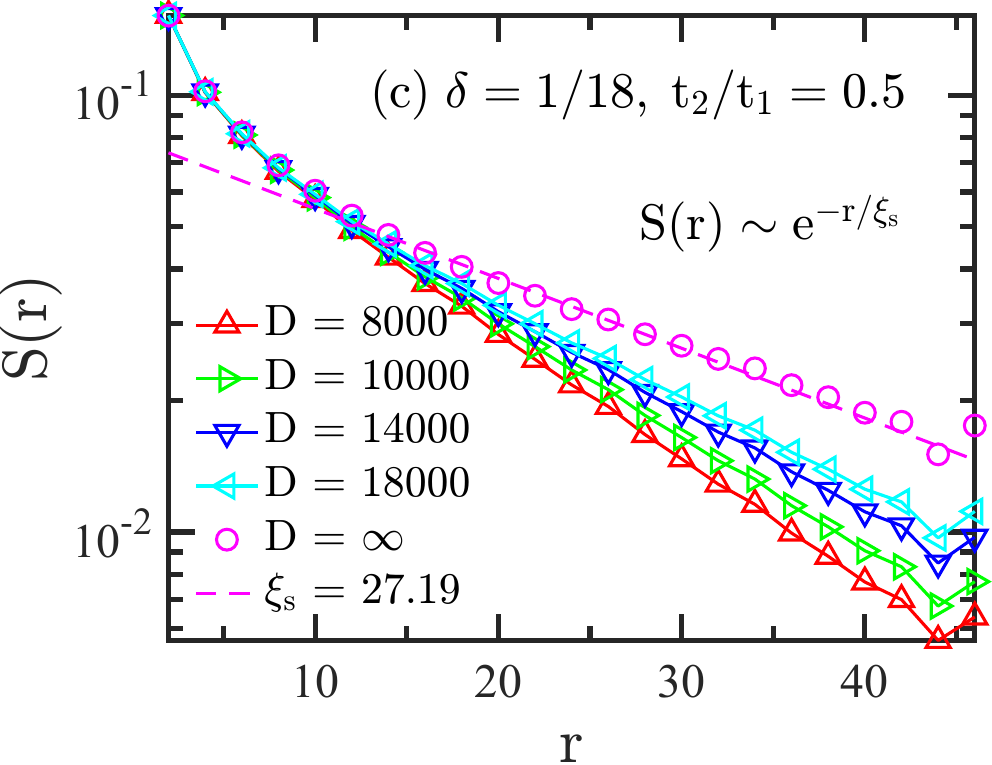} 
		\includegraphics[width=0.494\textwidth]{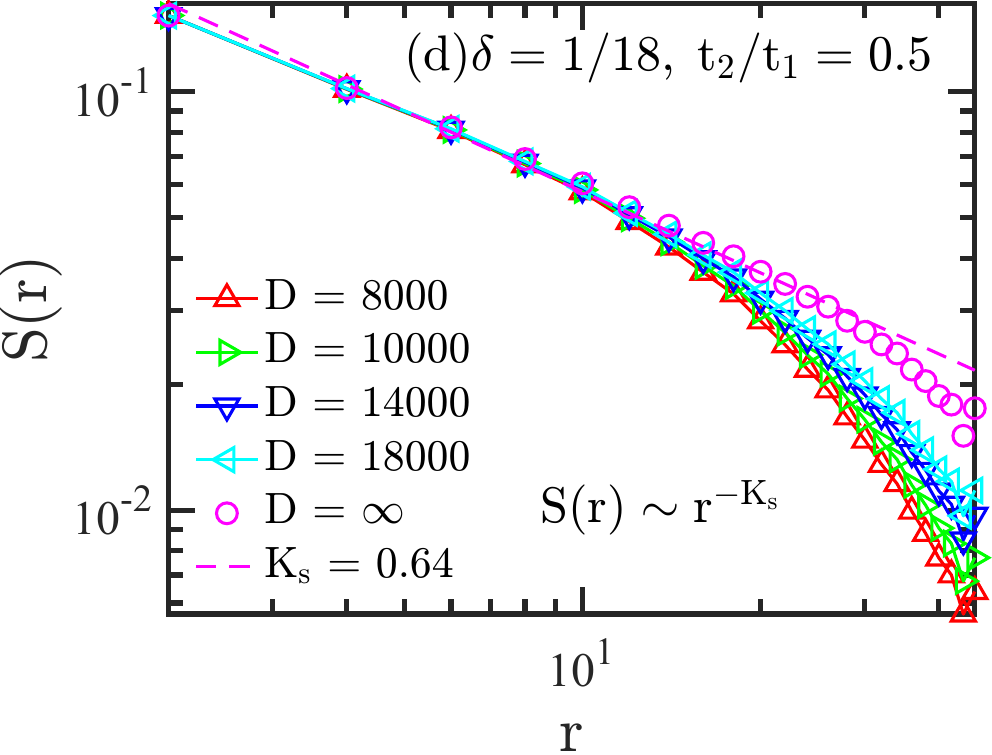} 
	\end{subfigure}
	\caption{\justifying Spin correlation $S(r)$ on the YC6 cylinder. Semi-logarithmic plot of  $S(r)$: (a) for different $t_2/t_1 $ values along the line $(t_2/t_1)^2 = J_2/J_1$ at $\delta = 1/18$, and  (b) for different doping levels and system sizes at $t_2/t_1 = 0.5$ and $J_2/J_1 = 0.25$. (c) Semi-logarithmic plot and (d) double-logarithmic plot of $S(r)$ for various bond dimensions in the range $D = 8000-18000$ at $\delta = 1/18$ for $t_2/t_1 = 0.5$ and $J_2/J_1=0.25$.}
    \label{Sr}
\end{figure}

Next, we examine the spin-singlet pairing correlation function $P_{\alpha\beta}(r)=\langle \Delta_\alpha^\dagger(i_0) \Delta_\beta(i_0 + r) \rangle$, where $\Delta_\alpha^\dagger(i)$ is the spin-singlet pair-field creation operator defined as $\Delta_\alpha^\dagger(i) = 1/\sqrt{2} ( c_{i\uparrow}^\dagger c_{i+\alpha\downarrow}^\dagger - c_{i\downarrow}^\dagger c_{i+\alpha\uparrow}^\dagger )$ with $\alpha$ denoting the bond direction [see Fig.~\ref{lattice_diagram}(a)], i.e., $\mathbf{a}$, $\mathbf{c}$, or $\mathbf{b}$, which are defined as $\mathbf{a} = \mathbf{e}_1/2$, $\mathbf{c} = \mathbf{e}_2/2$, and $\mathbf{b} = (\mathbf{e}_2 - \mathbf{e}_1)/2$. 
In Fig.~\ref{SC}(a), we show $P_{aa}$ with growing $t_2 / t_1$.
One can see a significant enhancement of the pairing correlation in the Fermi-liquid-like phase compared to the CDW phase, which has been reported in Ref.~\cite{kagome-tJlike-PCheng-2021}.
In the Fermi-liquid-like phase, we have examined the different NN bond pairing correlations $P_{\alpha \beta}$, and we find that $P_{aa}$, $P_{bb}$ and $P_{ba}$ are the same and much stronger than others, as shown in Appendix~\ref{app-2}.
Therefore, we only demonstrate $P_{aa}$ here.
We further show that the enhanced pairing correlation in the Fermi-liquid-like phase remains stable across different system lengths and doping ratios ($\delta = 1/36 - 1/18$) [Fig.~\ref{SC}(b)].
Next, we extrapolate the pairing correlations to the $D \rightarrow \infty$ limit using the polynomial function of bond dimension (see the details in Appendix~\ref{app-1}).
The extrapolated results fit the power-law decay quite well over a distance of $\sim 30$ sites [see Figs.~\ref{SC}(c) and \ref{SC}(d)], with the power exponent $K_{\rm sc} \approx 2.3$.
The slight deviation of the long-distance data ($r > 30$) from the fitting curve is attributed to the harder convergence of long-distance correlation, particularly the four-site pairing correlation~\cite{Square-tt'JJ'-gss-2021}.
To further clarify the nature of hole pairing, we compare $P_{aa}$ with the single-particle correlation square $(G(r)/2)^2$ and find that they are almost identical [Figs.~\ref{SC}(e) and \ref{SC}(f)], indicating the absence of hole pairing.

In the CDW phase, the spin correlation decays very fast.
With growing $t_2$ and $J_2$, spin correlation is also enhanced in the Fermi-liquid-like phase, as shown in Fig.~\ref{Sr}(a).
The strong spin correlation persists across different lengths $L_x$ and doping ratios $\delta=1/36-1/18$ [Fig.~\ref{Sr}(b)]. 
To obtain a better understanding of the decay behavior of spin correlation, we extrapolate the DMRG results to the $D \rightarrow \infty$ limit and fit the data using both semi-log and double-log plots [Figs.~\ref{Sr}(c) and \ref{Sr}(d)]. 
The data at short distance $r<30$ can be algebraically fitted quite well with a small power exponent $K_s\approx0.64$. At longer distance $r > 30$, the results continuously improve as the bond dimension increases, and the extrapolated data in the $D \rightarrow \infty$ limit are close to the fitting curve. These observations indicate that the decay behavior is more consistent with a power-law form.

\subsection{Momentum distribution and spin structure factor}

\begin{figure}[h] 
	\centering
    \begin{subfigure}[b]{0.48\textwidth}
		\includegraphics[width=\textwidth]{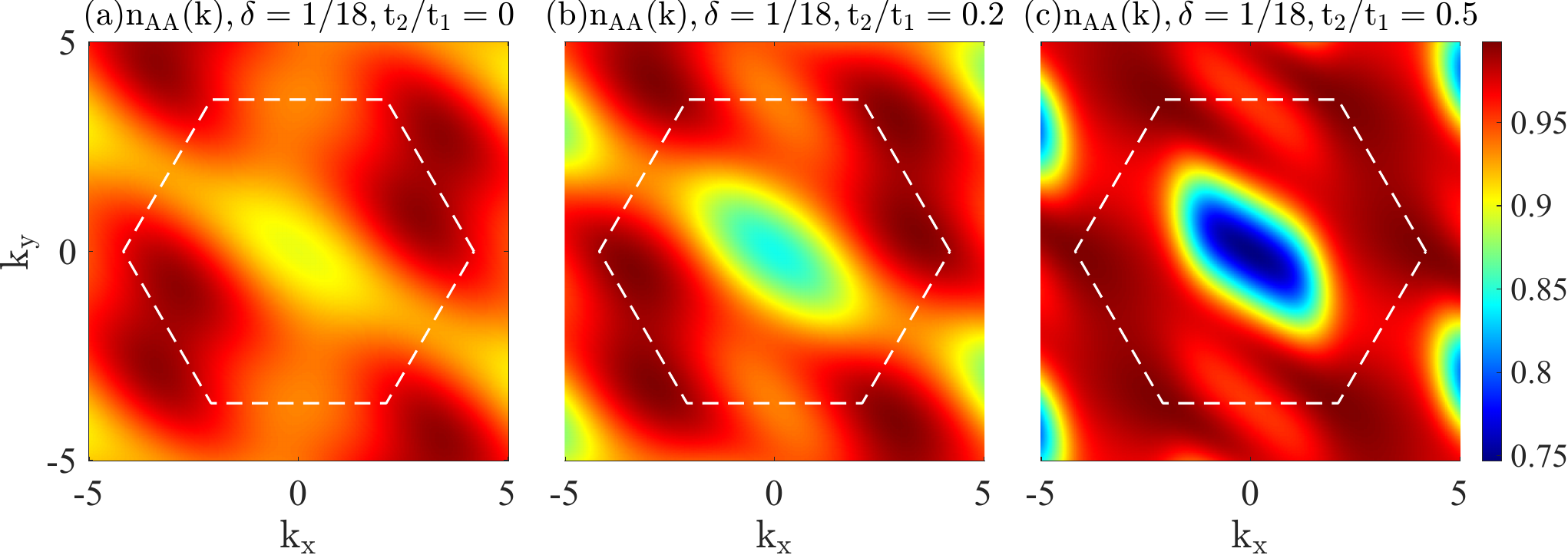} 
	\end{subfigure}
	\includegraphics[width=0.48\textwidth]{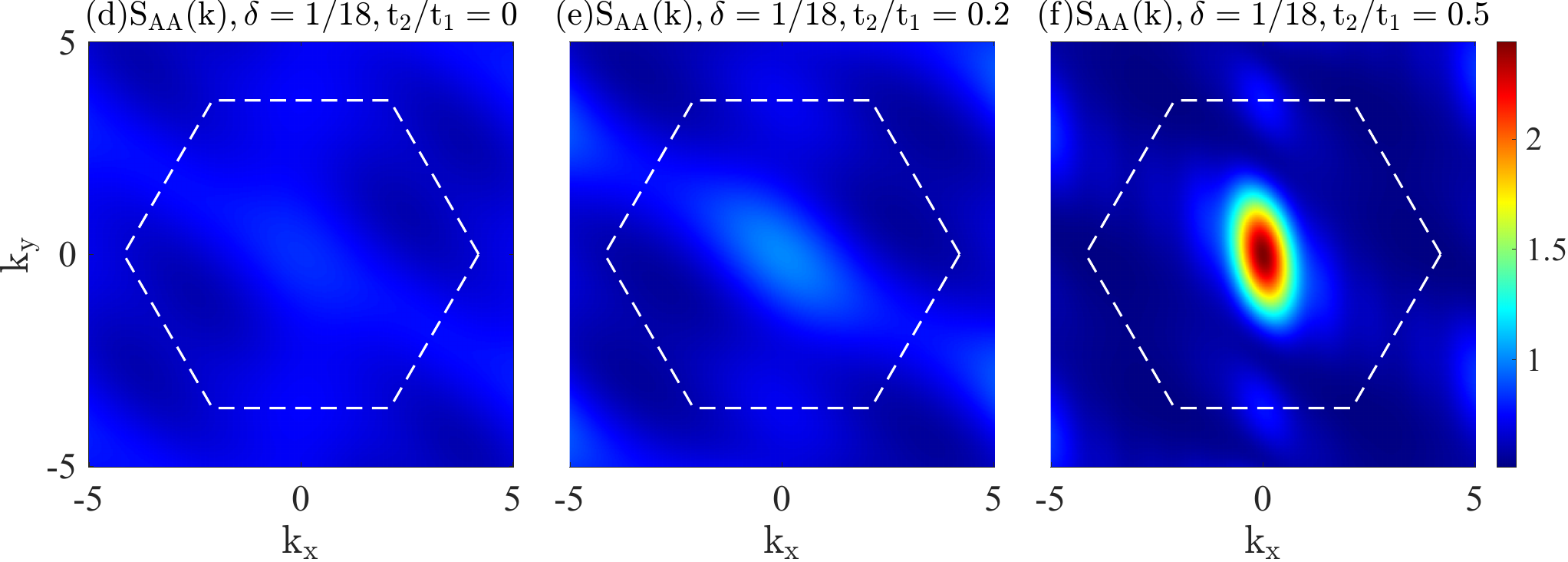} 
	\caption{\justifying Momentum distribution function $n_{AA}(\mathbf{k})$ and spin structure factor $S_{AA}(\mathbf{k})$ of the $A$ sublattice on the YC6 cylinder. $n_{AA}(\mathbf{k})$ for (a), (b) in the CDW phase, and (c) in the Fermi-liquid-like phase. $S_{AA}(\mathbf{k})$ for (d), (e) in the CDW phase, and (f) in the Fermi-liquid-like phase. The dashed hexagon is the first Brillouin zone of the kagome lattice. We denote the ${\bf k} = (0,0)$ point as the ${\bf \Gamma}$ point.
    }
    \label{nk_sk}
\end{figure}

We further present the results of the momentum distribution function and the spin structure factor.
We have examined the results for the three sublattices, which share similar features, and here we demonstrate the results of the $A$ sublattice in Fig.~\ref{nk_sk} as a representative.
The momentum distribution function is defined as $n_{AA}(\mathbf{k}) = (1/N_A)\sum_{i,j,\sigma}e^{i\mathbf{k} \cdot (\mathbf{r}_i-\mathbf{r}_j)}\langle \hat{c}_{i,\sigma}^\dagger \hat{c}_{j,\sigma}\rangle$, and the spin structure factor is given by $S_{AA}(\mathbf{k}) = (1/N_A) \sum_{i,j} \langle \mathbf{S}_i \cdot \mathbf{S}_j \rangle e^{i \mathbf{k} \cdot (\mathbf{r}_i - \mathbf{r}_j)}$, where the sites $i,j$ belong to the $A$ sublattice and $N_A$ is the number of sites.

In Figs.~\ref{nk_sk}(a)-\ref{nk_sk}(c), we show $n_{AA}(\bf k)$ at $\delta = 1/18$ with increasing $t_2$ and $J_2$.
In the CDW phase, the doped holes are relatively dispersed, though a small fraction of them are concentrated near the center of the Brillouin zone, i.e. around the ${\bf \Gamma}$ point.
In the Fermi-liquid-like phase, the holes further concentrate near the $\bf \Gamma$ point and form a hole pocket. 
Note that the jump of $n_{AA}(\bf k)$ across the Fermi surface, which reflects the quasi-particle weight in the Fermi liquid, is only of the order $0.1$. 
This is consistent with the fact that the quasi-particle weight in a doped Mott insulator usually scales with the doping ratio. 
For spin structure factor $S_{AA}(\bf k)$, it is featureless in the CDW phase, consistent with the very short spin correlation length shown in Fig.~\ref{Sr}(a).
With increasing $t_2$ and $J_2$, $S_{AA}(\bf \Gamma)$ grows gradually and becomes sharp in the Fermi-liquid-like phase, which agrees with the strong spin correlation in Fig.~\ref{Sr}.
We have also examined the spin correlations in real space, which are consistent with a three-sublattice structure and should originate from the ${\bf k} = (0,0)$ magnetic order of the $J_1$-$J_2$ kagome Heisenberg model~\cite{kagome-J1J2-Schollwock-2015,kagome-J1J2J3-ssg-2015}.
For a complete description of the results, we also show the momentum distributions and spin structure factors involving different sublattices in Appendix~\ref{app-3}, where the negative peaks of $S_{AB}$ and $S_{AC}$ at the ${\bf \Gamma}$ point agree with the ${\bf k} = (0,0)$ three-sublattice spin correlation.

\subsection{Entanglement entropy and central charge}

\begin{figure}[h] 
	\centering
		\includegraphics[width=0.236\textwidth]{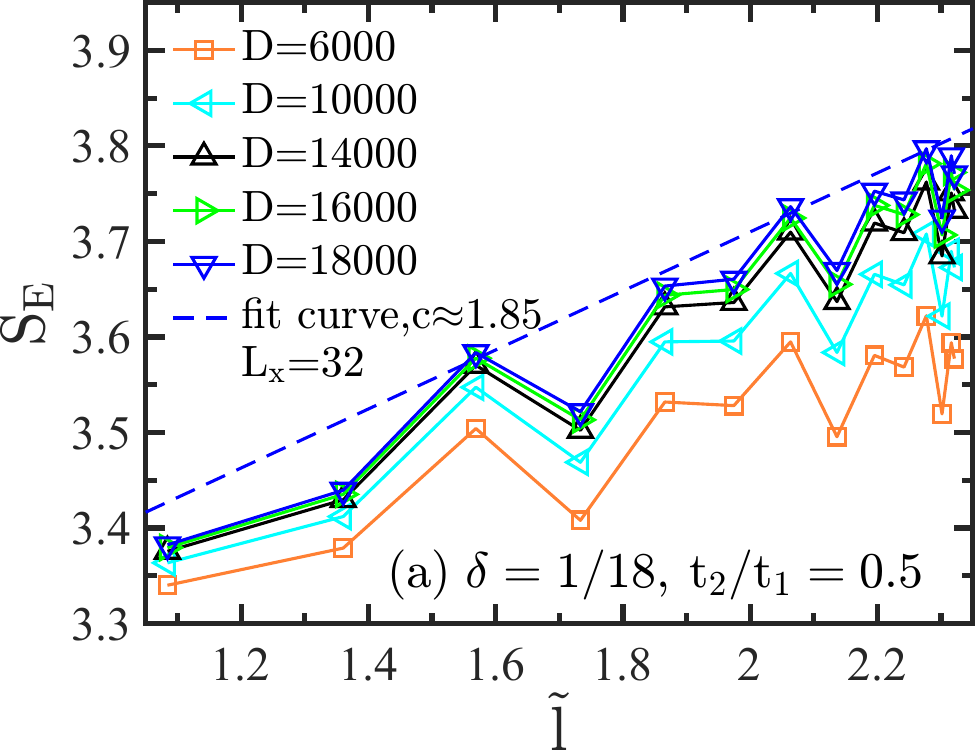} 
		\includegraphics[width=0.236\textwidth]{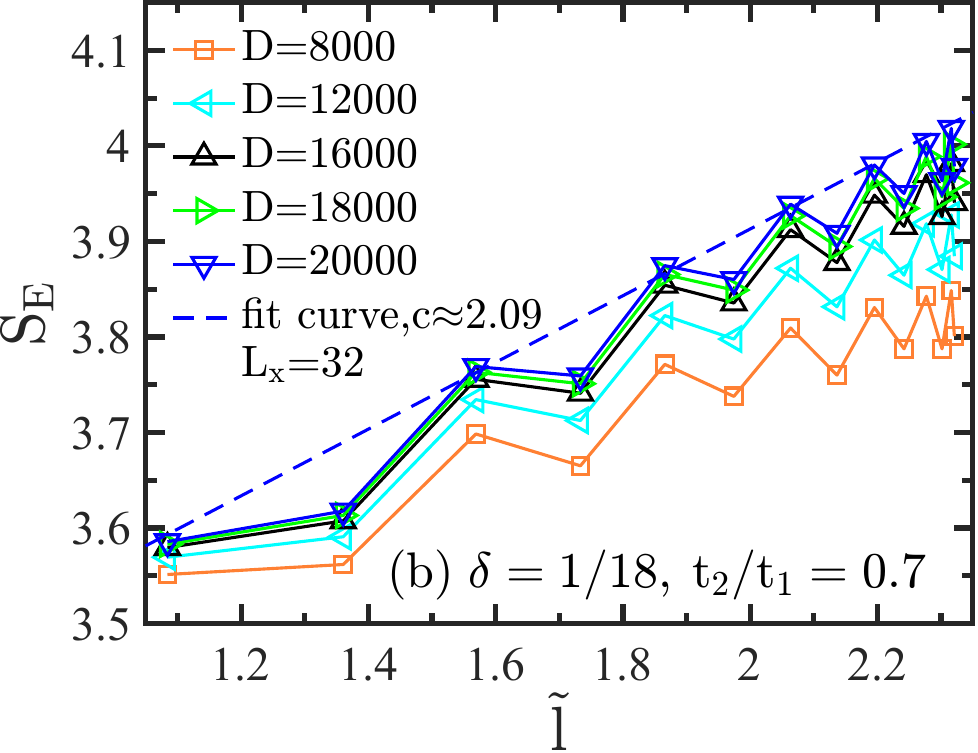} 
	\caption{\justifying Entanglement entropy and central charge for different bond dimensions in the $L_x=32$ YC6 system with $\delta = 1/18$. (a) and (b) show the results of $t_2/t_1=0.5,J_2/J_1=0.25$ and  $t_2/t_1=0.7,J_2/J_1=0.49$, respectively. The x-coordinate is given by the conformal distance: $\tilde{l} = \ln \left[(L_x/\pi) \sin \left(l \pi/L_x \right) \right]$, where $l$ represents the index of the unit cell in x-direction. The dashed line denotes the linear fit to $S_E = \frac{c}{6} \tilde{l} + \text{const.}$ with the results of maximum bond dimension.}
    \label{SE}
\end{figure}

In this subsection, we calculate the entanglement entropy and fit the central charge to detect the gapless nature of the Fermi-liquid-like phase. 
The entropy is defined as $S_E(l) = - \text{Tr}[\rho(l) \ln \rho(l)]$ ($1\leq l \leq L_x-1$), where $l$ and $L_x - l$ denote the lengths of the two subsystems when we divide the cylinder along the $\mathbf{e}_2$ direction, and $\rho(l)$ is the reduced density matrix of the subsystem $l$ obtained from the ground state of the whole system.
According to conformal field theory, for one-dimensional critical systems with open boundary conditions, the bipartite entanglement entropy $S_E$ is expected to scale linearly with the logarithmic conformal distance of the subsystem $l$:
$S_E = \frac{c}{6} \tilde{l} + g$, where $\tilde{l} = \ln \left[(L_x/\pi) \sin \left(l \pi/L_x \right) \right]$ \cite{Entanglement_entropy_2004}, where $g$ is a non-universal constant. 
In Fig.~\ref{SE}, we show the entropy $S_{E}$ versus $\tilde{l}$ in the Fermi-liquid-like phase including $t_2/t_1 = 0.5$ and $0.7$ at $\delta = 1/18$, $L_x = 32$.
With increased bond dimensions, $S_{E}$ continues to grow and approaches convergence. 
At the largest bond dimension, the central charge is fitted to $c \approx 2$, indicating the presence of gapless modes in the Fermi-liquid-like phase.

\subsection{Intermediate region}

\begin{figure}[h] 
	\centering
	\begin{subfigure}{0.48\textwidth}
		\includegraphics[width=0.494\textwidth]{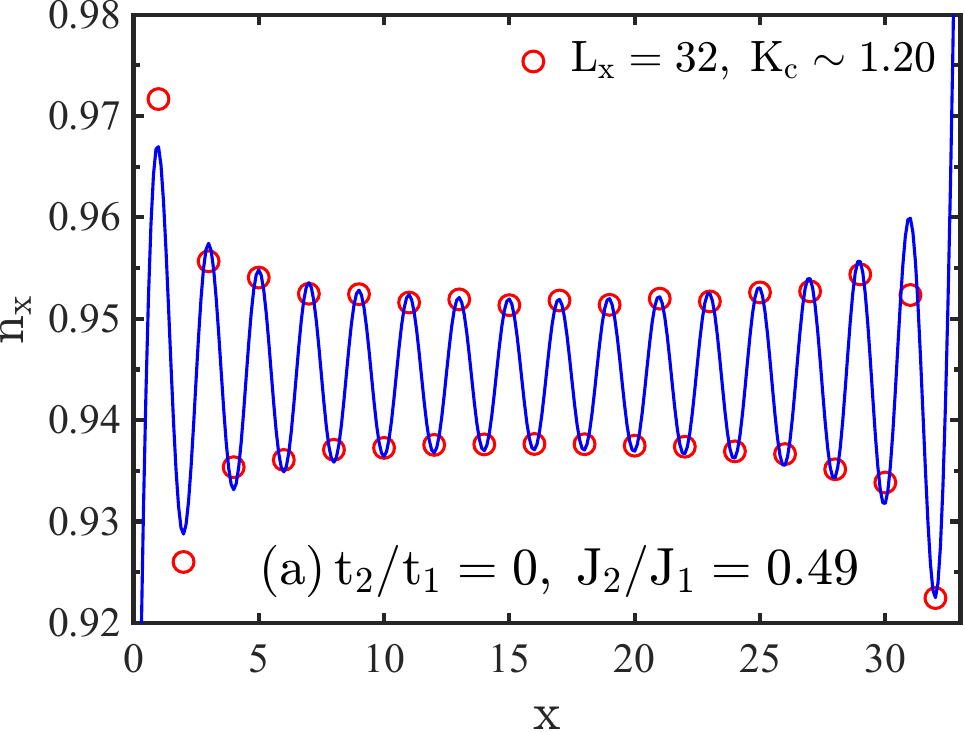} 
		\raisebox{1mm}{\includegraphics[width=0.48\textwidth]{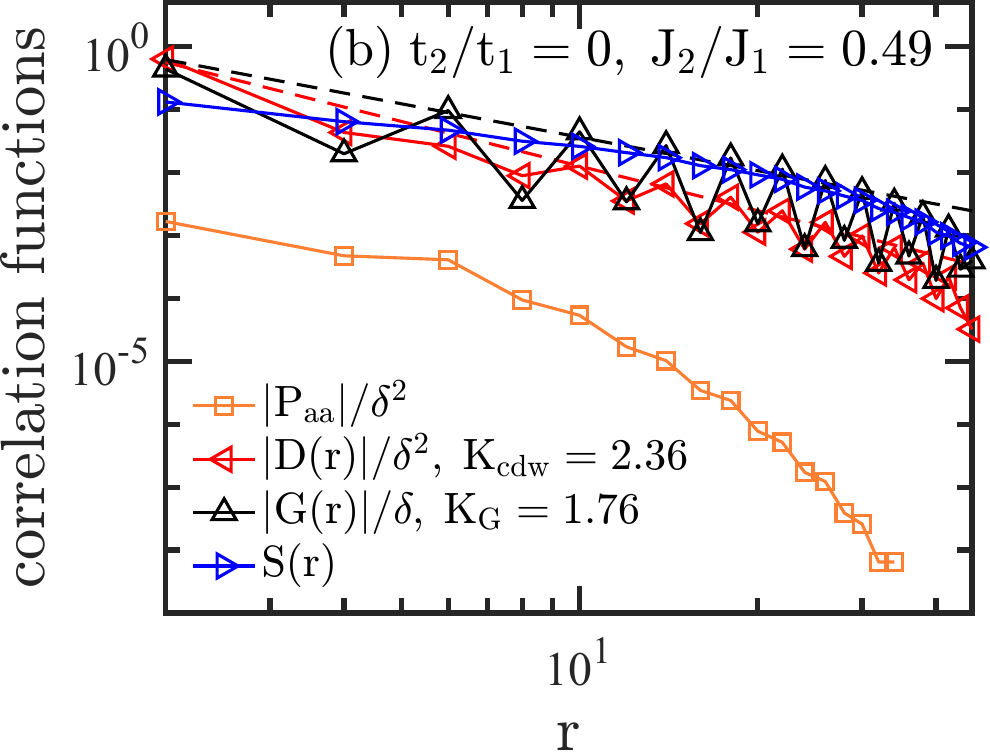}}
	\end{subfigure}
	\begin{subfigure}[b]{0.48\textwidth}
		\includegraphics[width=0.494\textwidth]{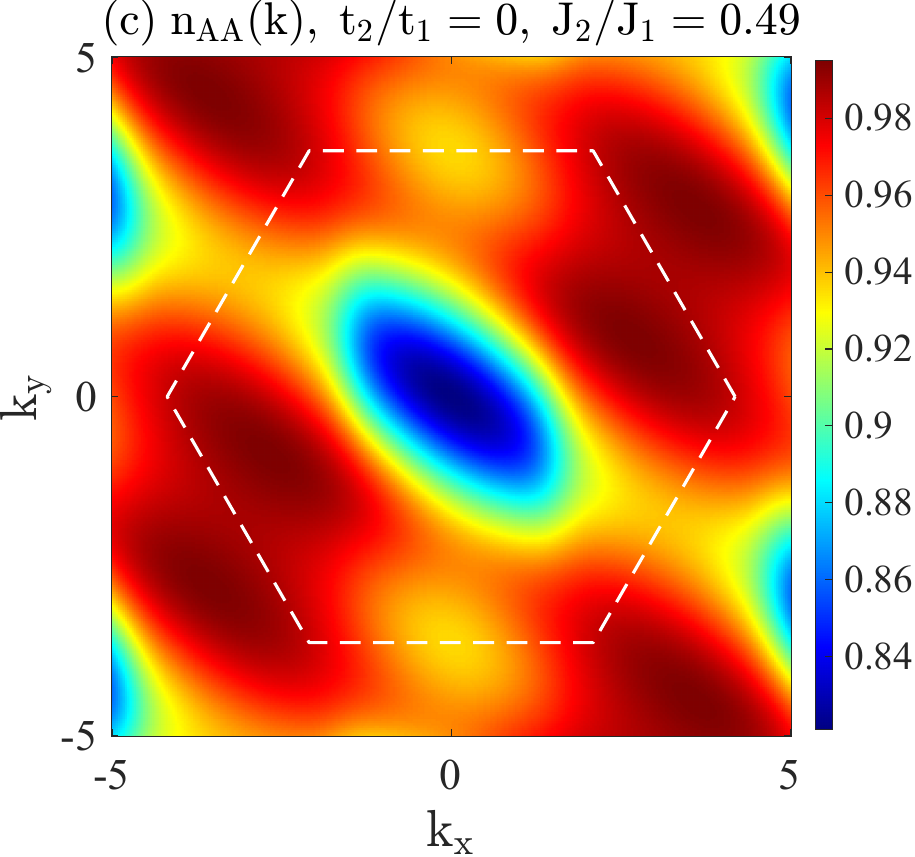}
        \includegraphics[width=0.488\textwidth]{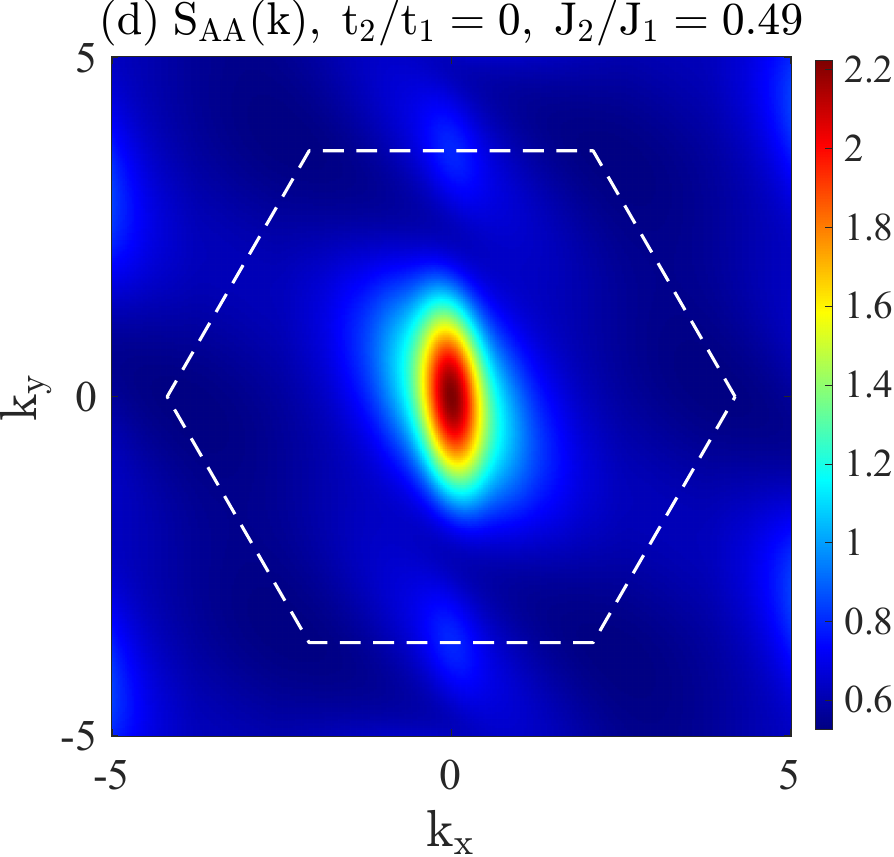}
	\end{subfigure}
	\caption{\justifying DMRG results for $t_2/t_1=0$, $J_2/J_1=0.49$, and $\delta=1/18$ on the YC6 cylinder. (a) Charge density profile $n_x$ and the corresponding fitting curve. (b) Comparison among the pairing correlation $P_{aa}$, density correlation $D(r)$, spin correlation $S(r)$, and single-particle Green’s function $G(r)$.
    The correlation functions are rescaled. (c) and (d) are the momentum distribution function $n_{AA}(\mathbf{k})$ and spin structure factor $S_{AA}(\mathbf{k})$, respectively.}
    \label{0049}
\end{figure}
\begin{figure}[h] 
	\centering
	\begin{subfigure}[b]{0.48\textwidth}
		\includegraphics[width=0.494\textwidth]{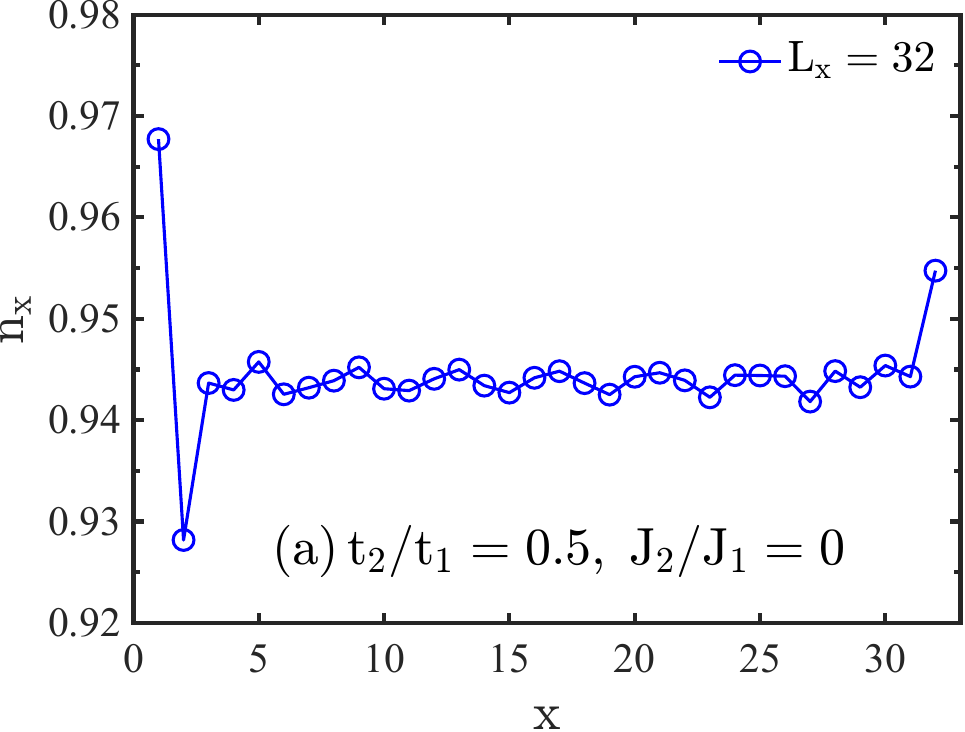} 
		\raisebox{1mm}{\includegraphics[width=0.48\textwidth]{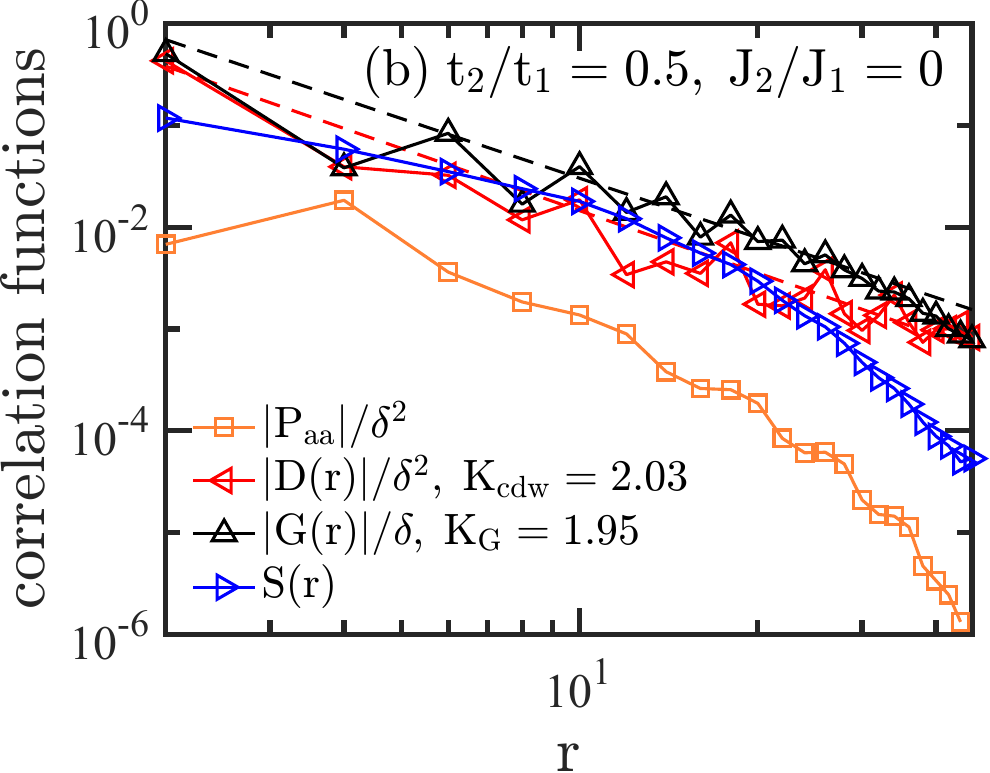}}
	\end{subfigure}
	\begin{subfigure}[b]{0.48\textwidth}
		\includegraphics[width=0.494\textwidth]{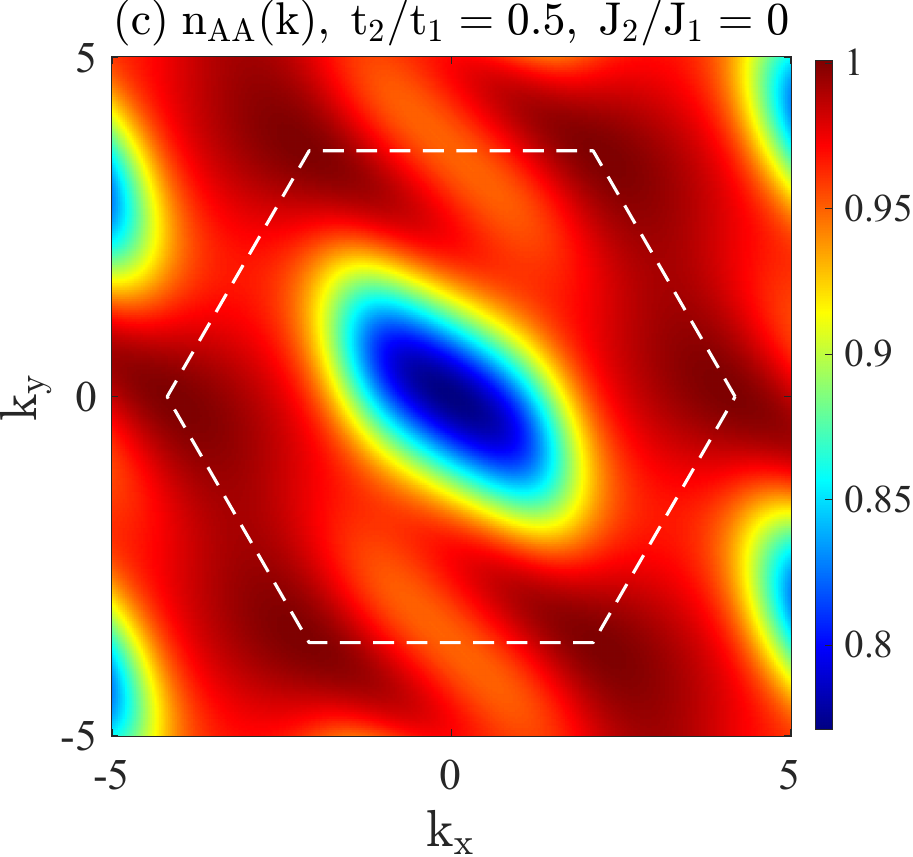}
        \includegraphics[width=0.488\textwidth]{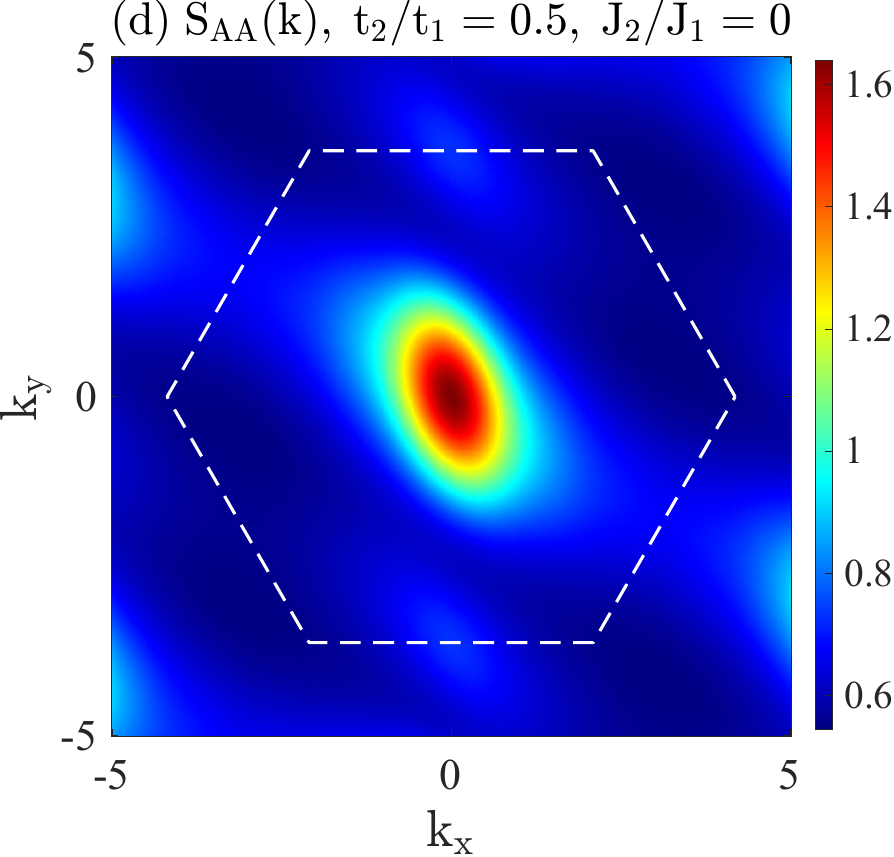}
	\end{subfigure}
	\caption{\justifying  DMRG results for $t_2/t_1=0.5$, $J_2/J_1=0$, and $\delta=1/18$ on the YC6 cylinder. (a) Charge density profile $n_x$. (b) Comparison among the pairing correlation $P_{aa}$, density correlation $D(r)$, spin correlation $S(r)$, and single-particle Green’s function $G(r)$. The correlation functions are rescaled. (c) and (d) are the momentum distribution function $n_{AA}(\mathbf{k})$ and spin structure factor $S_{AA}(\mathbf{k})$, respectively.}
    \label{0500}
\end{figure}

In the phase diagram Fig.~\ref{lattice_diagram}(b), besides the CDW and Fermi-liquid-like phase, we also find an intermediate regime (green color) with dominant $t_2$ or $J_2$.
In this regime, while most physical quantities are already similar to those of the Fermi-liquid-like phase, some quantities vary slowly with tuning couplings.
For example, at $t_2 =0$, $J_2/J_1 = 0.49$, $\delta = 1/18$ [Fig.~\ref{0049}], the density correlation, single-particle correlation, spin correlation, and spin structure factor are comparable to those in the Fermi-liquid-like phase, but the pairing correlation remains to decay exponentially, and the CDW still exhibits a moderate oscillation amplitude. 
In addition, although the electron momentum distribution $n_{AA}(\mathbf{k})$ shows the signal of a hole pocket at the $\bf \Gamma$ point, there are still visible hole occupations at other momentum points.
On the other hand, for $t_2/t_1 = 0.5$, $J_2=0$, $\delta = 1/18$ [Fig.~\ref{0500}], the CDW is significantly suppressed and the hole pocket of $n_{AA}(\bf k)$ is sharp, which are similar to those in the Fermi-liquid-like phase.
For spin correlation, although $S_{AA}(\bf k)$ shows a peak at the $\bf \Gamma$ point, the spin correlation is relatively weaker compared to the only $J_2$ case and appears to decay exponentially.
The different coupling dependence of these physical quantities in this intermediate regime may be attributed to the fact that $t_2$ and $J_2$ are directly coupled to the charge and spin degrees of freedom, respectively.

\section{Doping ratio dependence of the Fermi-liquid-like phase}

\begin{figure}[h] 
	\centering
	\begin{subfigure}[b]{0.48\textwidth}
		\includegraphics[width=0.494\textwidth]{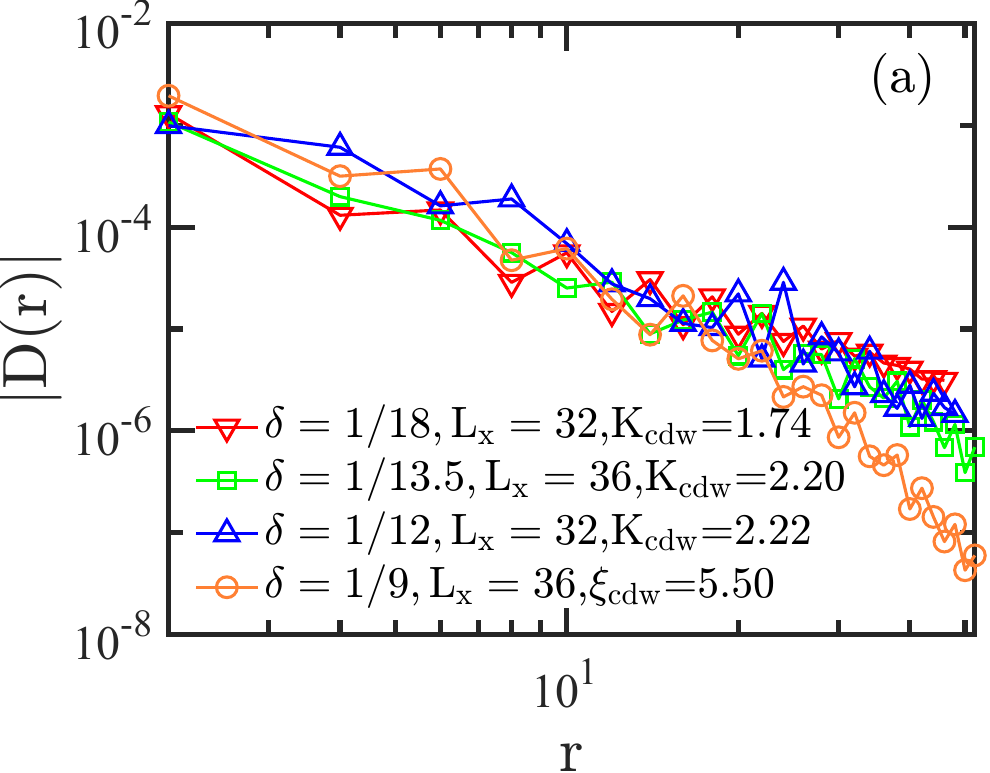} 
		\includegraphics[width=0.494\textwidth]{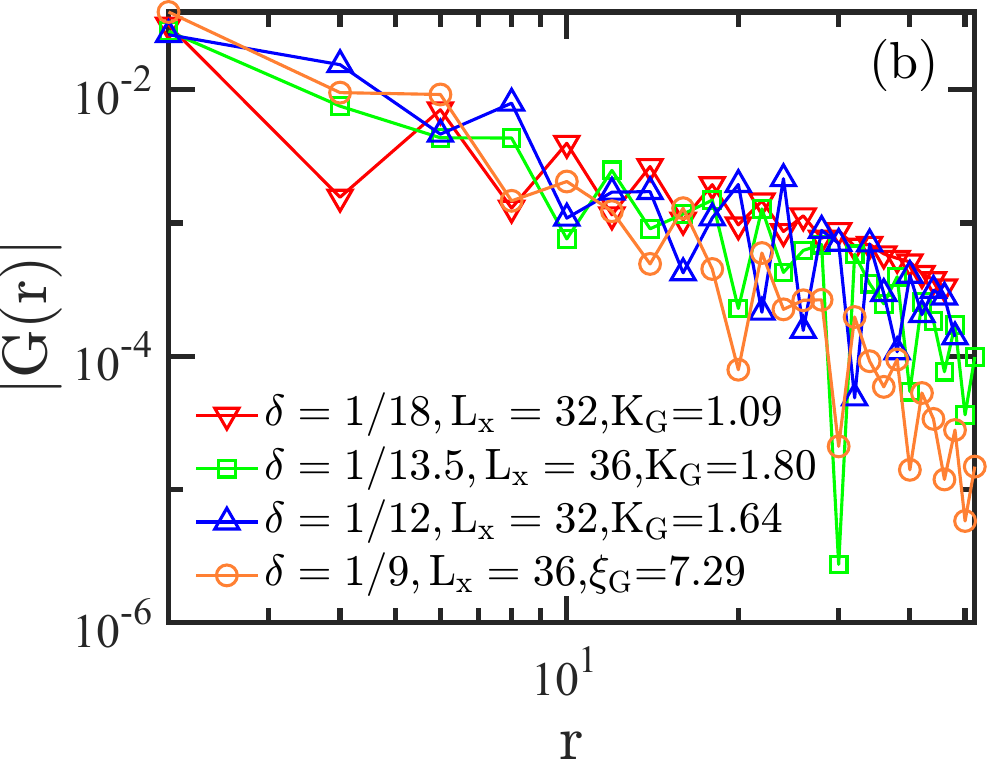} 
	\end{subfigure}
	\begin{subfigure}[b]{0.48\textwidth}
		\includegraphics[width=0.494\textwidth]{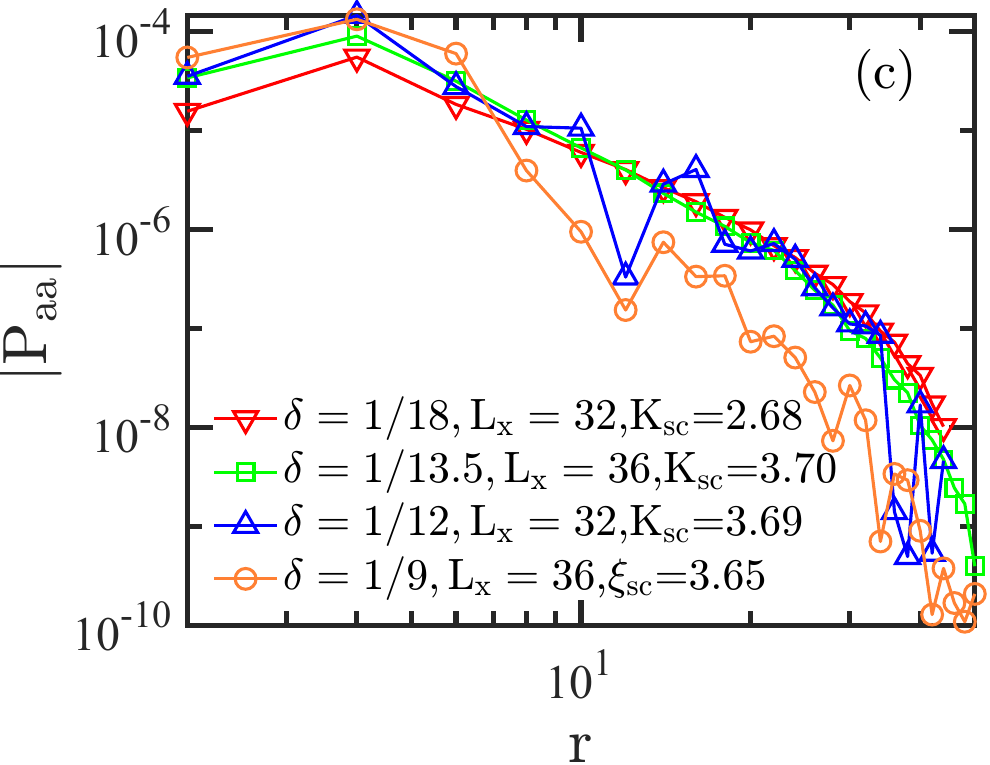} 
		\includegraphics[width=0.494\textwidth]{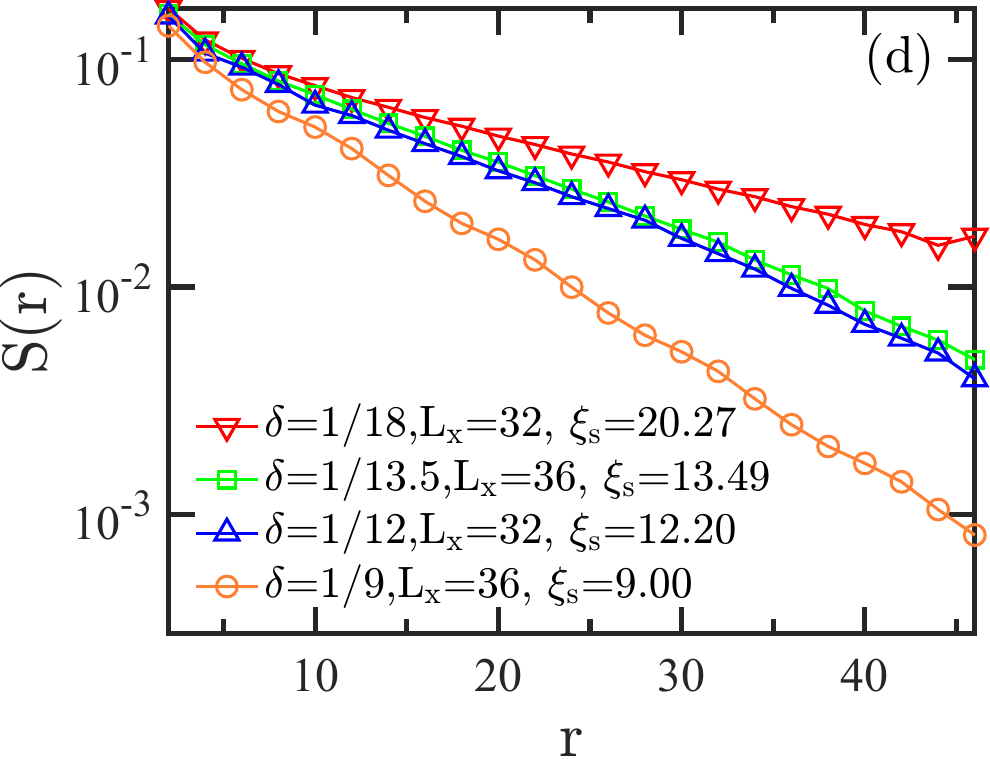} 
	\end{subfigure}
    \begin{subfigure}[b]{0.48\textwidth}
		\includegraphics[width=\textwidth]{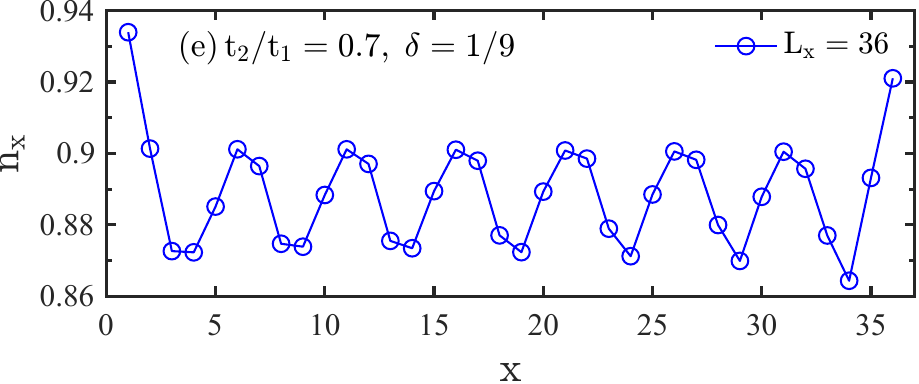} 
	\end{subfigure}
	\caption{\justifying DMRG results for $t_2/t_1=0.7$, $J_2/J_1=0.49$, and $\delta=1/18-1/9$ on the YC6 cylinder. (a), (b), (c) and (d) show the results of density correlation, single-particle Green’s function, pairing correlation, and spin correlation, respectively. (e) Charge density profile $n_x$ for the $L_x = 36$ system at $\delta = 1/9$.}
    \label{large_dp1}
\end{figure}

In the previous sections, we have shown the stable Fermi-liquid-like phase at doping levels $\delta = 1/36 - 1/18$.
Here, we further examine the system across a broader doping range $\delta = 1/18 - 1/9$ (with $t_2/t_1 = 0.7$, $J_2/J_1 = 0.49$) to investigate the doping ratio dependence of this Fermi-liquid-like phase on the YC6 cylinder.

We show the correlation functions in Figs.~\ref{large_dp1}(a)-\ref{large_dp1}(d).
The results at $\delta = 1/18 - 1/12$ are consistent, showing the existence of the Fermi-liquid-like phase up to $\delta = 1/12$.
However, at $\delta = 1/9$ the density correlation, the singlet-particle correlation and the pairing correlation all decay exponentially.
Meanwhile, the spin correlation also decays much faster.
We further examine the charge density profile $n_x$ at $\delta = 1/9$, as shown in Fig.~\ref{large_dp1}(e), which exhibits a strong oscillation. 
These observations suggest a CDW state at $\delta = 1/9$, characterizing a phase transition from the Fermi-liquid-like phase to a CDW phase near this doping level. 
We also notice that on the YC6 cylinder, this CDW state is not connected to the charge order state at $t_2 = J_2 = 0$~\cite{kagome-tJ-Jiang-2017}, since these two CDW states have different charge density distributions at $\delta = 1/9$.

\section{The Fermi-liquid-like state in the lightly doped YC8 system}

\begin{figure}[h] 
	\centering
	\begin{subfigure}[b]{0.48\textwidth}
		\includegraphics[width=0.494\textwidth]{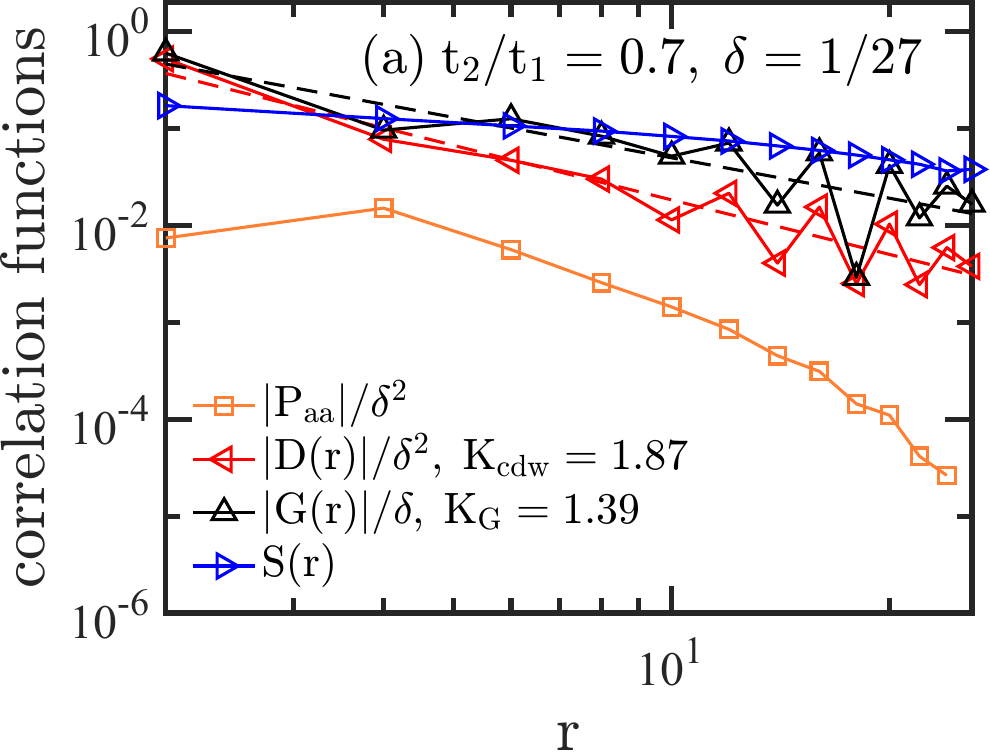} 
		\includegraphics[width=0.494\textwidth]{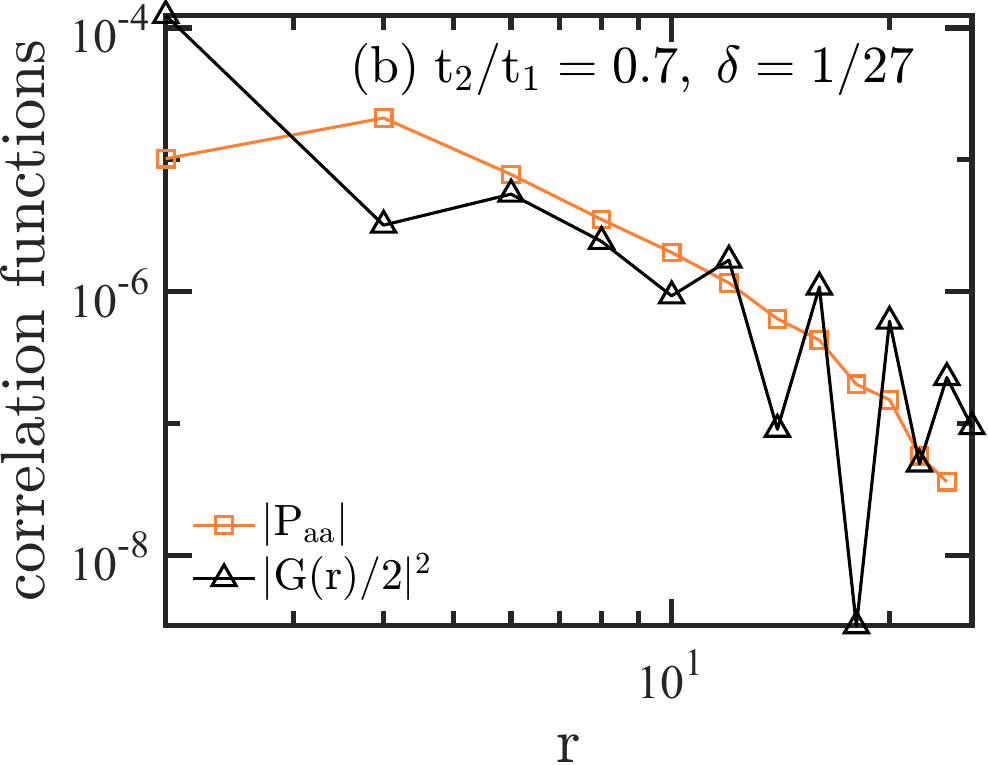} 
	\end{subfigure}
	\begin{subfigure}[b]{0.48\textwidth}
		\includegraphics[width=\textwidth]{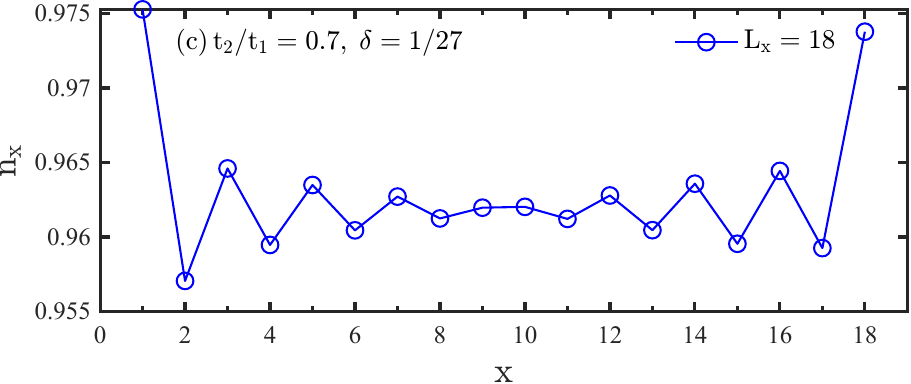} 
	\end{subfigure}
	\begin{subfigure}[b]{0.48\textwidth}
		\includegraphics[width=0.494\textwidth]{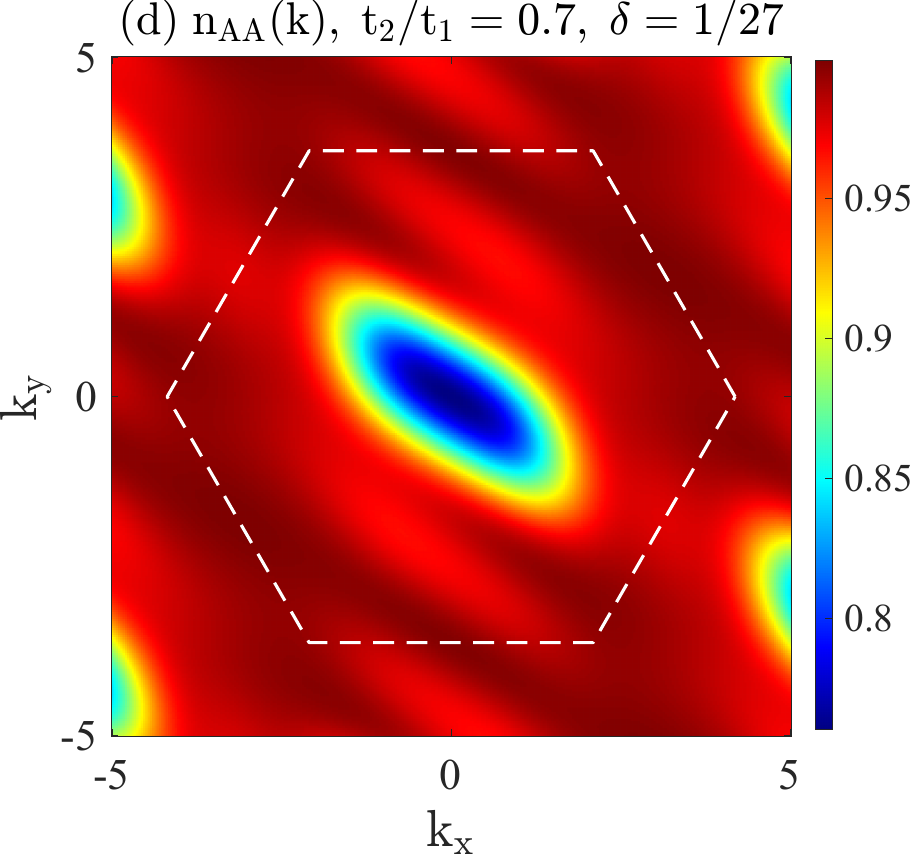}
        \includegraphics[width=0.488\textwidth]{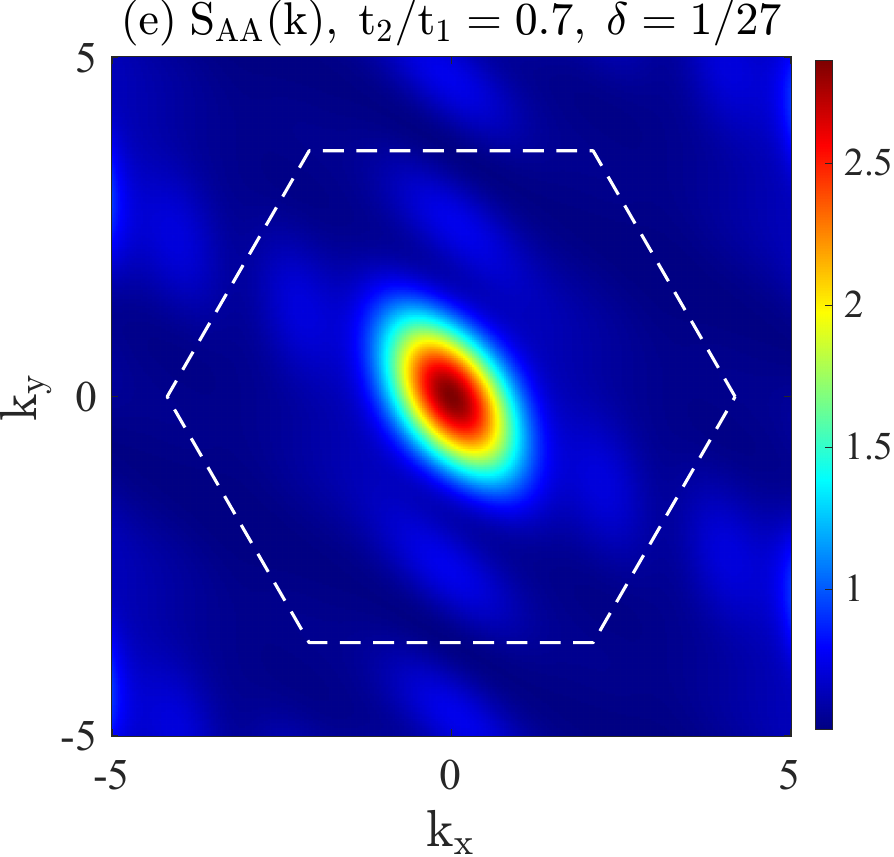}
	\end{subfigure}
	\caption{\justifying DMRG results for $t_2/t_1=0.7$, $J_2/J_1=0.49$, and $\delta = 1/27$ on a YC8 cylinder with $L_x = 18$. (a) Comparison among the pairing correlation $P_{aa}$, density correlation $D(r)$, spin correlation $S(r)$, and single-particle Green’s function $G(r)$. The correlation functions are rescaled. (b) Comparison of the pairing correlation $P_{aa}$ with the the square of single-particle Green’s function $G(r)$. (c) Charge density profile $n_x$. (d) and (e) are the momentum distribution function $n_{AA}(\mathbf{k})$ and spin structure factor $S_{AA}(\mathbf{k})$, respectively.}
    \label{YC8_27}
\end{figure}

Based on the phase diagram Fig.~\ref{lattice_diagram}(b) of the YC6 system, we further examine the Fermi-liquid-like state on the wider YC8 cylinder. 
We have tested doping levels at $\delta = 1/18$, $1/24$, and $1/27$. 
Since the simulations at $\delta = 1/18$ and $1/24$ are much harder to converge, here we only present the results at $\delta = 1/27$ (with $t_2/t_1 = 0.7$, $J_2/J_1 = 0.49$).
In Fig.~\ref{YC8_27}(a), we display correlation functions, where both single-particle and density correlations clearly maintain algebraic decay characteristics, with the power exponents close to those of the YC6 system.   
The spin correlation is also strong and $S_{AA}(\bf k)$ shows a peak at the $\Gamma$ point [Fig.~\ref{YC8_27}(e)].
Moreover, the magnitude of the pairing correlation remains comparable to the square of single-particle correlation, demonstrating the absence of hole pairing [Fig.~\ref{YC8_27}(b)]. 
For the charge density profile $n_x$, it exhibits a small oscillation amplitude and the oscillation decays very fast [Fig.~\ref{YC8_27}(c)], which appears like a Friedel oscillation. 
Similar to the Fermi-liquid-like state in the YC6 system, the momentum distribution function $n_{AA}(\mathbf{k})$ also shows a hole pocket near $\bf k = \Gamma$ [Fig.~\ref{YC8_27}(d)].
In Appendix~\ref{app-3}, we also show the $n({\bf k})$ and $S({\bf k})$ that involve different sublattices, which all agree with the results of the Fermi-liquid-like state in the YC6 system.

In addition, we have also examined the entanglement entropy of the YC8 system. Since entropy is harder to converge than local measurement, the entropy data of the YC8 system are not fully converged, which still keep growing with increasing bond dimensions. By fitting the YC8 entropy data, we can obtain a finite central charge that also grows with increasing bond dimensions, supporting the gapless nature of the state.
These consistent observations demonstrate the Fermi-liquid-like state in both YC6 and YC8 systems, which may extend to the lightly doped two-dimensional system.

\section{Summary and discussion}

Using DMRG calculations, we have studied an extended $t$-$J$ model on the kagome lattice with the additional NNN hopping $t_2$ and spin exchange interaction $J_2$.
We focus on the YC6 cylinder with $t_1/J_1 = 3$ and map out the quantum phase diagram by tuning $t_2 > 0, J_2 > 0$ at the doping ratio $\delta = 1/18$. 

With increased $t_2$ and $J_2$, the system shows a transition from the CDW phase at small $t_2, J_2$ to a Fermi-liquid-like phase.
In this Fermi-liquid-like phase, the charge density oscillation is significantly suppressed.
The pairing correlation, single-particle correlation, density correlation, and spin correlation are all greatly enhanced compared to those in the CDW phase.  
In particular, the single-particle correlation shows a good power-law decay with the exponent $K_{\rm G} \gtrsim 1$.
Although the pairing correlation also exhibits a power-law decay, the exponent $K_{\rm sc} > 2$ and the approximate equivalence between the pairing correlation and the squared single-particle correlation indicate the absence of hole pairing.
The gapless nature of this phase is further supported by a finite central charge. 
In the YC6 system, the Fermi-liquid-like phase can extend up to $\delta = 1/12$, and at the larger doping ratio such as $\delta = 1/9$, the system shows a transition to another CDW state.
We also examine the YC8 system with $t_2$ and $J_2$, which can also host this Fermi-liquid-like state at a small doping ratio. 
In both YC6 and YC8 systems, spin correlations are strong in this Fermi-liquid-like phase, which may persist in two dimensions and give rise to a Fermi liquid with the three-sublattice magnetic order at small doping level.

In the square-lattice $t$-$J$ model at small hole concentration, the increased $t_2 > 0, J_2 > 0$ can suppress the striped CDW order and give rise to a $d$-wave superconducting phase~\cite{Square-tt'JJ'-JiangHC-2021,Square-tt'JJ'-gss-2021,Square-tt'J-JiangShengtao-2021,Square-tt'JJ'-gss-2024,chen_2025,jiang_2023}.
The striped CDW state also features the existence of hole pairing, and increasing $t_2 > 0$ may play a crucial role to enhance coherence and give rise to a quasi-long-range superconducting order on finite-width systems~\cite{Square_tt'J-Luxin-2024}. 
In this kagome $t$-$J$ model, the absence of hole pairing in the CDW phase (near $t_2 = J_2 = 0$) may be the reason that increasing $t_2, J_2$ leads to a Fermi-liquid-like phase instead of a superconducting phase.

In the kagome lattice, the emergence of a superconducting phase appears to be challenging because of the absence of hole pairing.
Nevertheless, our study provides a foundation for further exploration of SC in kagome systems. 
It is possible that by testing different interaction mechanisms, a gap may open in the system, which can be explored in future studies.

\begin{acknowledgments}
X.~Y.~J. and S.~S.~G. were supported by the National Natural Science Foundation of China (No. 12274014), the Special Project in Key Areas for Universities in Guangdong Province (No. 2023ZDZX3054), and the Dongguan Key Laboratory of Artificial Intelligence Design for Advanced Materials. F.~Y. was supported by the National Natural Science Foundation of China under the Grant Nos. 12234016 and 12074031.  D.~N.~S. was supported by the U.S. Department of Energy, Office of Basic Energy Sciences under Grant No. DE-FG02-06ER46305 for DMRG studies of unconventional superconductivity.
The computational resources were supported by the SongShan Lake HPC Center (SSL-HPC) at Great Bay University (X. Y. J. and S. S. G.). 
The numerical simulation was in part supported by the US National Science Foundation instrument grant DMR-2406524 (D.~N.~S.).
\end{acknowledgments}

\appendix

\section{Extrapolation of correlation functions with growing bond dimension}
\label{app-1}

\begin{figure}[h] 
	\centering
	\begin{subfigure}[b]{0.48\textwidth}
		\includegraphics[width=0.494\textwidth]{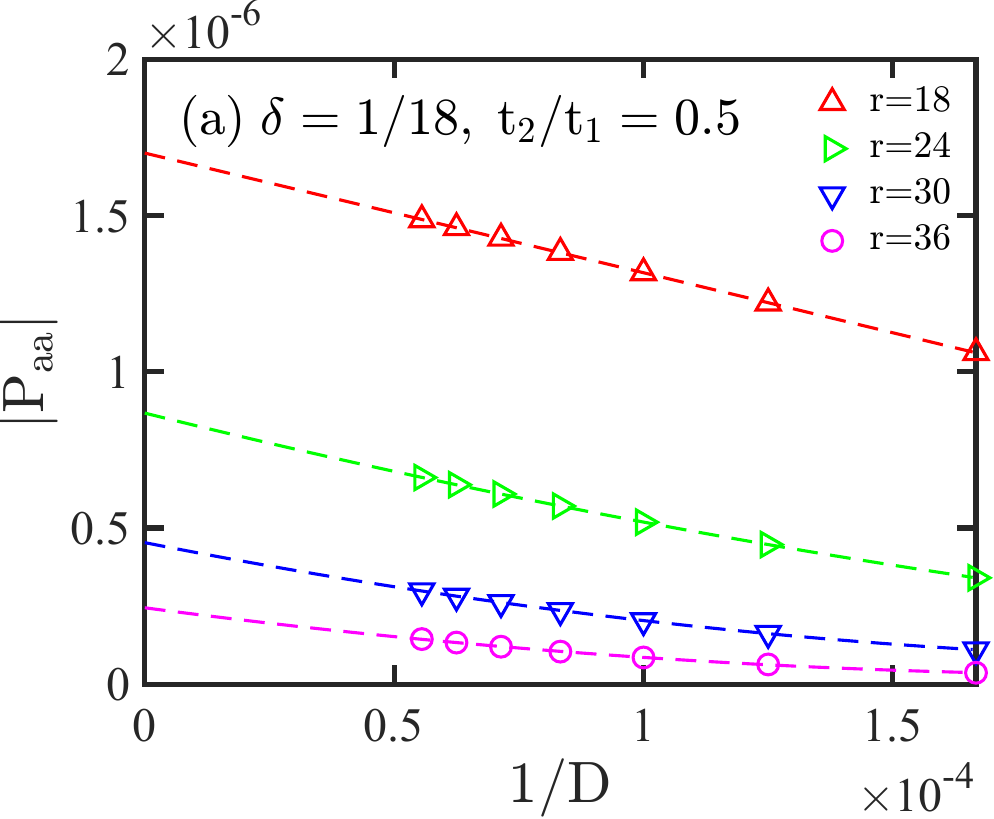} 
		\includegraphics[width=0.494\textwidth]{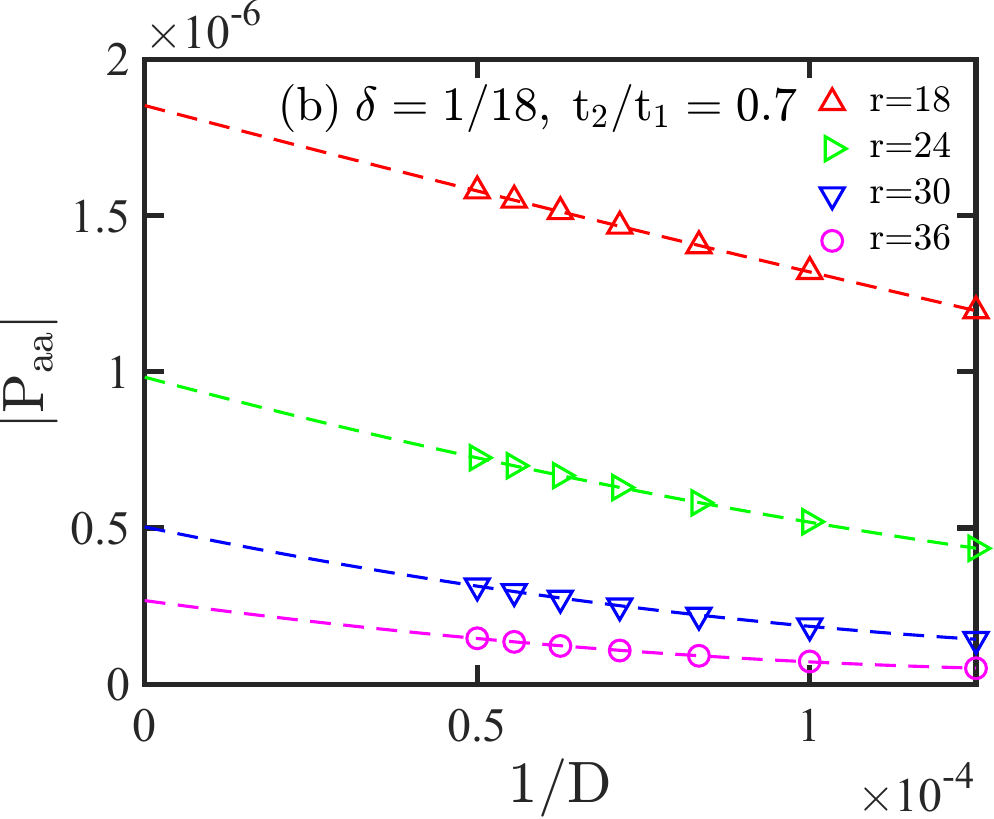} 
	\end{subfigure}
	\caption{\justifying  Extrapolations of pairing correlation functions $P_{aa}$ versus bond dimension at $\delta = 1/18$. (a) and (b) correspond to parameter sets $t_2/t_1=0.5,J_2/J_1=0.25$ and $t_2/t_1=0.7,J_2/J_1=0.49$, respectively. The $SU(2)$ bond dimension $D$ ranges from $6000$ to $18000$ in (a) and from $8000$ to $20000$ in (b). The different symbols denote the correlations at the different given distance $r$. For each given distance $r$, the correlations obtained by different bond dimensions are extrapolated by the polynomial function $C(1/D) = C(0) + a/D + b/D^2$.}
    \label{Extrapolate}
\end{figure}

In the DMRG simulations, it inevitably has the finite-bond-dimension effect. 
To eliminate this effect and extract the intrinsic physics, the extrapolated correlations are shown in the main text. 
Here we show the extrapolation detail.

We first obtain the correlation functions at different bond dimensions, and then perform a polynomial extrapolation for the data at different $D$ to extract the result in the infinite-$D$ limit.
We fit the data for a range of bond dimensions up to the largest $SU(2)$ bond dimensions $D = 20000$ (equivalent to about $60000$ $U(1)$ states). 
Two typical examples of data extrapolation are shown in Fig.~\ref{Extrapolate}. 
For each given distance $r$, the correlations obtained by at least five different bond dimensions are extrapolated by the polynomial function $C(1/D) = C(0) + a/D + b/D^2$, where $C(0)$, $a$, and $b$ are determined by fitting the DMRG data. 
The obtained $C(0)$ is the result in the infinite-$D$ limit.

\section{Comparison of different pairing correlations}
\label{app-2}

\begin{figure}[h] 
	\centering
	\begin{subfigure}[b]{0.48\textwidth}
		\includegraphics[width=0.494\textwidth]{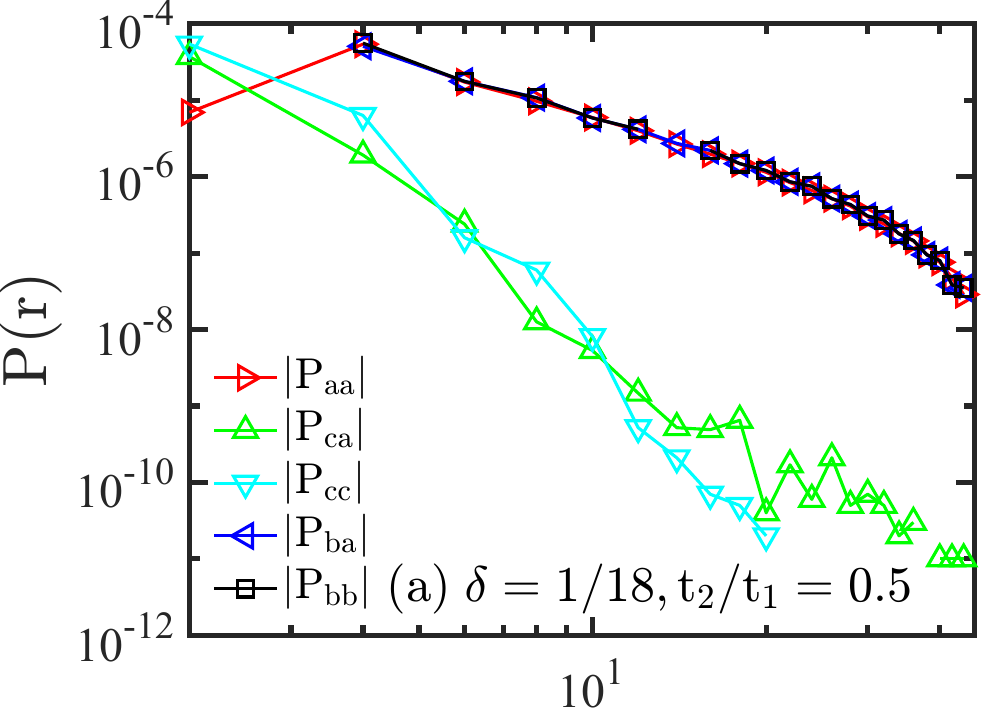} 
		\includegraphics[width=0.494\textwidth]{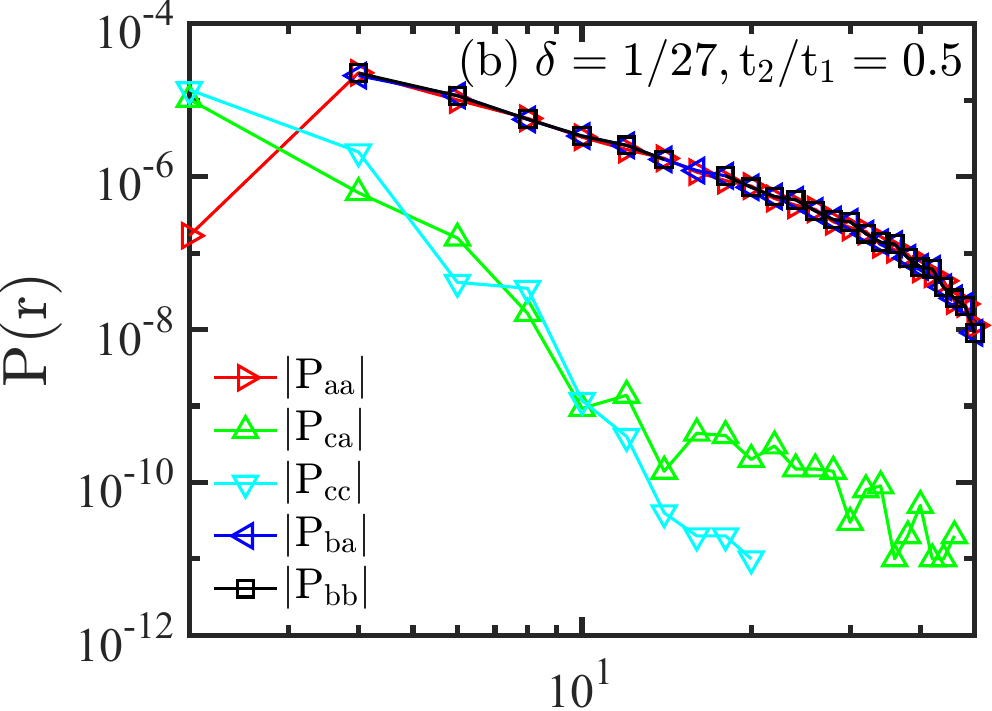} 
	\end{subfigure}
	\begin{subfigure}[b]{0.48\textwidth}
		\includegraphics[width=0.494\textwidth]{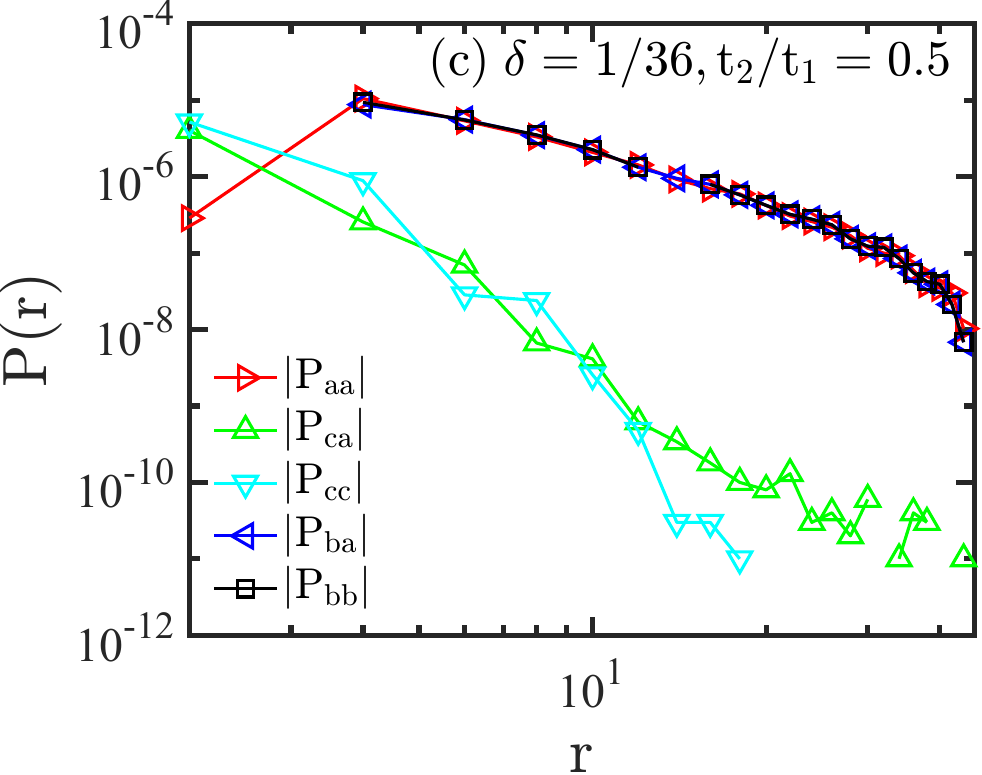} 
		\includegraphics[width=0.494\textwidth]{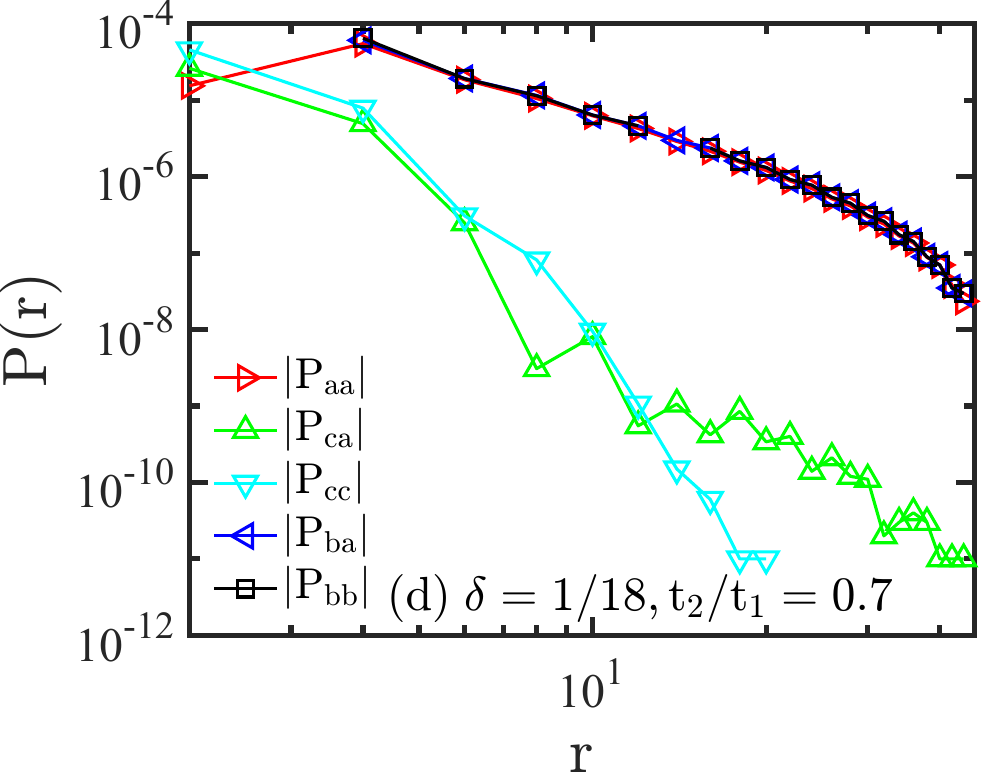} 
	\end{subfigure}
	\caption{\justifying  Pairing correlation functions $P_{\alpha,\beta}(r)$ of different bonds in the Fermi-liquid-like phase of the YC6 systems. (a), (b), and (c) are the double-logarithmic plots of $P_{\alpha,\beta}(r)$ for $t_2/t_1 = 0.5,J_2/J_1=0.25$ at $\delta = 1/18$, $1/27$, and $1/36$, respectively. (d) Double-logarithmic plot of $P_{\alpha,\beta}(r)$ for $t_2/t_1 = 0.7,J_2/J_1=0.49$ at $\delta = 1/18$.}
    \label{bond}
\end{figure}

We have examined the pairing correlation functions of different bonds in the Fermi-liquid-like phase, including $aa$, $ca$, $cc$, $ba$, and $bb$. 
As shown in Fig.~\ref{bond} of the results on the YC6 cylinder, while $P_{aa}$, $P_{ba}$, and $P_{bb}$ are essentially the same, $P_{ca}$ and $P_{cc}$ are much weaker.
Therefore, in the main text, we present the results of $P_{aa}$ as a representative example.

\section{Momentum distribution and spin structure factor involving different sublattices}
\label{app-3}

\begin{figure}[h] 
	\centering
	\begin{subfigure}[b]{0.48\textwidth}
		\includegraphics[width=0.494\textwidth]{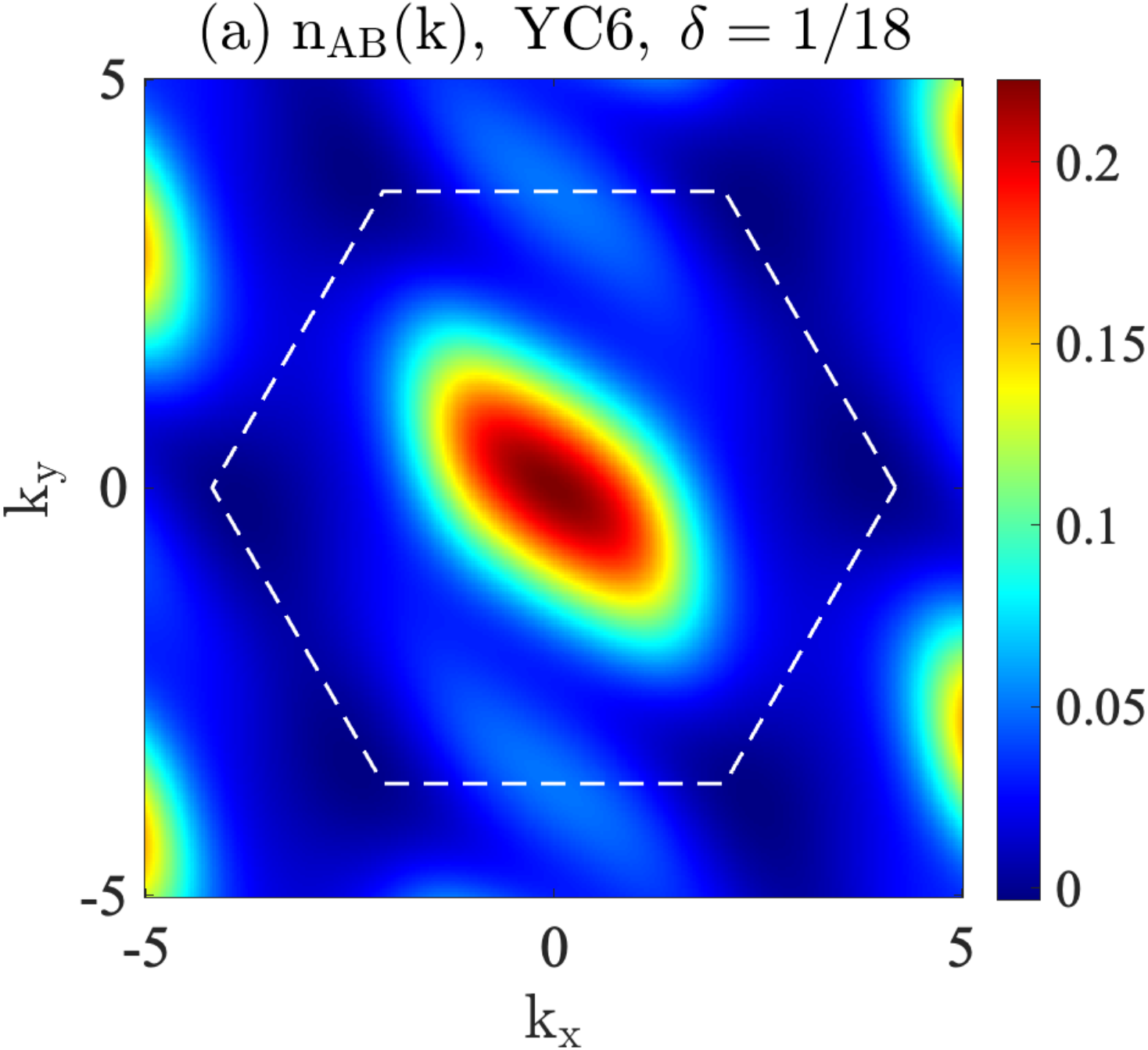} 
		\includegraphics[width=0.494\textwidth]{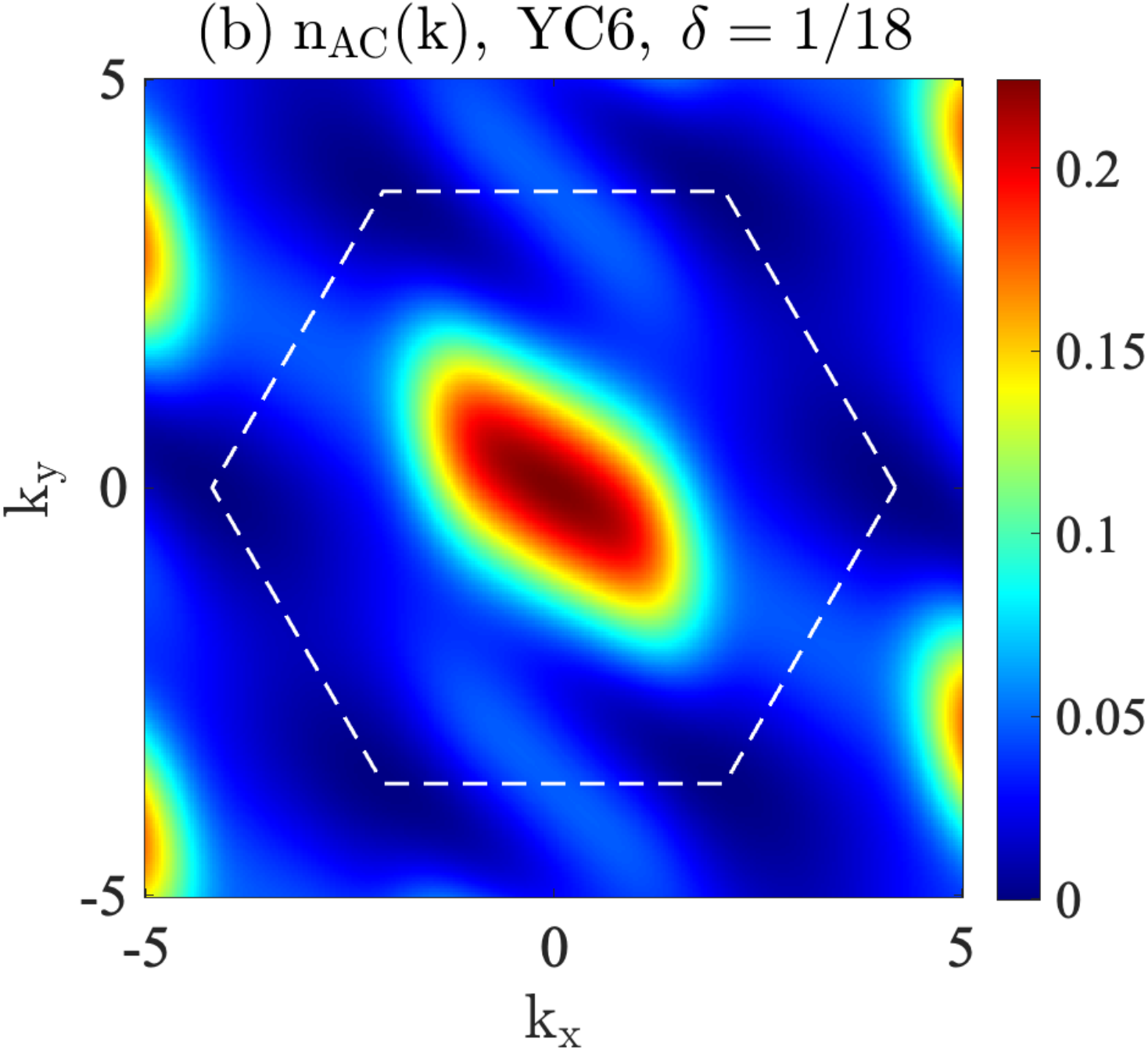} 
	\end{subfigure}
	\begin{subfigure}[b]{0.48\textwidth}
		\includegraphics[width=0.494\textwidth]{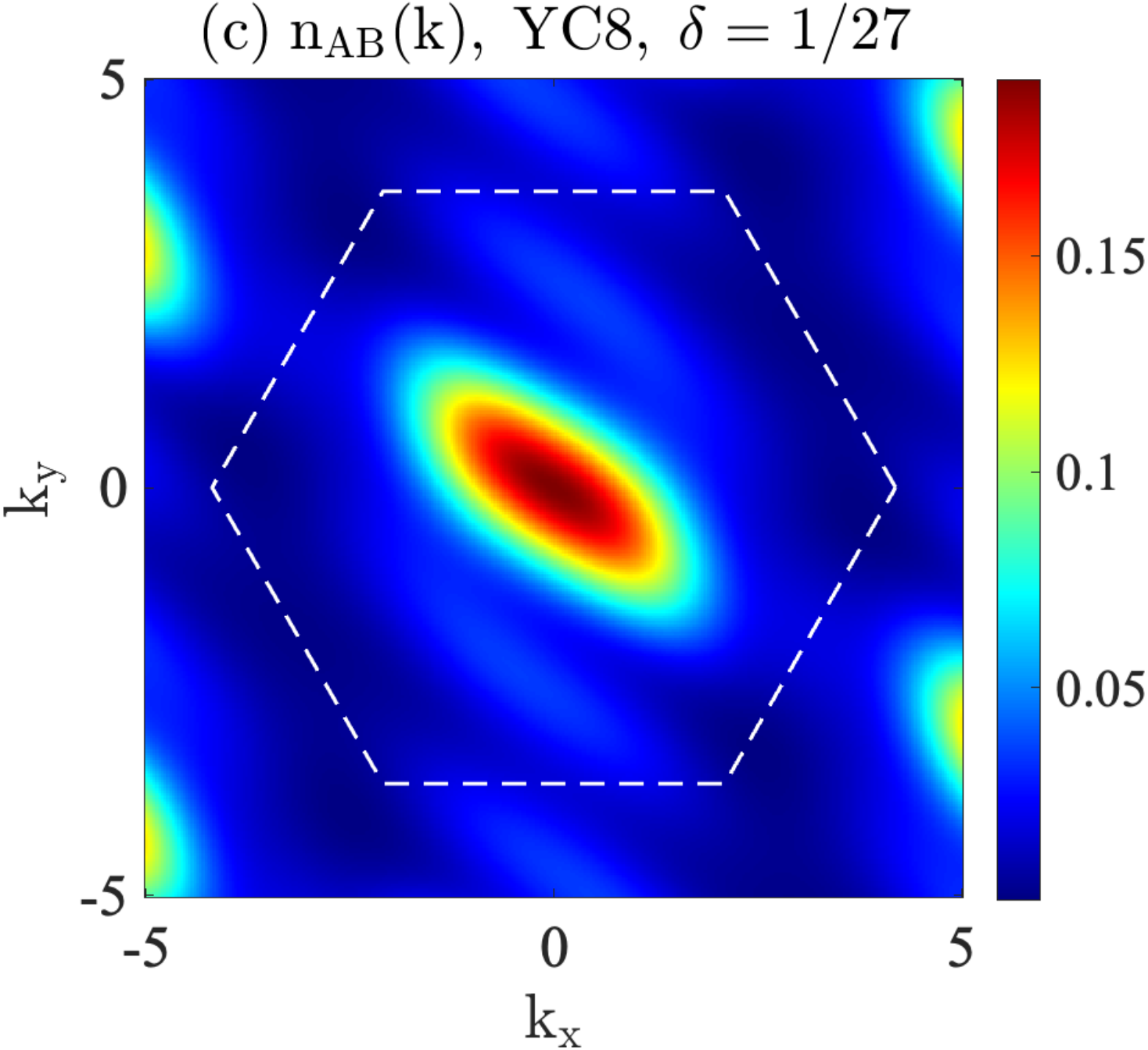} 
		\includegraphics[width=0.494\textwidth]{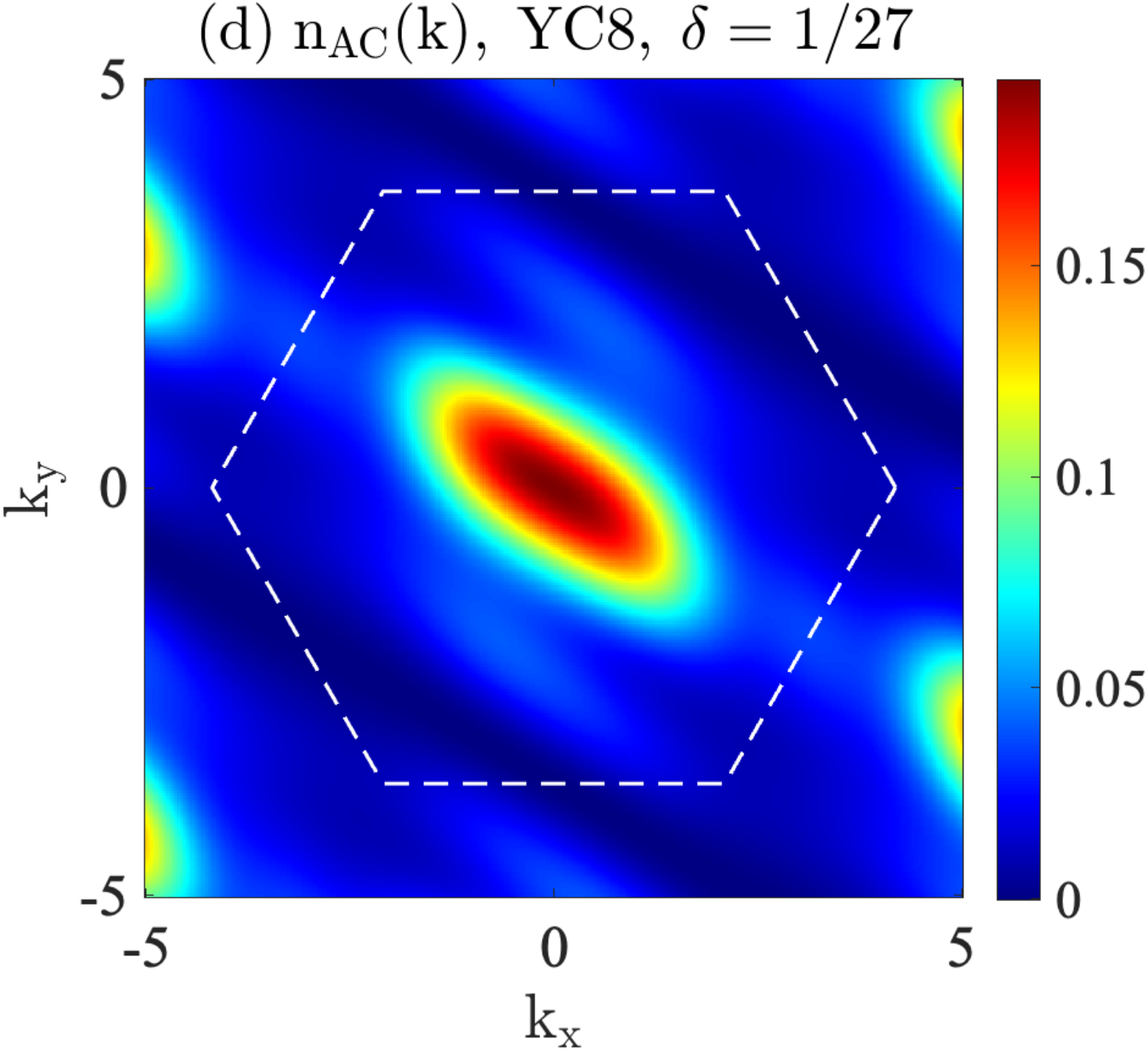} 
	\end{subfigure}
	\caption{\justifying  Momentum distribution functions $n_{AB}(\mathbf{k})$ and $n_{AC}(\mathbf{k})$ for $t_2/t_1 = 0.7,J_2/J_1=0.49$ in the Fermi-liquid-like phase. (a) $n_{AB}(\mathbf{k})$ on the YC6 cylinder at $\delta=1/18$. (b) $n_{AC}(\mathbf{k})$ on the YC6 cylinder at $\delta=1/18$. (c) $n_{AB}(\mathbf{k})$ on the YC8 cylinder at $\delta=1/27$. (d) $n_{AC}(\mathbf{k})$ on the YC8 cylinder at $\delta=1/27$.}
    \label{nk_ABAC}
\end{figure}

\begin{figure}[h] 
	\centering
	\begin{subfigure}[b]{0.48\textwidth}
		\includegraphics[width=0.494\textwidth]{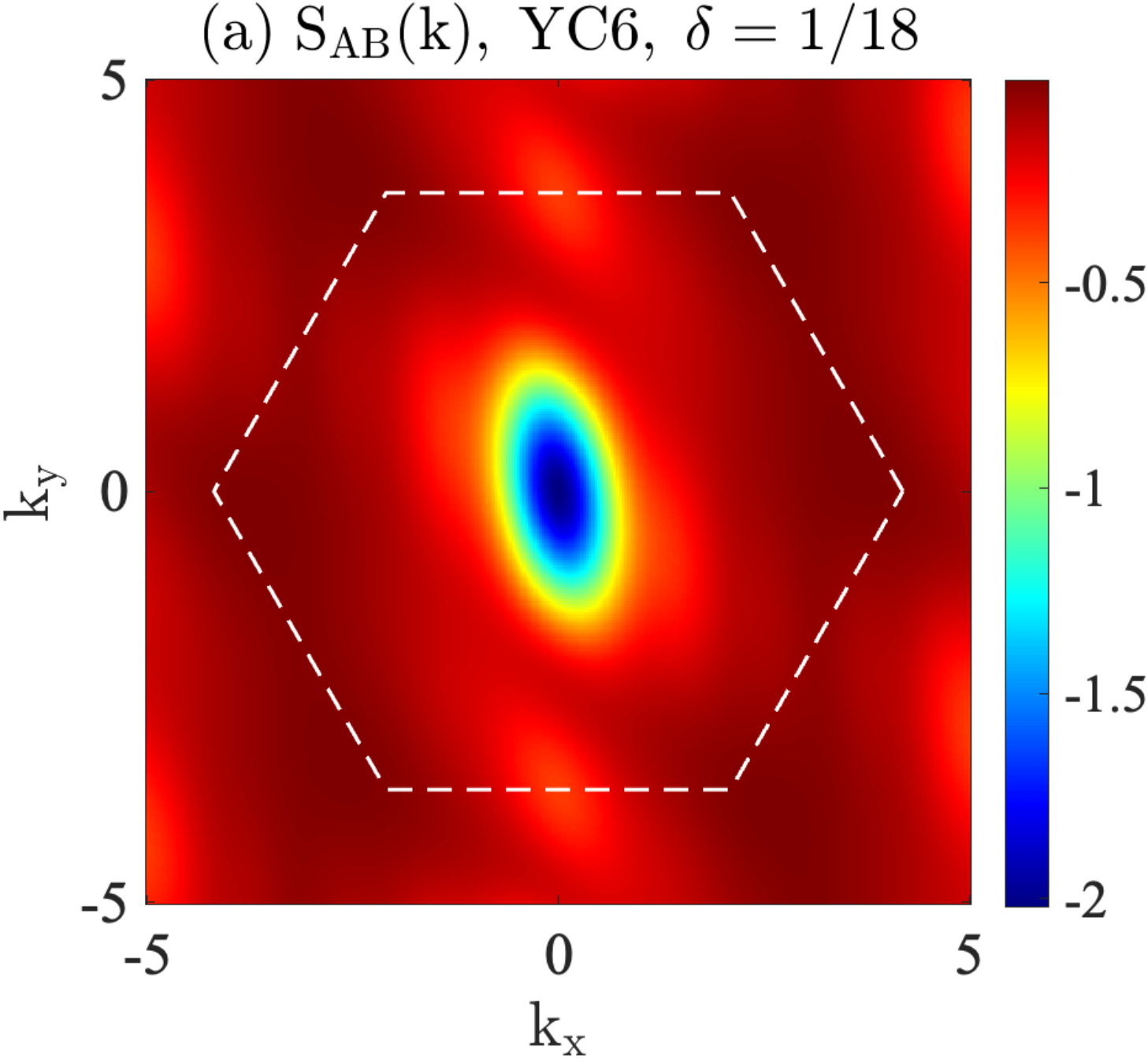} 
		\includegraphics[width=0.494\textwidth]{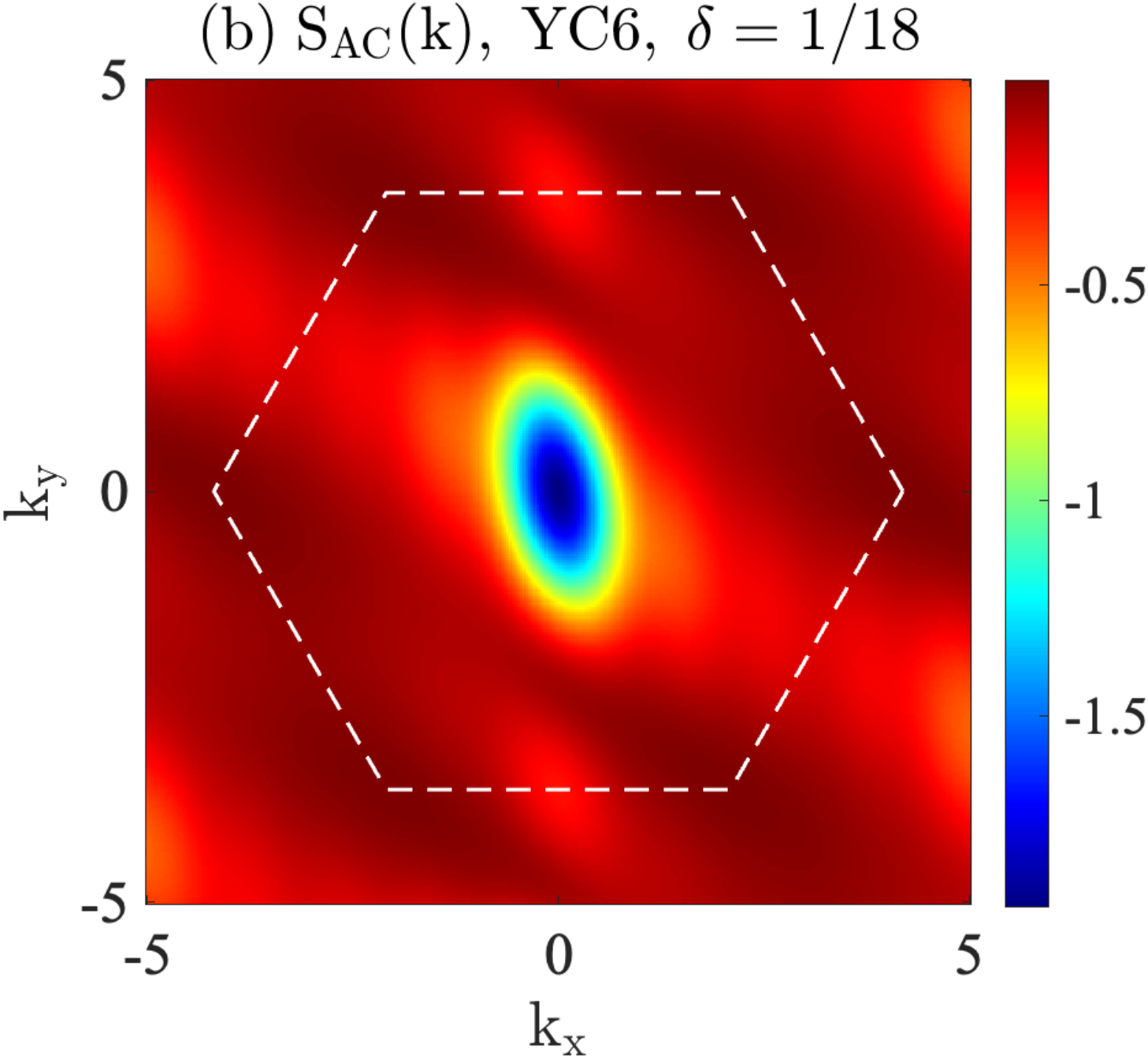} 
	\end{subfigure}
	\begin{subfigure}[b]{0.48\textwidth}
		\includegraphics[width=0.494\textwidth]{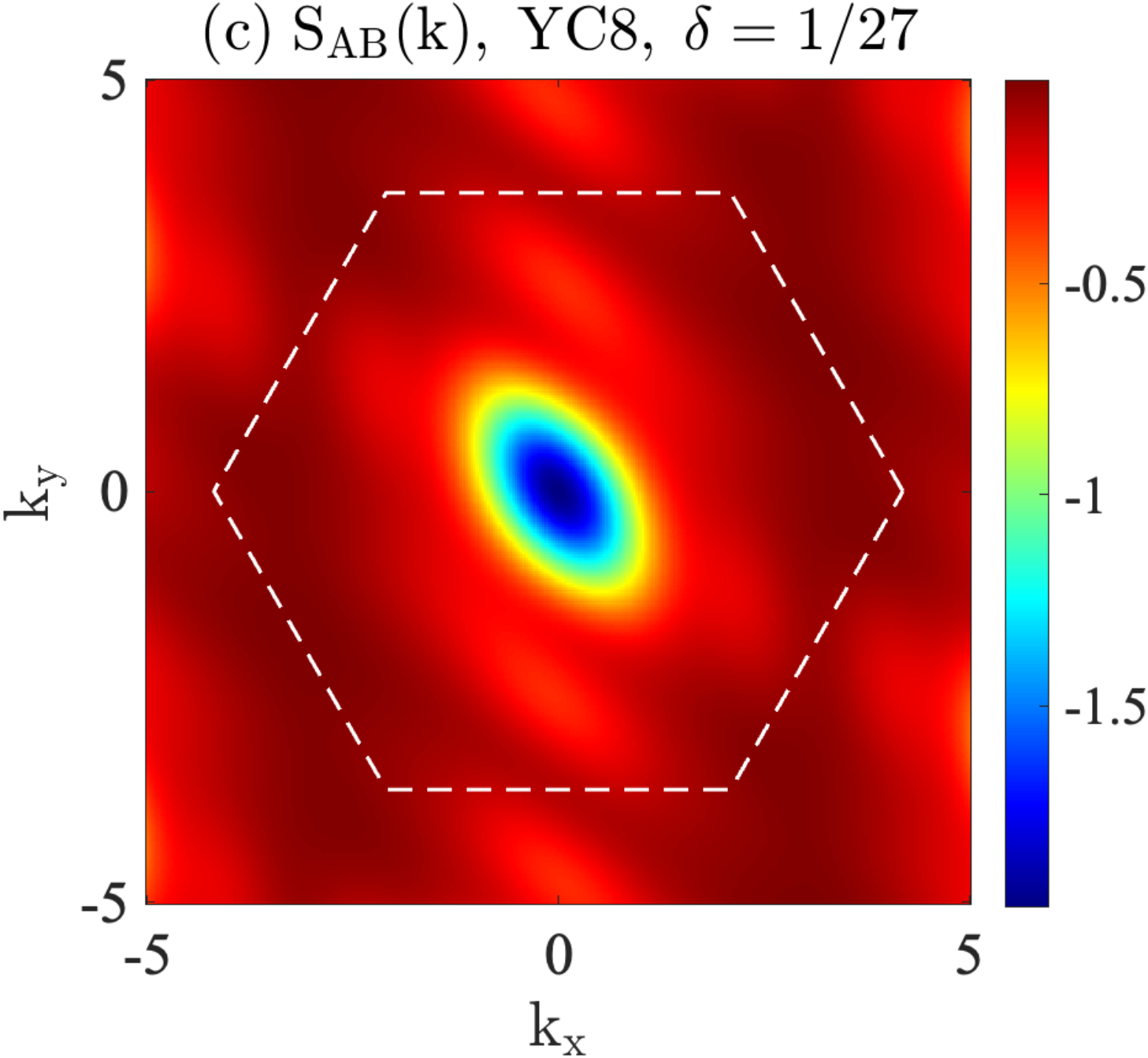} 
		\includegraphics[width=0.494\textwidth]{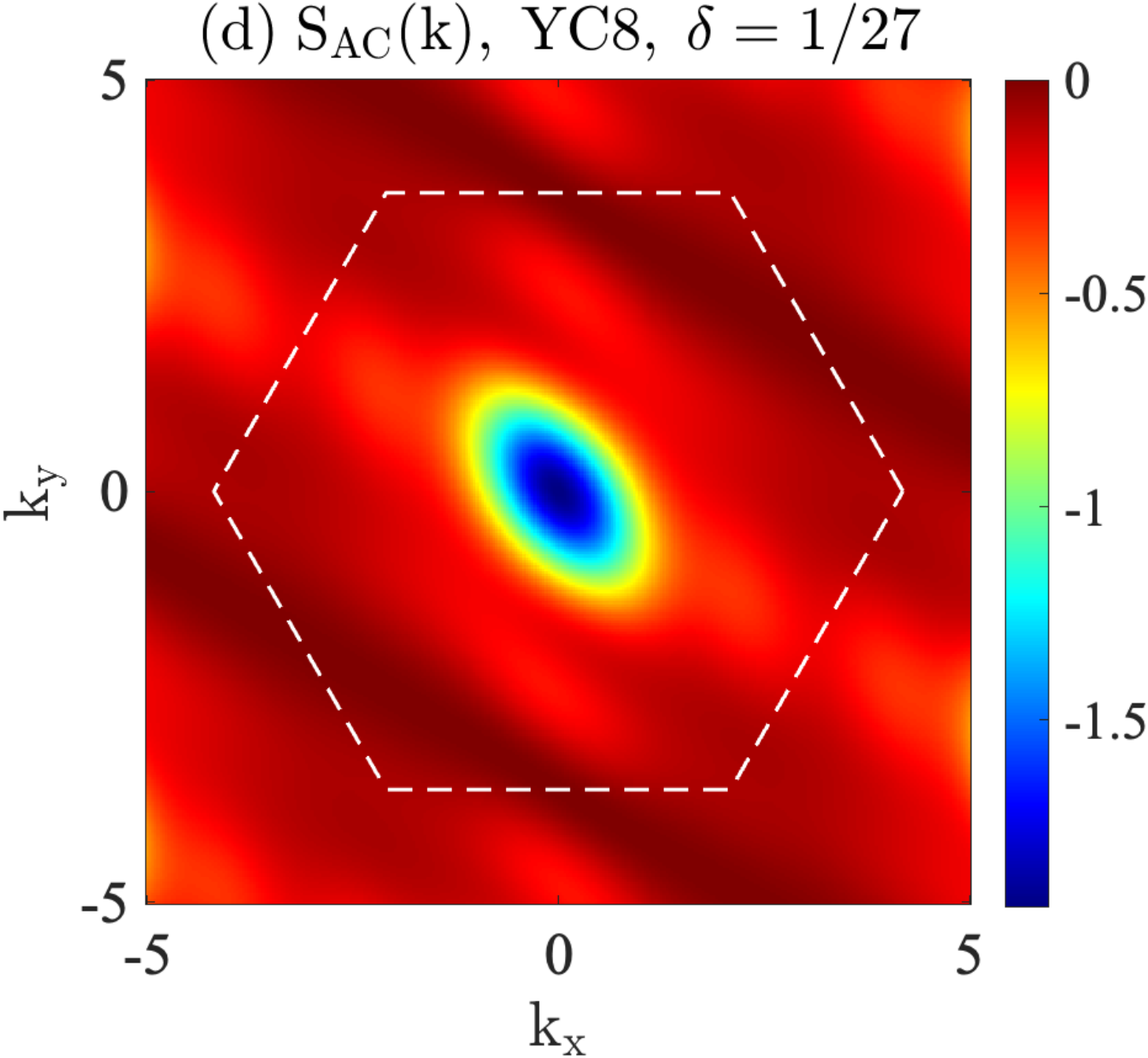} 
	\end{subfigure}
	\caption{\justifying  Spin structure factors $S_{AB}(\mathbf{k})$ and $S_{AC}(\mathbf{k})$ for $t_2/t_1 = 0.7,J_2/J_1=0.49$ in the Fermi-liquid-like phase. (a) $S_{AB}(\mathbf{k})$ on the YC6 cylinder at $\delta=1/18$. (b) $S_{AC}(\mathbf{k})$ on the YC6 cylinder at $\delta=1/18$. (c) $S_{AB}(\mathbf{k})$ on the YC8 cylinder at $\delta=1/27$. (d) $S_{AC}(\mathbf{k})$ on the YC8 cylinder at $\delta=1/27$.}
    \label{Sk_ABAC}
\end{figure}

In this section, we present the results of the momentum distribution functions [Fig.~\ref{nk_ABAC}] and the spin structure factors [Fig.~\ref{Sk_ABAC}] in the Fermi-liquid-like phase, which involve the different sublattices $A,B$ and $A,C$.
The momentum distribution function including $A$ and $B$ sublattices is defined as $n_{AB}(\mathbf{k}) = \frac{1}{\sqrt 2 N_A}\sum_{i,j,\sigma} ( e^{i\mathbf{k} \cdot (\mathbf{r}_i-\mathbf{r}_j)}\langle \hat{c}_{i,\sigma}^{\dagger} \hat{c}_{j,\sigma}\rangle + h.c.)$, and the spin structure factor $S_{AB}$ is given by $S_{AB}(\mathbf{k}) = \frac{1}{\sqrt 2 N_A} \sum_{i,j} ( \langle \mathbf{S}_{i} \cdot \mathbf{S}_{j} \rangle e^{i \mathbf{k} \cdot (\mathbf{r}_i - \mathbf{r}_j)}+h.c. )$, where the sites $i$ and $j$ belong to the $A$ and $B$ sublattices, respectively. 
$N_A$ is the number of the unit cell. 
The definitions of $n_{AC}$ and $S_{AC}$ are similar.
In this Fermi-liquid-like phase, the YC6 and YC8 systems have very consistent results. 
In Fig.~\ref{Sk_ABAC}, the negative peaks of $S_{AB}$ and $S_{AC}$ at the ${\bf \Gamma}$ point are consistent with the three-sublattice spin correlation.

\bibliography{kagome}

@article{kagome-RV-nature_reviews_physics-2023,
	author = {Wang, Yaojia and Wu, Heng and McCandless, Gregory T. and Chan, Julia Y. and Ali, Mazhar N.},
	date = {2023/11/01},
	doi = {10.1038/s42254-023-00635-7},
	id = {Wang2023},
	isbn = {2522-5820},
	journal = {Nature Reviews Physics},
	number = {11},
	pages = {635--658},
	title = {Quantum states and intertwining phases in kagome materials},
	url = {https://doi.org/10.1038/s42254-023-00635-7},
	volume = {5},
	year = {2023}}

@article{kagome-RV-nature-2022,
	author = {Yin, Jia-Xin and Lian, Biao and Hasan, M. Zahid},
	date = {2022/12/01},
	doi = {10.1038/s41586-022-05516-0},
	id = {Yin2022},
	isbn = {1476-4687},
	journal = {Nature},
	number = {7941},
	pages = {647--657},
	title = {Topological kagome magnets and superconductors},
	url = {https://doi.org/10.1038/s41586-022-05516-0},
	volume = {612},
	year = {2022}}

@article{kagome-AV3Sb5-nature_Review_material-2024,
	author = {Wilson, Stephen D. and Ortiz, Brenden R.},
	date = {2024/06/01},
	doi = {10.1038/s41578-024-00677-y},
	id = {Wilson2024},
	isbn = {2058-8437},
	journal = {Nature Reviews Materials},
	number = {6},
	pages = {420--432},
	title = {AV3Sb5 kagome superconductors},
	url = {https://doi.org/10.1038/s41578-024-00677-y},
	volume = {9},
	year = {2024}}

@article{kagome-AV3Sb5-National_Science_Review-2022,
    author = {Jiang, Kun and Wu, Tao and Yin, Jia-Xin and Wang, Zhenyu and Hasan, M Zahid and Wilson, Stephen D and Chen, Xianhui and Hu, Jiangping},
    title = {Kagome superconductors AV3Sb5 (A = K, Rb, Cs)},
    journal = {National Science Review},
    volume = {10},
    number = {2},
    pages = {nwac199},
    year = {2022},
    month = {09},
    issn = {2095-5138},
    doi = {10.1093/nsr/nwac199},
    url = {https://doi.org/10.1093/nsr/nwac199}
}

@article{kagome-Co3Sn2S2-Reviews_in_physics-2022,
    title = {Topological Co3Sn2S2 magnetic Weyl semimetal: From fundamental understanding to diverse fields of study},
    journal = {Reviews in Physics},
    volume = {8},
    pages = {100072},
    year = {2022},
    issn = {2405-4283},
    doi = {https://doi.org/10.1016/j.revip.2022.100072},
    url = {https://www.sciencedirect.com/science/article/pii/S2405428322000041},
    author = {M. Kanagaraj and Jiai Ning and Liang He}}

@article{QSL-RV-Leon-2010,
	author = {Balents, Leon},
	date = {2010/03/01},
	doi = {10.1038/nature08917},
	id = {Balents2010},
	isbn = {1476-4687},
	journal = {Nature},
	number = {7286},
	pages = {199--208},
	title = {Spin liquids in frustrated magnets},
	url = {https://doi.org/10.1038/nature08917},
	volume = {464},
	year = {2010}}

@article{kagome-Co3Sn2S2-sci-2019,
author = {Noam Morali  and Rajib Batabyal  and Pranab Kumar Nag  and Enke Liu  and Qiunan Xu  and Yan Sun  and Binghai Yan  and Claudia Felser  and Nurit Avraham  and Haim Beidenkopf },
title={Fermi-arc diversity on surface terminations of the magnetic Weyl semimetal Co3Sn2S2},
journal = {Science},
volume = {365},
number = {6459},
pages = {1286-1291},
year = {2019},
doi = {10.1126/science.aav2334},
URL = {https://www.science.org/doi/abs/10.1126/science.aav2334}}

@article{kagome-Co3Sn2S2-sci-2019-2,
author = {D. F. Liu  and A. J. Liang  and E. K. Liu  and Q. N. Xu  and Y. W. Li  and C. Chen  and D. Pei  and W. J. Shi  and S. K. Mo  and P. Dudin  and T. Kim  and C. Cacho  and G. Li  and Y. Sun  and L. X. Yang  and Z. K. Liu  and S. S. P. Parkin  and C. Felser  and Y. L. Chen },
title = {Magnetic Weyl semimetal phase in a Kagomé crystal},
journal = {Science},
volume = {365},
number = {6459},
pages = {1282-1285},
year = {2019},
doi = {10.1126/science.aav2873},
URL = {https://www.science.org/doi/abs/10.1126/science.aav2873}}

@article{kagome-Mn3Sn-nature-2015,
	author = {Nakatsuji, Satoru and Kiyohara, Naoki and Higo, Tomoya},
	date = {2015/11/01},
	doi = {10.1038/nature15723},
	id = {Nakatsuji2015},
	isbn = {1476-4687},
	journal = {Nature},
	number = {7577},
	pages = {212--215},
	title = {Large anomalous Hall effect in a non-collinear antiferromagnet at room temperature},
	url = {https://doi.org/10.1038/nature15723},
	volume = {527},
	year = {2015}}

@article{kagome-CsV3Sb5-nature-2021,
	author = {Zhao, He and Li, Hong and Ortiz, Brenden R. and Teicher, Samuel M. L. and Park, Takamori and Ye, Mengxing and Wang, Ziqiang and Balents, Leon and Wilson, Stephen D. and Zeljkovic, Ilija},
	date = {2021/11/01},
	doi = {10.1038/s41586-021-03946-w},
	id = {Zhao2021},
	isbn = {1476-4687},
	journal = {Nature},
	number = {7884},
	pages = {216--221},
	title = {Cascade of correlated electron states in the kagome superconductor CsV3Sb5},
	url = {https://doi.org/10.1038/s41586-021-03946-w},
	volume = {599},
	year = {2021}}

@article{kagome-CsV3Sb5-nature-2021-2,
	author = {Chen, Hui and Yang, Haitao and Hu, Bin and Zhao, Zhen and Yuan, Jie and Xing, Yuqing and Qian, Guojian and Huang, Zihao and Li, Geng and Ye, Yuhan and Ma, Sheng and Ni, Shunli and Zhang, Hua and Yin, Qiangwei and Gong, Chunsheng and Tu, Zhijun and Lei, Hechang and Tan, Hengxin and Zhou, Sen and Shen, Chengmin and Dong, Xiaoli and Yan, Binghai and Wang, Ziqiang and Gao, Hong-Jun},
	date = {2021/11/01},
	doi = {10.1038/s41586-021-03983-5},
	id = {Chen2021},
	isbn = {1476-4687},
	journal = {Nature},
	number = {7884},
	pages = {222--228},
	title = {Roton pair density wave in a strong-coupling kagome superconductor},
	url = {https://doi.org/10.1038/s41586-021-03983-5},
	volume = {599},
	year = {2021}}

@article{kagome-TbMn6Sn6-nature-2020,
	author = {Yin, Jia-Xin and Ma, Wenlong and Cochran, Tyler A. and Xu, Xitong and Zhang, Songtian S. and Tien, Hung-Ju and Shumiya, Nana and Cheng, Guangming and Jiang, Kun and Lian, Biao and Song, Zhida and Chang, Guoqing and Belopolski, Ilya and Multer, Daniel and Litskevich, Maksim and Cheng, Zi-Jia and Yang, Xian P. and Swidler, Bianca and Zhou, Huibin and Lin, Hsin and Neupert, Titus and Wang, Ziqiang and Yao, Nan and Chang, Tay-Rong and Jia, Shuang and Zahid Hasan, M.},
	date = {2020/07/01},
	doi = {10.1038/s41586-020-2482-7},
	id = {Yin2020},
	isbn = {1476-4687},
	journal = {Nature},
	number = {7817},
	pages = {533--536},
	title = {Quantum-limit Chern topological magnetism in TbMn6Sn6},
	url = {https://doi.org/10.1038/s41586-020-2482-7},
	volume = {583},
	year = {2020}}

@article{kagome-Herbertsmithite-RMP-2016,
  title = {Colloquium: Herbertsmithite and the search for the quantum spin liquid},
  author = {Norman, M. R.},
  journal = {Rev. Mod. Phys.},
  volume = {88},
  issue = {4},
  pages = {041002},
  numpages = {14},
  year = {2016},
  month = {Dec},
  publisher = {American Physical Society},
  doi = {10.1103/RevModPhys.88.041002},
  url = {https://link.aps.org/doi/10.1103/RevModPhys.88.041002}
}

@article{kagome-J1-DMRG-Sci-2011,
	author = {Simeng Yan  and David A. Huse  and Steven R. White },
	title={Spin-liquid ground state of the S= 1/2 kagome Heisenberg antiferromagnet},
	journal = {Science},
	volume = {332},
	number = {6034},
	pages = {1173-1176},
	year = {2011},
	doi = {10.1126/science.1201080},
	URL = {https://www.science.org/doi/abs/10.1126/science.1201080},
}

@article{kagome-J1-DMRG-PRL-2012,
  title = {Nature of the Spin-Liquid Ground State of the $S=1/2$ Heisenberg Model on the Kagome Lattice},
  author = {Depenbrock, Stefan and McCulloch, Ian P. and Schollw\"ock, Ulrich},
  journal = {Phys. Rev. Lett.},
  volume = {109},
  issue = {6},
  pages = {067201},
  numpages = {6},
  year = {2012},
  month = {Aug},
  publisher = {American Physical Society},
  doi = {10.1103/PhysRevLett.109.067201},
  url = {https://link.aps.org/doi/10.1103/PhysRevLett.109.067201}
}

@article{kagome-J1-VMC-2007,
  title = {Projected-Wave-Function Study of the Spin-$1/2$ Heisenberg Model on the Kagom\'e Lattice},
  author = {Ran, Ying and Hermele, Michael and Lee, Patrick A. and Wen, Xiao-Gang},
  journal = {Phys. Rev. Lett.},
  volume = {98},
  issue = {11},
  pages = {117205},
  numpages = {4},
  year = {2007},
  month = {Mar},
  publisher = {American Physical Society},
  doi = {10.1103/PhysRevLett.98.117205},
  url = {https://link.aps.org/doi/10.1103/PhysRevLett.98.117205}
}

@article{kagome-J1-VMC-2011,
  title = {Projected wave function study of ${\mathbb{Z}}_{2}$ spin liquids on the kagome lattice for the spin-$\frac{1}{2}$ quantum Heisenberg antiferromagnet},
  author = {Iqbal, Yasir and Becca, Federico and Poilblanc, Didier},
  journal = {Phys. Rev. B},
  volume = {84},
  issue = {2},
  pages = {020407},
  numpages = {4},
  year = {2011},
  month = {Jul},
  publisher = {American Physical Society},
  doi = {10.1103/PhysRevB.84.020407},
  url = {https://link.aps.org/doi/10.1103/PhysRevB.84.020407}
}

@article{kagome-J1-VMC-2013,
  title = {Gapless spin-liquid phase in the kagome spin-$\frac{1}{2}$ Heisenberg antiferromagnet},
  author = {Iqbal, Yasir and Becca, Federico and Sorella, Sandro and Poilblanc, Didier},
  journal = {Phys. Rev. B},
  volume = {87},
  issue = {6},
  pages = {060405},
  numpages = {5},
  year = {2013},
  month = {Feb},
  publisher = {American Physical Society},
  doi = {10.1103/PhysRevB.87.060405},
  url = {https://link.aps.org/doi/10.1103/PhysRevB.87.060405}
}

@article{kagome-J1-VMC-2014,
  title = {Vanishing spin gap in a competing spin-liquid phase in the kagome Heisenberg antiferromagnet},
  author = {Iqbal, Yasir and Poilblanc, Didier and Becca, Federico},
  journal = {Phys. Rev. B},
  volume = {89},
  issue = {2},
  pages = {020407},
  numpages = {5},
  year = {2014},
  month = {Jan},
  publisher = {American Physical Society},
  doi = {10.1103/PhysRevB.89.020407},
  url = {https://link.aps.org/doi/10.1103/PhysRevB.89.020407}
}

@article{kagome-J1-iDMRG-2017,
  title = {Signatures of Dirac Cones in a DMRG Study of the Kagome Heisenberg Model},
  author = {He, Yin-Chen and Zaletel, Michael P. and Oshikawa, Masaki and Pollmann, Frank},
  journal = {Phys. Rev. X},
  volume = {7},
  issue = {3},
  pages = {031020},
  numpages = {16},
  year = {2017},
  month = {Jul},
  publisher = {American Physical Society},
  doi = {10.1103/PhysRevX.7.031020},
  url = {https://link.aps.org/doi/10.1103/PhysRevX.7.031020}
}

@article{kagome-J1J2-Schollwock-2015,
  title = {Phase diagram of the ${J}_{1}\text{\ensuremath{-}}{J}_{2}$ Heisenberg model on the kagome lattice},
  author = {Kolley, F. and Depenbrock, S. and McCulloch, I. P. and Schollw\"ock, U. and Alba, V.},
  journal = {Phys. Rev. B},
  volume = {91},
  issue = {10},
  pages = {104418},
  numpages = {8},
  year = {2015},
  month = {Mar},
  publisher = {American Physical Society},
  doi = {10.1103/PhysRevB.91.104418},
}

@article{kagome-J1J2J3-mean_field-2012,
  title = {Kagome Antiferromagnet: A Chiral Topological Spin Liquid?},
  author = {Messio, Laura and Bernu, Bernard and Lhuillier, Claire},
  journal = {Phys. Rev. Lett.},
  volume = {108},
  issue = {20},
  pages = {207204},
  numpages = {5},
  year = {2012},
  month = {May},
  publisher = {American Physical Society},
  doi = {10.1103/PhysRevLett.108.207204},
  url = {https://link.aps.org/doi/10.1103/PhysRevLett.108.207204}
}

@article{kagome-J1J2J3-gss-2014,
	author = {Gong, Shou-Shu and Zhu, Wei and Sheng, D. N.},
	date = {2014/09/10},
	doi = {10.1038/srep06317},
	id = {Gong2014},
	isbn = {2045-2322},
	journal = {Scientific Reports},
	number = {1},
	pages = {6317},
	title = {Emergent Chiral Spin Liquid: Fractional Quantum Hall Effect in a Kagome Heisenberg Model},
	url = {https://doi.org/10.1038/srep06317},
	volume = {4},
	year = {2014}}

@article{kagome-J1J2J3-ssg-2015,
	title = {Global phase diagram of competing ordered and quantum spin-liquid phases on the kagome lattice},
	author = {Gong, Shou-Shu and Zhu, Wei and Balents, Leon and Sheng, D. N.},
	journal = {Phys. Rev. B},
	volume = {91},
	issue = {7},
	pages = {075112},
	numpages = {9},
	year = {2015},
	month = {Feb},
	publisher = {American Physical Society},
	doi = {10.1103/PhysRevB.91.075112},
	url = {https://link.aps.org/doi/10.1103/PhysRevB.91.075112}
}

@article{kagome-J1J2szJ3sz-heyinchen-2014,
  title = {Chiral Spin Liquid in a Frustrated Anisotropic Kagome Heisenberg Model},
  author = {He, Yin-Chen and Sheng, D. N. and Chen, Yan},
  journal = {Phys. Rev. Lett.},
  volume = {112},
  issue = {13},
  pages = {137202},
  numpages = {5},
  year = {2014},
  month = {Apr},
  publisher = {American Physical Society},
  doi = {10.1103/PhysRevLett.112.137202},
  url = {https://link.aps.org/doi/10.1103/PhysRevLett.112.137202}
}

@article{kagome-J1Jchihz-Ncom-2014,
	author = {Bauer, B. and Cincio, L. and Keller, B. P. and Dolfi, M. and Vidal, G. and Trebst, S. and Ludwig, A. W. W.},
	date = {2014/10/10},
	doi = {10.1038/ncomms6137},
	id = {Bauer2014},
	isbn = {2041-1723},
	journal = {Nature Communications},
	number = {1},
	pages = {5137},
	title = {Chiral spin liquid and emergent anyons in a Kagome lattice Mott insulator},
	url = {https://doi.org/10.1038/ncomms6137},
	volume = {5},
	year = {2014}}

@article{kagome-J1J2J3/J1Jchi-VMC-2015,
  title = {Variational Monte Carlo study of a chiral spin liquid in the extended Heisenberg model on the kagome lattice},
  author = {Hu, Wen-Jun and Zhu, Wei and Zhang, Yi and Gong, Shoushu and Becca, Federico and Sheng, D. N.},
  journal = {Phys. Rev. B},
  volume = {91},
  issue = {4},
  pages = {041124},
  numpages = {5},
  year = {2015},
  month = {Jan},
  publisher = {American Physical Society},
  doi = {10.1103/PhysRevB.91.041124},
  url = {https://link.aps.org/doi/10.1103/PhysRevB.91.041124}
}

@article{kagome-tJ-Jiang-2017,
	title = {Holon Wigner Crystal in a Lightly Doped Kagome Quantum Spin Liquid},
	author = {Jiang, Hong-Chen and Devereaux, T. and Kivelson, S. A.},
	journal = {Phys. Rev. Lett.},
	volume = {119},
	issue = {6},
	pages = {067002},
	numpages = {5},
	year = {2017},
	month = {Aug},
	publisher = {American Physical Society},
	doi = {10.1103/PhysRevLett.119.067002},
	url = {https://link.aps.org/doi/10.1103/PhysRevLett.119.067002}
}

@article{kagome-tJlike-PCheng-2021,
	author = {Peng, Cheng and Jiang, Yi-Fan and Sheng, Dong-Ning and Jiang, Hong-Chen},
	title = {Doping Quantum Spin Liquids on the Kagome Lattice},
	journal = {Advanced Quantum Technologies},
	volume = {4},
	number = {3},
	pages = {2000126},
	doi = {https://doi.org/10.1002/qute.202000126},
	year = {2021}
}

@article{kagome-tJ-VMC-2011,
  title = {Doping on the kagome lattice: A variational Monte Carlo study of the $t$-$J$ model},
  author = {Guertler, Siegfried and Monien, Hartmut},
  journal = {Phys. Rev. B},
  volume = {84},
  issue = {17},
  pages = {174409},
  numpages = {4},
  year = {2011},
  month = {Nov},
  publisher = {American Physical Society},
  doi = {10.1103/PhysRevB.84.174409},
}

@article{kagome-tJ-VMC-2013,
  title = {Unveiling the Physics of the Doped Phase of the $t\ensuremath{-}J$ Model on the Kagome Lattice},
  author = {Guertler, Siegfried and Monien, Hartmut},
  journal = {Phys. Rev. Lett.},
  volume = {111},
  issue = {9},
  pages = {097204},
  numpages = {5},
  year = {2013},
  month = {Aug},
  publisher = {American Physical Society},
  doi = {10.1103/PhysRevLett.111.097204},
}

@article{kagome-tJ-VMC-2021,
	title = {Possible Superconductivity with a Bogoliubov Fermi Surface in a Lightly Doped Kagome U(1) Spin Liquid},
	author = {Jiang, Yi-Fan and Yao, Hong and Yang, Fan},
	journal = {Phys. Rev. Lett.},
	volume = {127},
	issue = {18},
	pages = {187003},
	numpages = {6},
	year = {2021},
	month = {Oct},
	publisher = {American Physical Society},
	doi = {10.1103/PhysRevLett.127.187003},
}

@misc{kagome-tJ-GuZC-2024,
	author={Zheng-Tao Xu and Zheng-Cheng Gu and Shuo Yang},
    title={Global phase diagram of doped quantum spin liquid on the Kagome lattice},
	year={2024},
	eprint={2404.05685},
	archivePrefix={arXiv},
	primaryClass={cond-mat.str-el},
	url={https://arxiv.org/abs/2404.05685}, 
}

@article{DMRG-White-1992,
	author  = {S. R. White},
	title   = {Density matrix formulation for quantum renormalization groups},
	journal = {Physical Review Letters},
	volume  = {69},
	number  = {19},
	pages   = {2863--2866},
	year    = {1992},
	doi     = {10.1103/PhysRevLett.69.2863}
}

@article{Entanglement_entropy_2004,
doi = {10.1088/1742-5468/2004/06/P06002},
year = {2004},
month = {jun},
publisher = {},
volume = {2004},
number = {06},
pages = {P06002},
author = {Pasquale Calabrese and John Cardy},
title = {Entanglement entropy and quantum field theory},
journal = {Journal of Statistical Mechanics: Theory and Experiment}
}

@article{kalmeyer_1987,
  title = {Equivalence of the resonating-valence-bond and fractional quantum Hall states},
  author = {Kalmeyer, V. and Laughlin, R. B.},
  journal = {Phys. Rev. Lett.},
  volume = {59},
  issue = {18},
  pages = {2095--2098},
  numpages = {0},
  year = {1987},
  month = {Nov},
  publisher = {American Physical Society},
  doi = {10.1103/PhysRevLett.59.2095},
  url = {https://link.aps.org/doi/10.1103/PhysRevLett.59.2095}
}

@article{anderson_1987,
author = {P. W. Anderson },
title={The resonating valence bond state in La2CuO4 and superconductivity},
journal = {Science},
volume = {235},
number = {4793},
pages = {1196-1198},
year = {1987},
doi = {10.1126/science.235.4793.1196},
URL = {https://www.science.org/doi/abs/10.1126/science.235.4793.1196}
}

@article{CSL-Laughlin-1988,
  title = {Superconducting Ground State of Noninteracting Particles Obeying Fractional Statistics},
  author = {Laughlin, R. B.},
  journal = {Phys. Rev. Lett.},
  volume = {60},
  issue = {25},
  pages = {2677--2680},
  numpages = {0},
  year = {1988},
  month = {Jun},
  publisher = {American Physical Society},
  doi = {10.1103/PhysRevLett.60.2677},
  url = {https://link.aps.org/doi/10.1103/PhysRevLett.60.2677}
}

@article{CSL-xiaogang-1989,
  title = {Chiral spin states and superconductivity},
  author = {Wen, X. G. and Wilczek, Frank and Zee, A.},
  journal = {Phys. Rev. B},
  volume = {39},
  issue = {16},
  pages = {11413--11423},
  numpages = {0},
  year = {1989},
  month = {Jun},
  publisher = {American Physical Society},
  doi = {10.1103/PhysRevB.39.11413},
  url = {https://link.aps.org/doi/10.1103/PhysRevB.39.11413}
}

@article{CSL-Fisher-1989,
  title = {Anyon superconductivity and the fractional quantum Hall effect},
  author = {Lee, Dung-Hai and Fisher, Matthew P. A.},
  journal = {Phys. Rev. Lett.},
  volume = {63},
  issue = {8},
  pages = {903--906},
  numpages = {0},
  year = {1989},
  month = {Aug},
  publisher = {American Physical Society},
  doi = {10.1103/PhysRevLett.63.903},
  url = {https://link.aps.org/doi/10.1103/PhysRevLett.63.903}
}

@article{vHS-kagome-1,
	author = {Park, Pyeongjae and Ortiz, Brenden R. and Sprague, Milo and Sakhya, Anup Pradhan and Chen, Si Athena and Frontzek, Matthias D. and Tian, Wei and Sibille, Romain and Mazzone, Daniel G. and Tabata, Chihiro and Kaneko, Koji and DeBeer-Schmitt, Lisa M. and Stone, Matthew B. and Parker, David S. and Samolyuk, German D. and Miao, Hu and Neupane, Madhab and Christianson, Andrew D.},
	date = {2025/05/12},
	doi = {10.1038/s41467-025-59460-4},
	id = {Park2025},
	isbn = {2041-1723},
	journal = {Nature Communications},
	number = {1},
	pages = {4384},
	title = {Spin density wave and van Hove singularity in the kagome metal CeTi3Bi4},
	url = {https://doi.org/10.1038/s41467-025-59460-4},
	volume = {16},
	year = {2025}}

@article{vHS-kagome-2,
  title = {Competing electronic orders on kagome lattices at van Hove filling},
  author = {Wang, Wan-Sheng and Li, Zheng-Zhao and Xiang, Yuan-Yuan and Wang, Qiang-Hua},
  journal = {Phys. Rev. B},
  volume = {87},
  issue = {11},
  pages = {115135},
  numpages = {8},
  year = {2013},
  month = {Mar},
  publisher = {American Physical Society},
  doi = {10.1103/PhysRevB.87.115135},
  url = {https://link.aps.org/doi/10.1103/PhysRevB.87.115135}
}

@article{vHS-kagome-3,
  title = {Unconventional Fermi Surface Instabilities in the Kagome Hubbard Model},
  author = {Kiesel, Maximilian L. and Platt, Christian and Thomale, Ronny},
  journal = {Phys. Rev. Lett.},
  volume = {110},
  issue = {12},
  pages = {126405},
  numpages = {5},
  year = {2013},
  month = {Mar},
  publisher = {American Physical Society},
  doi = {10.1103/PhysRevLett.110.126405},
  url = {https://link.aps.org/doi/10.1103/PhysRevLett.110.126405}
}

@article{vHS-kagome-4,
  title = {Electronic instabilities of kagome metals: Saddle points and Landau theory},
  author = {Park, Takamori and Ye, Mengxing and Balents, Leon},
  journal = {Phys. Rev. B},
  volume = {104},
  issue = {3},
  pages = {035142},
  numpages = {20},
  year = {2021},
  month = {Jul},
  publisher = {American Physical Society},
  doi = {10.1103/PhysRevB.104.035142},
  url = {https://link.aps.org/doi/10.1103/PhysRevB.104.035142}
}

@article{Flatband-HTC-1,
  title = {High-temperature surface superconductivity in topological flat-band systems},
  author = {Kopnin, N. B. and Heikkil\"a, T. T. and Volovik, G. E.},
  journal = {Phys. Rev. B},
  volume = {83},
  issue = {22},
  pages = {220503},
  numpages = {4},
  year = {2011},
  month = {Jun},
  publisher = {American Physical Society},
  doi = {10.1103/PhysRevB.83.220503},
  url = {https://link.aps.org/doi/10.1103/PhysRevB.83.220503}
}

@article{Flatband-FQHE,
  title = {High-Temperature Fractional Quantum Hall States},
  author = {Tang, Evelyn and Mei, Jia-Wei and Wen, Xiao-Gang},
  journal = {Phys. Rev. Lett.},
  volume = {106},
  issue = {23},
  pages = {236802},
  numpages = {4},
  year = {2011},
  month = {Jun},
  publisher = {American Physical Society},
  doi = {10.1103/PhysRevLett.106.236802},
  url = {https://link.aps.org/doi/10.1103/PhysRevLett.106.236802}
}

@article{Flatband-Wignercrystal,
	author = {Shayegan, Mansour},
	date = {2022/04/01},
	doi = {10.1038/s42254-022-00444-4},
	id = {Shayegan2022},
	isbn = {2522-5820},
	journal = {Nature Reviews Physics},
	number = {4},
	pages = {212--213},
	title = {Wigner crystals in flat band 2D electron systems},
	url = {https://doi.org/10.1038/s42254-022-00444-4},
	volume = {4},
	year = {2022}}

@article{huang_2022,
  title = {Topological Chiral and Nematic Superconductivity by Doping Mott Insulators on Triangular Lattice},
  author = {Huang, Yixuan and Sheng, D. N.},
  journal = {Phys. Rev. X},
  volume = {12},
  issue = {3},
  pages = {031009},
  numpages = {11},
  year = {2022},
  month = {Jul},
  publisher = {American Physical Society},
  doi = {10.1103/PhysRevX.12.031009},
  url = {https://link.aps.org/doi/10.1103/PhysRevX.12.031009}
}

@article{jiang_2020,
  title = {Topological Superconductivity in the Doped Chiral Spin Liquid on the Triangular Lattice},
  author = {Jiang, Yi-Fan and Jiang, Hong-Chen},
  journal = {Phys. Rev. Lett.},
  volume = {125},
  issue = {15},
  pages = {157002},
  numpages = {6},
  year = {2020},
  month = {Oct},
  publisher = {American Physical Society},
  doi = {10.1103/PhysRevLett.125.157002},
  url = {https://link.aps.org/doi/10.1103/PhysRevLett.125.157002}
}

@article{chen_2025,
  title={Global phase diagram of D-wave superconductivity in the square-lattice tJ model},
  author={Chen, Feng and Haldane, FDM and Sheng, DN},
  journal={Proceedings of the National Academy of Sciences},
  volume={122},
  number={12},
  pages={e2420963122},
  year={2025},
  publisher={National Academy of Sciences}
}

@article{savary_2017,
doi = {10.1088/0034-4885/80/1/016502},
url = {https://dx.doi.org/10.1088/0034-4885/80/1/016502},
year = {2016},
month = {nov},
publisher = {IOP Publishing},
volume = {80},
number = {1},
pages = {016502},
author = {Savary, Lucile and Balents, Leon},
title = {Quantum spin liquids: a review},
journal = {Reports on Progress in Physics}
}

@article{zhou_2017,
  title = {Quantum spin liquid states},
  author = {Zhou, Yi and Kanoda, Kazushi and Ng, Tai-Kai},
  journal = {Rev. Mod. Phys.},
  volume = {89},
  issue = {2},
  pages = {025003},
  numpages = {50},
  year = {2017},
  month = {Apr},
  publisher = {American Physical Society},
  doi = {10.1103/RevModPhys.89.025003},
  url = {https://link.aps.org/doi/10.1103/RevModPhys.89.025003}
}

@article{broholm_2020,
author = {C. Broholm  and R. J. Cava  and S. A. Kivelson  and D. G. Nocera  and M. R. Norman  and T. Senthil },
title = {Quantum spin liquids},
journal = {Science},
volume = {367},
number = {6475},
pages = {eaay0668},
year = {2020},
doi = {10.1126/science.aay0668},
URL = {https://www.science.org/doi/abs/10.1126/science.aay0668}}

@article{jiang_2008,
  title = {Density Matrix Renormalization Group Numerical Study of the Kagome Antiferromagnet},
  author = {Jiang, H. C. and Weng, Z. Y. and Sheng, D. N.},
  journal = {Phys. Rev. Lett.},
  volume = {101},
  issue = {11},
  pages = {117203},
  numpages = {4},
  year = {2008},
  month = {Sep},
  publisher = {American Physical Society},
  doi = {10.1103/PhysRevLett.101.117203},
  url = {https://link.aps.org/doi/10.1103/PhysRevLett.101.117203}
}

@article{liao_2017,
  title = {Gapless Spin-Liquid Ground State in the $S=1/2$ Kagome Antiferromagnet},
  author = {Liao, H. J. and Xie, Z. Y. and Chen, J. and Liu, Z. Y. and Xie, H. D. and Huang, R. Z. and Normand, B. and Xiang, T.},
  journal = {Phys. Rev. Lett.},
  volume = {118},
  issue = {13},
  pages = {137202},
  numpages = {6},
  year = {2017},
  month = {Mar},
  publisher = {American Physical Society},
  doi = {10.1103/PhysRevLett.118.137202},
  url = {https://link.aps.org/doi/10.1103/PhysRevLett.118.137202}
}

@article{lauchli_2019,
  title = {$S=\frac{1}{2}$ kagome Heisenberg antiferromagnet revisited},
  author = {L\"auchli, Andreas M. and Sudan, Julien and Moessner, Roderich},
  journal = {Phys. Rev. B},
  volume = {100},
  issue = {15},
  pages = {155142},
  numpages = {7},
  year = {2019},
  month = {Oct},
  publisher = {American Physical Society},
  doi = {10.1103/PhysRevB.100.155142},
  url = {https://link.aps.org/doi/10.1103/PhysRevB.100.155142}
}

@article{sun_2024,
	author = {Sun, Rong-Yang and Jin, Hui-Ke and Tu, Hong-Hao and Zhou, Yi},
	date = {2024/02/03},
	doi = {10.1038/s41535-024-00627-5},
	id = {Sun2024},
	isbn = {2397-4648},
	journal = {npj Quantum Materials},
	number = {1},
	pages = {16},
	title = {Possible chiral spin liquid state in the S = 1/2 kagome Heisenberg model},
	url = {https://doi.org/10.1038/s41535-024-00627-5},
	volume = {9},
	year = {2024}}

@article{zhu_2018,
    author = {Wei Zhu  and Xiao Chen  and Yin-Chen He  and William Witczak-Krempa},
    title = {Entanglement signatures of emergent Dirac fermions: Kagome spin liquid and quantum criticality},
    journal = {Science Advances},
    volume = {4},
    number = {11},
    pages = {eaat5535},
    year = {2018},
    doi = {10.1126/sciadv.aat5535},
    URL ={https://www.science.org/doi/abs/10.1126/sciadv.aat5535}
}

@article{pinaki_2025,
  title = {Spin-$1/2$ Kagome Heisenberg Antiferromagnet: Machine Learning Discovery of the Spinon Pair-Density-Wave Ground State},
  author = {\DJ{}uri\ifmmode \acute{c}\else \'{c}\fi{}, Tanja and Chung, Jia Hui and Yang, Bo and Sengupta, Pinaki},
  journal = {Phys. Rev. X},
  volume = {15},
  issue = {1},
  pages = {011047},
  numpages = {22},
  year = {2025},
  month = {Mar},
  publisher = {American Physical Society},
  doi = {10.1103/PhysRevX.15.011047},
  url = {https://link.aps.org/doi/10.1103/PhysRevX.15.011047}
}

@article{jiang_2024,
  title = {Ground-state phase diagram and superconductivity of the doped Hubbard model on six-leg square cylinders},
  author = {Jiang, Yi-Fan and Devereaux, Thomas P. and Jiang, Hong-Chen},
  journal = {Phys. Rev. B},
  volume = {109},
  issue = {8},
  pages = {085121},
  numpages = {6},
  year = {2024},
  month = {Feb},
  publisher = {American Physical Society},
  doi = {10.1103/PhysRevB.109.085121},
  url = {https://link.aps.org/doi/10.1103/PhysRevB.109.085121}
}

@article{jiang_2023,
  title = {Superconducting valence bond fluid in lightly doped eight-leg $t\text{\ensuremath{-}}J$ cylinders},
  author = {Jiang, Hong-Chen and Kivelson, Steven A. and Lee, Dung-Hai},
  journal = {Phys. Rev. B},
  volume = {108},
  issue = {5},
  pages = {054505},
  numpages = {12},
  year = {2023},
  month = {Aug},
  publisher = {American Physical Society},
  doi = {10.1103/PhysRevB.108.054505},
  url = {https://link.aps.org/doi/10.1103/PhysRevB.108.054505}
}

@article{kagome_spin_model_zw,
	author = {Zhu, W. and Gong, Shou-Shu and Sheng, D. N.},
	date = {2025/08/26},
	doi = {10.1007/s44214-025-00084-6},
	id = {Zhu2025},
	isbn = {2731-6106},
	journal = {Quantum Frontiers},
	number = {1},
	pages = {11},
	title = {Quantum spin liquids in frustrated Kagome Heisenberg model},
	url = {https://doi.org/10.1007/s44214-025-00084-6},
	volume = {4},
	year = {2025}}

\end{document}